%% file: ms.tex
\documentclass[preprint]{aastex}

\def\teff{T$_{\rm{eff}}$}
\def\teq{T$_{\rm{eq}}$}

\def\etal{{\it et~al.\,}}

\def\mj{$M_{\rm J}\,$}

\def\Dwa{$\,$\uppercase\expandafter{\romannumeral5}$\,$}
\def\mic{$\mu$m$\,$}

\def\sless{\lower2pt\hbox{$\buildrel {\scriptstyle <}
   \over {\scriptstyle\sim}$}}
\def\sgreat{\lower2pt\hbox{$\buildrel {\scriptstyle >}
   \over {\scriptstyle\sim}$}}

\def\apjl{Astrophys.~J.~Letters\ }

\begin{document}

\title{Theoretical Spectra and Atmospheres of Extrasolar Giant Planets}

\author{David Sudarsky\altaffilmark{1}, Adam Burrows\altaffilmark{1}, \&  Ivan Hubeny\altaffilmark{2,3}}

\altaffiltext{1}{Department of Astronomy and Steward Observatory, 
                 The University of Arizona, Tucson, AZ \ 85721}
\altaffiltext{2}{NASA Goddard Space Flight Center, Greenbelt, MD 20771}
\altaffiltext{3}{NOAO, Tucson, AZ 85726}
\begin{abstract}

We present a comprehensive theory of the spectra and atmospheres of irradiated
extrasolar giant planets.  We explore the dependences on stellar type, orbital distance,
cloud characteristics, planet mass, and surface gravity.  Phase-averaged spectra for
specific known extrasolar giant planets that span a wide range of the relevant parameters
are calculated, plotted, and discussed.  The connection between atmospheric composition
and emergent spectrum is explored in detail.  Furthermore, we calculate the effect of stellar insolation
on brown dwarfs.  We review a variety of representative observational techniques and programs 
for their potential for direct detection, in light of our theoretical expectations,
and we calculate planet-to-star flux ratios as a function of wavelength.  Our
results suggest which spectral features are most diagnostic of giant planet atmospheres
and reveal the best bands in which to image planets of whatever physical and
orbital characteristics.

\end{abstract}

\section{Introduction}

In the last seven years, more than 100 extrasolar giant planets (EGPs) 
have been discovered by precision radial-velocity measurements of the telltale
wobble of their central stars\footnote{see J. Schneider's Extrasolar Planet Encyclopaedia at http://www.obspm.fr/encycl/encycl.html
for an up-to-date compilation},
and their discovery constitutes a major turning
point in both planetary science and astronomy.  The unanticipated
variety of these extrasolar planetary systems (most completely unlike our own) 
is drawing an increasing fraction of the world's astronomers into the technological,
theoretical, and observational programs designed to study and understand them in more detail.  

Because the indirect Doppler technique does not reveal the
inclination ($i$) of the planet's orbit, only the minimum or projected mass
of the planet (M$_p$sin($i$)) can be inferred.  Currently,
the M$_p$sin($i$)s for the known EGPs range from less than half a Saturn mass to
over 10 Jupiter masses (\mj). One object, HD 209458b, transits its primary,
revealing its edge-on orbit and, hence, its mass of 0.69\mj.  However,
with the majority of EGPs we are not so fortunate and accurate astrometric
measurements will be needed to obtain orbital inclinations.

The orbital distances (semi-major axes) of the known EGPs 
cover the wide range from 0.038 AU to
$\sim$5.5 AU, a range that is expected to expand outward as detection
techniques and instrumentation improve.  Orbital eccentricities
are near zero for close-in objects, as expected due to the tidal dissipation
of their orbits.  But for EGPs at larger orbital distances,
eccentricities are spread from $\sim$0.03 to $\sim$0.9.  The most eccentric
known EGP, HD 80606b, has an eccentricity of 0.927.

The central stars around which giant planets have been found range from main
sequence spectral type F7 to M4, but due to the 
biased nature of the radial-velocity searches most are of spectral type G or K.  
The metallicities of the primaries vary significantly, with [Fe/H] ranging from $\sim$ -0.6 to +0.4;
the majority are above solar metallicity ([Fe/H] = 0).  

Although the radial-velocity programs have launched the modern study of EGPs,  
it is only by making direct measurements of the planet's spectrum 
that its physical characteristics can be fully characterized.  Hence, the next
stage after the initial ground-breaking Doppler campaigns in the study
of EGPs will be their direct imaging and spectroscopic measurement. 
This paper has been written to support this burgeoning effort.
The theoretical modeling of EGP atmospheres is important for at least
two major reasons. First, since the emergent spectra of EGPs are
determined by the chemistry and physics of their outer atmospheres,
when direct detection of EGPs is achieved and spectra are obtained,
theoretical models will be essential in the
interpretation of the data and in the extraction of essential physical
information such as radius, gravity, temperature, and composition.
Second---and this is the
more pressing of the two---theoretical spectral models
are important in guiding current and upcoming direct EGP searches.  Observers need
to know which regions of the spectrum provide the greatest chance
for detection, while avoiding those regions in which attempts are
likely to be futile.

A host of exciting ground-based and space-based missions for the detection,
characterization, and imaging of EGPs are well underway
or under serious consideration (\S\ref{ch_future}).  Ground-based interferometric
endeavors include astrometry with the VLT Interferometer
\citep{Paresce01}, astrometry and differential phase methods with the
Keck Interferometer \citep{vanBelleVasisht98, AkesonSwain00},
and nulling interferometry with the LBT Interferometer
\citep{Hinz01}.
Ground-based differential direct imaging with ARIES on
the MMT \citep{Freed02} and with TRIDENT \citep{Marois02} will
attempt to ``subtract out'' starlight by differencing images in two
or more narrow bands in which only the planetary spectrum is expected
to differ substantially.  Such imaging techniques may 
work well in conjunction with adaptive optics systems on large
ground-based telescopes. 

EGP masses themselves, rather than just lower limits, can be determined 
if transits are detected.  Many ground-based transit surveys are in progress, 
such as STARE \citep{BrownCharbonneau99},
OGLE \citep{Udalski02}, and TEMPEST \citep{BaliberCochran01}.  Space-based instruments,
such as Kepler \citep{Koch98}, MOST \citep{Matthews01}, COROT
\citep{AntonelloRuiz02}, and MONS \citep{Christensen00} will be able to detect
transits with very high photometric precision.  Additionally, such sensitive
instruments should be able to discern reflected light from EGPs that
do $not$ transit their primaries, depending upon the details of the
planetary phase functions.  The eventual
microarcsecond astrometry of the Space Interferometry Mission (SIM)
\citep{UnwinShao00} could
reveal the orbital inclinations and, hence, the masses of
virtually all known EGPs within the next decade.
Perhaps of even greater import are
space-based coronagraphic imaging missions, which may
include $Eclipse$ \citep{Trauger00, Trauger01}, the Jovian Planet Finder
\citep{Clampin01}, the Extra-Solar Planet Imager \citep{Melnick01}, 
and NIRcam on the NGST \citep{Rieke02,Boccaletti99}.  Such advanced
instruments are specifically designed for high-contrast imaging and/or
spectrophotometry of EGPs.  These missions should provide fundamental
EGP atmospheric data, which in combination with theoretical models
is certain to advance our understanding of these systems.

While there are many similarities with brown dwarfs,
external stellar irradiation can have an overwhelming
influence on EGP atmospheres, particularly for the very close-in
objects, while most known brown dwarfs
experience little external radiation.
At the very least, the heating of an EGP atmosphere by such irradiation
alters its equilibrium chemical composition and, hence, its
emergent spectrum.  Furthermore, the short-wavelength spectrum of an EGP will
differ from that of a brown dwarf due to the reflection of starlight by condensates
in its outer atmosphere and/or due to Rayleigh scattering, phenomena largely
responsible for the visible and near-infrared spectra of the giant
(and other) of our Solar System.
While one cannot cleanly split the emergent spectrum of an
EGP into reflected light and emitted light,
the vast majority of the flux ($F$) in the visible region
for an EGP of moderate to large orbital distance is reflected starlight.
Distant EGPs are not hot enough to emit large thermal fluxes in this region
of the spectrum.

Early investigations of EGP atmospheres and spectra neglected thermal emission 
and concentrated on albedo and reflection spectra for a range of EGP
compositions, using planar radiative transfer codes.
Marley \etal (1999), utilizing temperature-pressure (T-P) profiles of isolated
brown dwarfs, explored some of the major controls on EGP spectra, such as Rayleigh
scattering in the UV/blue, molecular absorption in the infrared, and the
sensitivity of the reflected spectra to cloud particle sizes and vertical
extent.  Sudarsky \etal (2000) used both isolated and more nearly isothermal
profiles to bracket the effects of compositional differences on the albedo
and reflection spectra of EGPs.
Albedo and reflection models provide useful spectral information for the
full range of EGPs, although they are most accurate for objects at
moderate to large orbital distances.

The radiative-convective equilibrium modeling
of irradiated planets began with Jupiter and Saturn in the 1960s.  Although
the earliest theoretical work actually
ignored the effects of radiation on thermal structure, Trafton (1967)
broke ground with non-gray, radiative-convective equilibrium models.
Subsequently, improved computational capability and $Pioneer$
and $Voyager$ data allowed for more detailed models of Jupiter and
Saturn (Cess and Khetan 1973; Cess and Chen 1975; Appleby and Hogan 1984).   
The first fully self-consistent atmosphere
models of close-in irradiated EGPs
were by Seager \& Sasselov (1999) and Goukenleuque \etal
(2000) using modified planar stellar atmosphere codes to model 51 Peg b.  With incoming fluxes at
both boundaries (the outer due to the central star and the inner due
to the internal luminosity of the EGP), the T-P profile is adjusted
iteratively until the net flux at every depth zone is the same (or nearly
so).  Relative to a brown dwarf atmosphere of similar temperature, the
resulting profile is more isothermal due to outer atmospheric heating
by the incident stellar flux.  Seager \& Sasselov explored the differences
between cloud-free and dusty atmospheres and used basic equilibrium gas
abundances without the alkali metals.  A similar approach was taken by
Goukenleuque \etal, except that their deep silicate cloud layer acted as
a reflecting layer at the lower boundary; radiative processes
within and below the cloud were not considered.  

Additional self-consistent modeling of EGPs was completed by
Seager, Whitney, and Sasselov (2000, SWS) and Barman, Hauschildt, and
Allard (2001, BHA).  SWS used a 
planar atmosphere code to model close-in EGPs, as in
their previous work, but they also produced theoretical optical
photometric light curves and polarization curves
using phase functions from their separate
3-D Monte Carlo calculations.  The effects of various condensates and
particle sizes on the variability were investigated.  BHA used an
atmosphere code to model a range of EGP orbital distances from generic
close-in objects to those at 1 AU.  Both cloud-free and dusty atmospheric
compositions were explored.  Both groups assumed condensate layers with
infinite scale heights (i.e. extending vertically to the lowest pressures
of their models).  Such models provide upper limits to the effects of
condensates on the emergent spectra of EGPs, and along with the cloud-free
models, they likely bracket the actual emergent spectra.

Unlike brown dwarfs, whose temperatures are not expected to vary
significantly as a function of latitude or longitude, EGPs have a
day side and a night side.  Hence, an EGP atmospheric T-P profile may
be a strong function of the angle from the substellar point, particularly
for slow-rotating or close-in objects.  This dependence
will be a function of the planet's rotational period, although
a proper understanding would require a radiative-hydrodynamic
treatment, since winds will transport material from the
higher-temperature day side to the night side of the planet
\citep{Cho02, ShowmanGuillot02}.  It is also worth noting that
the latitudinal T-P variation of Jupiter is minimal
because the incident Solar energy penetrates into the convective
regions.  While
such complex meteorology is beyond the scope of the present work,
it is possible to compute the equilibrium temperature (\teq) of the planet,
the effective temperature it would have in the $absence$ of an intrinsic luminosity
(inner boundary flux),
based on a few parameters: the stellar luminosity ($L_*$),
the Bond albedo of the planet ($A_B$), and the orbital distance of
the planet ($a$).  The result is:
\begin{equation}
T_{\textrm{eq}} = \left[{(1-A_B)L_{*}\over {16\pi\sigma a^{2}f}}\right]^{1/4}
\label{equilequation}
\end{equation}
where $\sigma$ is the Stefan-Boltzmann constant and the factor $f$ is equal to
unity for re-emission by the planet over
its entire surface or equal to 0.5 for re-emission from only the
illuminated hemisphere \citep{Saumon96}.  This factor may tend toward unity for a
planet that is rotating very quickly if the heat is
well redistributed.  For a slow rotator, this factor might
be closer to 0.5 since re-radiation from
the day side would be significantly greater than that from the
night side.

The spectra of EGPs strongly depend upon their outer
atmospheric chemical compositions, which in turn depend upon the
run of temperature and pressure with atmospheric depth.  Because
of qualitative similarities in the compositions and spectra of
objects within several broad $T_{\textrm{eq}}$ ranges, EGPs fall naturally
into groups, or composition classes \citep{Sudarsky00}.  Since $T_{\textrm{eq}}$
is a function of stellar luminosity and orbital distance 
(eq. \ref{equilequation}), an EGP's class will depend
upon these quantities as well.  
Such a classification scheme, however preliminary,
brings a degree of order to the rich variety of EGP systems known
to exist today.

A ``Jovian'' class of EGPs (Class I) exists at low temperatures
($\sless$ 150 K).
Such planets are characterized by tropospheric ammonia clouds
and strong gaseous methane absorption and include Jupiter and Saturn themselves. 
Somewhat warmer Class II, or ``water class'' EGPs 
are affected by condensed H$_2$O, as well as water and methane vapor.
When the outer atmosphere of an EGP is too hot for water to condense,
radiation generally penetrates more deeply. 
In these objects, designated Class III, ``clear'' or ``gaseous,''
due to a lack of
condensation in their outer atmospheres, ro-vibrational molecular
absorption is very effective.
In those EGPs with exceedingly small orbital
distance and outer atmospheric temperatures in excess of $\sim$900 K
(Class IV), strongly pressure-broadened resonance
lines of sodium and potassium are expected to dominate the visible
spectrum, while molecular absorption in the infrared remains strong.  
The very hot Class V EGPs
orbit even closer to their primaries ($\sim$0.05 AU) than Class IV EGPs.
These extremely close-in ``roasters'' are hot enough and/or have a
low enough surface gravity that silicate and/or iron clouds condense
high in the outer atmosphere and can have substantial effects on the
reflected spectrum.

The present work is a detailed study of EGP atmospheres and spectra,
including individual models representative of the full range of systems known
today.  Using a self-consistent planar atmosphere code
\citep{HubenyLanz95} along with the latest atomic and molecular
cross sections, cloud models, Mie theory treatment of grain
scattering and absorption, and incident stellar fluxes
\citep{Kurucz94} (themselves a function of orbital 
distance), we produce an extensive set of theoretical EGP
model atmospheres and emergent spectra.

In \S\ref{ch_numerical}, we begin our study of EGP spectra 
and atmospheres with a discussion of atmosphere
modeling, including numerical techniques and assumptions.
In \S\ref{ch_composition}, we review some basic equilibrium
chemistry and explore the compositions of EGPs,
discussing gas-phase species as well as condensates.
In \S\ref{opacity}, we discuss some of the relevant absorption and scattering opacities.
Section \ref{ch_classes} begins our presentation of EGP model atmospheres
and spectra.  In particular, this section defines and describes the
five different classes of EGPs, from the cold Class I Jovian-types to
the hot ``roasters'' with orbital periods of only a few days,
and presents their generic temperature-pressure structures and emergent spectra.
In \S\ref{sec_genericseq}, we present some systematic results as a function of orbital distance,
inner flux boundary condition, surface gravity, and cloud particle size.
This section provides a sense of the overall parameter dependences
of the irradiated EGP spectra and atmospheres.
Section \ref{ch_specific} is an extensive discussion of a number of
representative, known EGP systems.  Atmosphere models, emergent spectra, and
planet-to-star flux ratios, of central importance to any campaign
to detect EGPs directly, are provided.
This is followed in \S\ref{ch_future} by a discussion of current and future
ground-based and space-based missions,
in light of our theoretical results.
We conclude in \S\ref{sec_conclusion} with a summary 
of our most important findings.

\section{Numerical Techniques for Modeling Extrasolar Giant Planets} 
\label{ch_numerical}

Numerical atmosphere modeling entails the combination of a radiative transfer
technique with atmospheric structure and boundary constraints.  Because
the atmospheric temperature-pressure structure is almost always strongly
coupled to the transfer of radiation, most methods converge on a
self-consistent solution through an iterative process.  The minimum
requirements for a converged model include the satisfaction of hydrostatic
equilibrium and a constant net flux of radiation throughout the atmosphere
(constant to within some very small fraction; we choose 0.1\%).

To calculate fully consistent, radiative equilibrium atmospheres
and spectra we have employed the computer program TLUSTY (Hubeny 1988; 
Hubeny \& Lanz 1995).  TLUSTY uses the very efficient Complete 
Linearization/Accelerated Lambda Iteration (CL/ALI)
hybrid method (Hubeny \& Lanz 1995) that combines the 
advantages of Complete Linearization  
(Auer \& Mihalas 1969), where only a few iterations are 
needed to obtain a converged model, with the ALI 
scheme's modest time requirements per iteration (Hubeny 1992).

TLUSTY allows for several radiative transfer formal 
solvers - the original Feautrier method (Feautrier 1964; 
Mihalas 1978), the 4-th order variant (Auer 1976), 
and the Discontinuous Finite Element (DFE) scheme 
(Castor, Dykema, \& Klein 1992).  However, we have found that 
for irradiated EGPs the DFE method is preferable.  In 
the case of strong irradiation, the Feautrier 
scheme can often lead to negative emergent fluxes at the 
shortest wavelengths.  This is because the Feautrier 
method does not evaluate the specific intensity  
$I(\nu,\mu)$ itself, but rather its symmetric part, 
$j(\nu,\mu) = [I(\nu,\mu)+ I(\nu,-\mu)]/2$.
At the surface, the emergent flux is evaluated as
\begin{equation}
H^{\rm out}(\nu) = \int_0^1 j(\nu,\mu) \mu \,d\mu -
(1/2) \int_0^1 I^{\rm ext}(\nu,-\mu) \mu \, d\mu]
\equiv \int_0^1 j(\nu,\mu) \mu \,d\mu - H^{\rm ext}(\nu)  \, .
\end{equation}
If the thermal source function at the surface, $S_\nu(0)$, 
is very small compared to the incoming radiation, the 
emergent flux is given as a difference between two 
quantities of almost the same value; the difference thus becomes 
quite inaccurate. If the ratio $S_\nu(0)/I^{\rm ext}(\nu)$ is 
smaller than machine accuracy (roughly $10^{-15}$), 
which may indeed happen at optical wavelengths for
EGPs irradiated by a solar-type star, in evaluating $j(\nu,\mu)$
using the Feautrier scheme the difference is completely 
dominated by rounding errors. In contrast, the DFE method 
is a first-order method that evaluates the specific 
intensity $I(\nu,\mu)$ directly and is not vulnerable 
to such rounding errors.

Because the outer atmospheres of brown dwarfs and EGPs are thin compared
with the radii of these objects, we use a planar geometry for modeling both
isolated and irradiated objects.  A one-dimensional planar atmosphere
code is the simplest means for the self-consistent modeling of EGPs.
Unlike an isolated brown dwarf, it is
likely that the atmospheric temperature-pressure structure of an EGP
will vary substantially as a function of latitude and longitude.
While a planar code cannot account for such variation, an approximate
means of accounting for the different amounts of external flux received
by a planet along its surface is to take a suitable average.  No
such adjustment would be necessary if it were a plane that was being
irradiated.  In that case, the total power received by the
plane would simply be $\pi R^2F$, where $F$ is the incident flux and
$R$ is the radius of the plane (or disk).  If this same flux were
received instead by a hemisphere, the energy would be distributed unevenly
over an area of 2$\pi R^2$.  In this case, the average energy received per unit
area would be $F/2$, not $F$.  When using a planar code, one can
account for this difference by weighting the incident flux by a
factor of 1/2.  Finally, if we want the average energy per time per unit
area received by the full sphere (i.e. giving equal weighting to the night
side of the planet), the adjustment is $F/4$.  Such ``redistribution'' over
the full sphere may be appropriate for a very fast rotator.  Unless
stated otherwise, in this work we choose to redistribute the incident
energy over the front hemisphere of the planet only.  Some
researchers have chosen to redistribute over the full sphere, while others
have chosen not to redistribute at all.  We will point out these differences
when comparing our models with those of others.  Furthermore, the spectra
we show are the so-called ``phase-averaged" spectra for which we distribute
uniformly in solid angle the total energy reradiated and coming from the 
inner boundary.  In this way, energy $in$ equals energy $out$ and energy is conserved. The
phase functions, however, are not calculated and we defer this to another paper. 
Since the variation with phase is generally smaller than the variation from model to model
(SWS) and as a function of wavelength, and since the inclination angles 
of the orbits of most known EGPs are not known, given our goal of mapping out the entire 
model space for irradiated EGPs, the choice of phase-averaged spectra is convenient and sensible. 

There is a rich history of stellar atmosphere modeling, and these same
basic techniques are used for EGPs and brown dwarfs.  But unlike most
stars, save the latest M-dwarfs, the outer atmospheres of substellar
objects commonly contain condensed species, such as grains in relatively
hot objects or water or ammonia clouds in cool EGPs.  The usual treatment
of the scattering of radiation in stellar atmospheres assumes forward-backward
scattering symmetry (and often simply isotropic scattering), but condensates
in EGPs and brown dwarfs typically scatter strongly in the forward direction.
Relative to isotropic scattering, forward scattering reduces the albedo
of an atmosphere, because on average more scattering events would be required
for a photon to emerge from the atmosphere before being absorbed.
A useful approximation that
allows one to maintain symmetric scattering numerically is to account for
the asymmetry in the scattering of radiation
by reducing the actual scattering cross section by $1-g$, where $g$
is the averaged cosine of the scattering angle (Chamberlain \& Hunten 1987).
In an optically
thick medium, this so-called ``transport cross section'' mimics the
effects of asymmetric scattering quite well \citep{Sudarsky00} and we use it 
in the calculations we present in this work.

\section{Atmospheric Compositions}
\label{ch_composition}

The emergent spectra of EGPs are determined mainly
by their outer atmospheric chemical compositions.  But unlike
stars, whose atmospheres are dominated by gaseous species (grain
formation occurs only in the coolest M dwarfs) most EGPs
also contain condensed species that contribute
substantially to the opacities.  Some condensates, such
as water ice or iron grains, are formed homogeneously (i.e. from a single
species), while the formation of others such as forsterite
or gehlenite is heterogeneous, resulting in the depletion
of several gas-phase species.  The largely unknown meteorology in
EGP atmospheres and condensate sedimentation, or
``rainout,'' makes the depth-dependent composition modeling
of EGPs a difficult task.  

Any method useful in determining compositions of EGP atmospheres must
be capable of treating the gaseous,
liquid, and solid phase species simultaneously.  Additionally,
the rainout of species must be handled appropriately,
because the removal of a given species to
deeper layers by such sedimentation can
alter chemical abundances drastically.
We use results from the chemical equilibrium code of Burrows and Sharp (1999),
which includes a prescription to account for the rainout of species in
a gravitational field.
Our fiducial models assume an Anders and Grevesse (1989) solar
composition of 27 elements (H, He, Li, C, N, O, F, Ne, Na, Mg,
Al, Si, P, S, Cl, Ar, K, Ca, Ti, V, Cr, Mn, Fe, Co, Ni, Rb, Cs),
resulting in approximately 300 molecular gas-phase species and
over 100 condensates.  From this large set of species,
only some have the requisite abundances to have significant effects on the emergent
spectra and/or temperature-pressure (T-P) structure.

\subsection{Chemical Abundances in EGPs and Brown Dwarfs}
\label{chemequil}

The equilibrium gas-phase and condensed-phase abundances of
species at a given pressure, temperature, and elemental
composition are obtained by minimizing the total free energy
of the system.  The abundances of
most species are highly dependent on temperature, as well
as moderately dependent on pressure.  As an example,  consider
the balance of CO and methane, the dominant equilibrium forms
of carbon that are governed by the process:
\begin{equation}
CO + 3H_2 \longleftrightarrow H_2O + CH_4.
\label{methanebalance}
\end{equation}
For exothermic processes, such as this one, an increase in temperature
at a given pressure favors the reactants, so equilibrium favors
CO over CH$_4$ with increasing temperature.  But at increased
pressure (for a fixed temperature),
the side of the reaction with fewer molecules is favored
(Le Chatelier's Principle), which results
in a greater CH$_4$ abundance at the expense of CO.  The analogous
process that governs N$_2$ and NH$_3$ abundances, the dominant
equilibrium forms of nitrogen, is
\begin{equation}
N_2 + 3H_2 \longleftrightarrow 2 NH_3.
\label{ammoniabalance}
\end{equation}
At high temperatures, N$_2$ is favored, but the equilibrium
shifts toward NH$_3$ with increasing pressure (cf. the Haber process).

These temperature and pressure dependences have direct implications
for the atmospheric compositions of EGPs and brown dwarfs.  
In brown dwarfs, CH$_4$ is the primary carrier of carbon (number fraction: 
$\sim$$6 \times 10^{-4}$) in the outer atmosphere,
but a transition to CO occurs at depth, due to increasing
temperature.  Similarly, the outer atmospheres of EGPs at large orbital
distances tend to be dominated by methane, due to their low atmospheric
temperatures.  In contrast, in the low-pressure outer atmospheres of close-in
EGPs ($\sim$ 0.05 AU), CO is dominant due to strong irradiation (and heating) by 
the central star.  Analogous transitions between NH$_3$ and N$_2$ as
the dominant carrier of nitrogen ($\sim$$2 \times 10^{-4}$) occur
as well, but at lower temperatures than those for CH$_4$ and CO.

At the temperatures and pressures of EGPs and brown dwarfs,
H$_2$O is the main reservoir of oxygen and is the
third most abundant species, with a mixing ratio of
$\sim 10^{-3}$,
behind H$_2$ ($\sim$ 0.83) and He ($\sim$ 0.16).
The H$_2$O abundance is reduced somewhat at high atmospheric
temperatures because CO competes for oxygen (eq. \ref{methanebalance})
and because silicates form.

TiO and VO, important molecular species in the atmospheres of
M dwarfs and some brown dwarfs, form at temperatures greater than $\sim$2000 K, as do CrH and FeH.  The alkali
metals persist to much lower temperatures ($\sim$ 1000 K)
and sodium and potassium can be particularly important both in brown dwarfs
and in the hotter of the EGP classes.
Although their abundances are relatively low, Na
($\sless$$3 \times 10^{-6}$) and K ($\sless$$2 \times 10^{-7}$) are
strong absorbers in the visible region of the spectrum.  For most
EGPs and brown dwarfs, the Na and K abundances are negligible in the
outermost atmosphere, but at depth the mixing ratios of both can rise rapidly.  In the case of
a strongly-irradiated, close-in ($\sim$0.05 AU) EGP, Na and K abundances
can reach their maximum even at the lowest atmospheric pressures.  
The other alkali metals, lithium, rubidium, and cesium,
are present in EGPs and brown dwarfs as well, albeit in lower abundance.  Sulfur appears
in the form of H$_2$S at low-to-moderate abundance over a large
range in temperature and pressure, while phosphorus is in the
form of PH$_3$.

\subsection{Condensation and Rainout}
\label{sec_condandrainout}

The condensation and gravitational settling of species
results in compositions that differ significantly from those
for true equilibrium (Fegley \& Lodders 1996; Burrows \& Sharp 1999;
Lodders \& Fegley 2002).  Condensation sequesters most
of the heavier elements, such as silicon, magnesium, calcium,
aluminum, and iron, in compounds that settle, or rain out,
leaving the upper atmosphere depleted of
species that would otherwise contribute to the molecular
chemistry.  Condensates are very important in their own right
as well, often providing substantial absorption and scattering
opacity in the visible and infrared spectral regions.

Some of the more refractory species expected to exist in EGP
(and brown dwarf) atmospheres are calcium-aluminum compounds,
such as Ca$_2$Al$_2$SiO$_7$ (gehlenite), and aluminum oxides,
such as Al$_2$O$_3$ (corundum).  Since both calcium and
aluminum are less abundant than silicon or oxygen, the formation
and gravitational settling of these condensates
should consume most of the calcium and aluminum,
thereby depleting the upper atmosphere of these metals.  The
atmospheric temperature and pressure at which this occurs depends
on the details of the particular T-P profile.
For a given mass, the T-P profile of an irradiated 
object (as opposed to an isolated object) will
intersect the condensation curve of the refractory species
at a lower temperature and pressure, so the condensate will settle
higher (at lower pressure) in the atmosphere.

Another refractory species, TiO, can condense into the calcium
titanates \citep{Lodders02} and/or into titanium oxides in EGP atmospheres.
Iron is expected to condense mainly in a homogeneous manner rather than as FeS
or some other heterogeneous species, because the condensation
temperature for pure iron is quite high ($\sim$1500-2300 K,
depending strongly on pressure).
In fact, the existence of H$_2$S in Jupiter's
atmosphere supports a scenario where iron condenses homogeneously and
settles at depth.  If this were not the case, FeS, not H$_2$S, would be
the dominant atmospheric reservoir of sulfur (Fegley \& Lodders 1996).
In contrast with iron condensation, the homogeneous
condensation of VO is unlikely because vanadium condenses into solid solution
with titanium-bearing condensates \citep{Lodders02}. 

The silicates are somewhat less refractory, so in general,
they will appear higher in the atmosphere ({\it i.e.} at lower
temperature and pressure).  The exact composition of the silicates is
unknown, although some common species include MgSiO$_3$
(enstatite) and Mg$_2$SiO$_4$ (forsterite).  Of these two
possibilites, forsterite is more likely to exist in EGP and
brown dwarf atmospheres because it has a higher condensation
temperature and so will consume most of the magnesium, the
limiting element.  And although Mg$_2$AlO$_4$ (magnesium-aluminum
spinel) condenses at even higher temperatures, it is unlikely
to form in abundance because most of the aluminum will be
consumed by the formation of corundum and/or other aluminum compounds.  In brown dwarf and giant
planet atmospheres, forsterite will condense within the $\sim$
1500-2000 K range, depending upon atmospheric pressure.
We should mention that forsterite and enstatite are only two
species in the olivine (Mg$_x$Fe$_{1-x}$SiO$_3$) and
pyroxene (Mg$_{2x}$Fe$_{2-2x}$SiO$_4$) sequences, respectively. The
optical properties of other such species have been explored by Dorschner \etal
(1995).  If iron has not been fully removed by homogeneous
condensation and rain out---and it is unlikely that homogeneous
condensation will consume all the iron---then some of these iron-bearing
species are likely to form.

The alkali metal abundances are higher in a rainout scenario than
they are in the true equilibrium case.  In the true equilibrium case,
NaAlSi$_3$O$_8$ (high albite) and KAlSi$_3$O$_8$ (high sanidine)
can condense, lowering the gas-phase
abundance of sodium and potassium \citep{BurrowsSharp99}.  But in the more realistic rainout
scenario, the aluminum is consumed by other species that condense
at higher temperature, such as Ca$_2$Al$_2$SiO$_7$ (gehlenite).
Gravitational settling then removes the aluminum from the upper
atmosphere, thereby inhibiting the formation of sodium- and
potassium-bearing condensates.

At cooler temperatures ($\sim$700-1100 K), sulfide
and chloride condensates are expected to sequester the alkali
metals.  Sodium will end up in Na$_2$S and/or NaCl \citep{Lodders99},
while potassium is likely to be in the form of KCl.  Due to the
low abundances of the alkalis, thick clouds are unlikely, but ``cirrus-like''
layers are certainly possible.

At lower temperatures, H$_2$O will condense homogeneously for
atmospheric temperatures below $\sim$250 K.  Hence, many
EGPs at large orbital distances will contain water clouds, as will
cool brown dwarfs.  Even if only a small fraction of
H$_2$O condenses, as governed by the degree of
supersaturation \citep{Cooper02}, the high abundance
of H$_2$O and its large reflectivity in ice or liquid form indicates
that condensed H$_2$O is important to the emergent spectra of EGPs.
   
At very low temperatures ($\sim$ 150-200 K), NH$_3$ and NH$_4$SH condense.
NH$_4$SH condenses at somewhat higher temperatures than ammonia, so
a layer of this heterogeneous material is expected to appear between
a water cloud and an ammonia cloud in cold EGPs.  Such is the case
for the standard model of Jupiter.  However,
we do not model an NH$_4$SH layer because the optical constants for
this material over a broad wavelength range are not available.

\section{Opacities}
\label{opacity}

The total gaseous absorption opacities are calculated by weighting
the atomic and molecular opacity of each species by its
abundance, all as a function of temperature and pressure.  The resulting
product is a large opacity table within which interpolation can yield
the frequency-dependent total opacity at any point in the EGP or brown
dwarf atmosphere.

Because both the gaseous abundances and their opacities vary
substantially with temperature and pressure, the spectral region
that dominates the opacity will differ for objects of different
temperatures.  To illustrate this fact, Figure \ref{fig_comboopacs}
shows the P = 1 bar total gas opacity at two very different temperatures,
200 K and 1600 K.  For a hot atmosphere, the visible region
is overwhelmed by the sodium and potassium resonance lines, yet for
a cool atmosphere, the visible is canvassed by weak methane absorption.
These visible opacities (in cm$^2$ g$^{-1}$) differ by up to several
orders of magnitude, but there are also spectral regions in the near infrared where the
opacity of the cool atmosphere is greater than that of the hot atmosphere.
In general, the bulk of the outgoing flux ultimately will find its way out of
the object between the peaks in the opacity.

\subsection{Rayleigh Scattering}

Rayleigh scattering is a conservative scattering process by atoms
and molecules.  Although strong in the ultraviolet/blue, the scattering
cross sections quickly fade toward the red region of the spectrum
($\propto \lambda^{-4}$).  Rayleigh scattering has little effect on the
spectra of isolated brown dwarfs, but irradiated EGPs reflect a
fraction of the incident intensity in the ultraviolet/blue.

The Rayleigh scattering cross sections are derived from polarizabilities,
which are in turn derived from refractive indices:
\begin{equation}
\sigma_{Ray} = {8\over 3}\pi k^4 \left({{n-1}\over{2\pi L_0}}
\right)^2\, ,
\end{equation}
where $k$ is the wavenumber
(2$\pi/\lambda$) and $L_0$ is Loschmidt's number, the number of molecules
per cubic centimeter at STP (= 2.687$\times$10$^{19}$).

\subsection{Scattering and Absorption by Condensates in EGP Atmospheres}
\label{ch_condensates}

Condensed species in substellar objects range from ammonia ice
in low temperature EGPs
to silcate, iron, and aluminum compounds at high temperatures.  Some of
the condensates relevant
to EGP atmospheres include NH$_3$ ($\sless$ 150--200 K), NH$_4$SH
($\sless$ 200 K), H$_2$O ($\sless$ 250--300 K), low-abundance
sulfides and chlorides ($\sless$ 700--1100 K),
silicates such as Mg$_2$SiO$_4$ and/or MgSiO$_3$ ($\sless$ 1600--2000 K),
iron or iron-rich compounds
($\sless$ 1700--2200 K), and aluminum or calcium compounds
($\sless$ 1700--2200$^+$ K).  
Additionally, photochemical processes in
the upper atmosphere due to irradiation by a central star can produce
non-equilibrium condensates.
Stratospheric hazes may be composed of
polyacetylene \citep{Bar-Nun88} and other aerosols.  Chromophores,
those non-equilibrium
species which cause the coloration of Jupiter and Saturn, might include
P$_4$ \citep{Noy81} or
organic species similar to Titan tholin \citep{KhareSagan84}.  Brown
Dwarf atmospheres can contain a similar host of silicate, iron,
and aluminum compounds.  However, only the lowest temperature brown dwarfs
will contain condensed water, since the temperatures are simply too
high in most objects.

Condensates can have large effects on the emergent spectra of EGPs
and brown dwarfs.
The optical properties of most condensates are relatively featureless
in comparison with molecular bands and
atomic lines.  Hence, the presence of condensates tends to ``gray'' the
spectrum by shallowing the troughs and lowering the spectral peaks,
providing opacity at the wavelengths at which the radiation would
otherwise escape.  In EGPs, condensates are the principal source of
scattering of incident irradiation, and if located in the outer
atmosphere, their presence increases the reflected flux in the
visible and near-infrared regions of the spectrum.
One exception is the presence of non-equilibrium photochemical species
with absorptive optical properties in the UV/blue, a region of the
spectrum that in the absence of these species can be quite reflective
due to the Rayleigh scattering of incident radiation.  It is such compounds
that are thought to be responsible for the absorption in the blue in
Jupiter and Saturn and their consequent reddish coloration.

Of course, those condensates that are higher in the atmosphere
generally will have a greater effect on emergent spectra than those
that reside more deeply, unless the high condensed layer is very optically
thin.  The presence and location of a particular
condensed species is determined
largely by an object's T-P profile, and by the tendency of the condensate
to settle (rainout) to a depth in the atmosphere near the region where
the T-P profile crosses the condensation curve \citep{BurrowsSharp99}.
Hence, a given low-temperature EGP atmosphere
(T$_{\textrm{eq}}$ $\sless$ 150 K)
might consist of an ammonia cloud deck high in the troposphere and a water
cloud deck somewhat deeper, with purely gaseous regions above, beneath,
and between the clouds.  Similarly, a higher-temperature atmosphere
(T$_{\textrm{eq}}$ $\sim$1200 K) might consist of a 
tropospheric silicate cloud deck above a deeper iron cloud deck.  Depending
upon the wavelength region and the amount of condensate in the upper
cloud, the presence of deeper clouds may or may not have
a recognizable effect on the emergent spectrum.  If for a given wavelength
the cloud deck resides at $\tau_{\lambda} \gg 1$, it will have little effect 
on the emergent spectrum. 

The scattering and absorption of electromagnetic radiation by condensed
species in planetary and brown dwarf atmospheres is a very complex problem.
The scattering properties of ices, grains, and droplets of
various sizes, shapes, and compositions cannot be described
accurately by simple
means.  Most often, these scattering properties are approximated by Mie
Theory, which describes the solution of Maxwell's
equations inside and outside a homogeneous sphere with a given complex
refractive index.

The principal condensates to which we have applied the Mie theory (roughly
in order of increasing condensation temperature) include NH$_3$ ice,
H$_2$O ice and liquid, MgSiO$_3$ (enstatite),
Mg$_2$SiO$_4$ (forsterite), MgAl$_2$O$_4$ (magnesium-aluminum spinel), iron,
Al$_2$O$_3$ (corundum), and Ca$_2$Al$_2$SiO$_7$ (gehlenite).

The optical constants (complex indices of refraction) for ammonia ice,
H$_2$O ice, and H$_2$O liquid were obtained from Martonchik \etal (1984),
and Warren (1984 and 1991), respectively.  The constants for
enstatite, magnesium-aluminum spinel, iron, and corundum were obtained from
Dorschner \etal (1995), Tropf \& Thomas (1990),
and Gervais (1990), respectively.  The indices for iron were provided
by Aigen Li (private communication, 2002).  Those for forsterite
shortward of $\sim$33 $\mu$m were taken from Scott \& Duley (1996); longward
of this wavelength, the indices were provided by Aigen Li.  For
gehlenite, the real index
was taken from Burshtein \etal (1993) and an absorbance spectrum
\citep{RossmanTaran01} was used to derive the imaginary index.

Some of these condensates have relatively strong absorption
features.
H$_2$O ice has broad absorption
features at $\sim$ 1.5, 2, and 3, and 4.5 $\mu$m.  Those of water in liquid
form mirror these quite well, with only small displacements, mainly
to shorter wavelengths.  The silicates, MgSi$_2$O$_4$ and MgSiO$_3$
have broad features near 10 and $\sim$ 20 $\mu$m and at $\sless$ 0.2
and 0.35 $\mu$m, respectively.  MgAl$_2$O$_4$ absorbs strongly only
in the infrared, and especially around 15 and 20 $\mu$m.  
Ammonia ice, a principal
condensate at low temperatures, exhibits little relative absorption
in the blue and visible spectral regions.   However, it can provide
absorption in the infrared, with particular strength near 2.0, 2.25,
3, 4.3, 6.2, 9.5 micorns.  
  
P$_4$ and tholin are
chromophore candidates for the coloration of Jupiter and Saturn,
due to their large imaginary indices of refraction in the ultraviolet/blue
and their plausibility of production.
A somewhat yellowish allotrope of phosphorus, P$_4$ was produced in
the laboratory by Noy et al.
(1981) by ultraviolet irradiation of an H$_2$/PH$_3$ gaseous
mixture, a process that may be responsible for its production in Jupiter.
Tholin is a dark-reddish organic solid
(composed of over 75 compounds) synthesized by Khare and Sagan (1984) by
irradiation of gases in a 
simulated Titan atmosphere.  A tholin-like solid may be
produced similarly in giant planet atmospheres.  Polyacetylenes, polymers of
C$_2$H$_2$, were investigated by Bar-Nun et al. (1988) and likely are an optically
dominant species in the photochemical stratospheric hazes of the
Solar System giant planets,
where hydrocarbons are abundant \citep{Edgington98, Noll86}.  Due to
uncertainties concerning their production in EGPs,
we do not include P$_4$, tholin, polyacetylenes, or
other nonequilibrium species in our baseline models.

Cloud particle sizes are not easily modeled and are a strong function of
the unknown meteorology in EGP atmospheres.  Inferred particle sizes in
Solar System giant planet atmospheres can guide the construction of EGP models, although they range
widely from fractions of a micron to tens of microns.

We have investigated various particle size distributions.  A commonly used
distribution is
\begin{equation}
n(a) \propto \left({a\over a_0}\right)^6 \exp\left[-6\left({a\over a_0}\right)
\right],
\label{cloud}
\end{equation}
which reproduces the distribution in cumulus water clouds in Earth's
atmosphere fairly well if the
peak of the distribution is
$a_0$ $\sim$ 4 $\mu$m \citep{Deirmendjian64,Deirmendjian69}.  

In the present work, we use this cloud distribution with various modal
particle sizes.  Our fiducial modal particle radius for water and ammonia is
5 $\mu$m, as inferred from Deirmendjian (1964) and measurements 
of Jupiter's atmosphere.  For grains such as silicates and iron, our fiducial modal
particle size is  10 $\mu$m, a value in broad agreement with recent cloud
models for EGPs and brown dwarfs \citep{Cooper02, AckermannMarley01}.  As
these authors have shown, the modal particle size and vertical cloud
extent likely are functions of surface gravity and convective flux, but for
consistency and ease of calculation, we have chosen simple particle size
distributions and a vertical extent of one pressure scale height.

\section{Spectra and Atmospheric Profiles for the Different Composition Classes of Extrasolar Giant Planets}
\label{ch_classes}

Due to qualitative
similarities in the compositions and spectra of objects within
several broad atmospheric temperature ranges, EGPs fall naturally
into groups, or ``composition classes''\citep{Sudarsky00}. 
Figure \ref{fig_classcurves} depicts sample temperature-pressure
structures for Classes I through V.  Also shown are the condensation
curves for the principal condensates, ammonia, water, forsterite
(a representative silicate), and iron.
The intersection of a T-P profile with a
condensation curve indicates the approximate position of the base
of a condensate cloud.  (For clarity, the T-P profiles of cloud-free
models are shown.)  The Class I profile intersects the ammonia condensation
curve at roughly one bar.  At greater pressures, the temperature is
too high for ammonia condensation to occur.  This profile also intersects the
water condensation curve at nearly 10 bars, indicating the presence
of water clouds at this depth.   Of course, a whole variety of Class I
objects with differing T-P profiles will exist, and the ammonia
and water condensation curves will be intersected at different pressures
for each object,
but by definition Class I EGPs have an ammonia cloud that is well
above a water cloud.  The warmer Class II profile crosses only the water
condensation curve, indicating the existence of a water cloud in the
upper atmosphere.  Classes IV and V intersect the silicate and iron
condensation curves, but at quite different pressures.  Figure
\ref{fig_classcurves} is a general guide to the T-P structures and
approximate positions of the various principal condensate clouds in the
atmospheres of a full range of EGPs.  

\subsection{Jovian Class I EGPs}
\label{sec_class1}

Class I EGPs orbit at distances of at least a few AU from their central
stars.  With T$_{\textrm{eq}}$ $\sless$ 150 K,
the spectrum of such a cool, Jovian-like gas giant at 10 parsecs is shown
in Figure \ref{fig_class1}.
Methane and ammonia are the dominant
gaseous species, because at these low temperatures, carbon is
predominantly in the form of CH$_4$ rather than CO, and nitrogen is
in NH$_3$, as opposed to N$_2$.  The visible and near-infrared spectrum
is comprised mostly of reflected stellar light, since such an object
is far too cool to emit significant thermal radiation at short wavelengths.
This reflection is provided mostly by the upper ammonia cloud, with some
additional contribution in the blue due to Rayleigh scattering.
Some ammonia ice absorption
features are also indicated in Figure \ref{fig_class1} by the
\{NH$_3$\} symbols; for ammonia ice, the imaginary index of refraction is
rather large at $\sim$1.5, 1.65, 2, and 2.2 $\mu$m.
The upper atmosphere is largely depleted of H$_2$O, which has
condensed and settled into a cloud layer at
several bars pressure.  Provided that
the upper ammonia cloud is optically thick (here we assumed a particle
size distribution peaked at 5 microns, with a condensate abundance at 10\% of
the total ammonia abundance), one should not
expect to see substantial effects due to 
gaseous or condensed H$_2$O in the visible or near infrared emergent
spectrum.

In our present-day Solar System, both Jupiter and Saturn are Class I
objects, but it is important to note that not all EGPs orbiting at several AU
will be Class I planets.  Young and/or massive objects may have
high enough inner flux temperatures 
(defined as $(F/\sigma)^{1/4}$
for an object in $isolation$) that ammonia will remain in
gaseous form.  Therefore, those hotter objects will be Class II
(or even Class III).

\subsection{Water Class II EGPs}

Class II EGPs generally orbit at a distances of $\sim$1 to 2 AU, or somewhat
farther out for an object orbiting an early-type central star.  Outer
atmospheric temperatures ($\sless$ 250 K)
are below that of the water condensation curve,
producing a tropospheric cloud layer of H$_2$O.
Figure \ref{fig_class2} shows an example spectrum of a Class II EGP.
Reflection in the visible and near-infrared is provided primarily by the water
cloud, which we have assumed to be comprised of a particle size
distribution peaked at 5 microns.  While scattering by this
H$_2$O condensate layer elevates the emergent spectrum, the spectrum
is dominated by the gaseous absorption features of water, methane, and to
a lesser extent, ammonia.

\subsection{Gaseous Class III EGPs}

Class III EGPs are too warm to contain condensed H$_2$O, but are
too cool for silicate or iron grains to exist in their outer
atmospheres.  In the absence of principle condensates, we label
this class the ``gaseous'' or ``clear'' class.  Typically orbiting
at distances less than about 1 AU, such objects have equilibrium temperatures
between $\sim$ 350 K and 800 K.  Although an isolated
object with T$_{\textrm{eq}}$ $\sim$ 350 K is likely to contain
water clouds, stellar irradiation keeps the temperature
of a Class III EGP high enough that its profile never reaches
the condensation curve.

A purely gaseous EGP is expected to have a very low albedo \citep{Sudarsky00}.  Rayleigh
scattering does have an impact
in the blue and visible regions of the spectrum, but the
majority of the spectrum is made up almost entirely of absorption and emission.
Figure \ref{fig_class3} shows an example spectrum of a Class III EGP.
Gaseous water and methane absorption are strong, while ammonia
absorption is weaker than for a Class II.  The alkali metals,
particularly the sodium and potassium resonance lines at $\sim$ 0.6
$\mu$m and $\sim$ 0.77 $\mu$m, respectively, appear with modest intensity.
Due to the absence of clouds, the
incident radiation penetrates deeply between the near-infrared
ro-vibrational absorption bands.  Here, collision-induced absorption
(CIA) by H$_2$, a continuum absorption that is strong near $\sim$0.8, 1.2, and 2.4
$\mu$m, keeps the flux peaks lower than they would be otherwise. 

The slope of the visible spectrum is a combined consequence of the
$\lambda^{-4}$ dependence of Rayleigh scattering and the incident
stellar spectrum.  The later the stellar spectral type, the flatter
the slope will be if the spectral energy distribution
of the star falls off rapidly with decreasing wavelength.

Figure \ref{fig_bd3TP} provides a comparison of the T-P structure
of a Class III model with that of a cloud-free brown dwarf of the
same gravity and integrated emergent flux.  As expected, the EGP profile
is more isothermal in the outer atmosphere, while it is somewhat
cooler than the brown dwarf deeper in the atmosphere.
Figure \ref{fig_bd3spec} depicts the
emergent spectra associated with these profiles out to 30 $\mu$m.  By
construction,
both integrated emergent fluxes are identical, but the individual spectral energy
distributions differ significantly.  Most striking is the visible region.
Here, Rayleigh scattering of the incident radiation keeps the EGP
flux high relative to that of the brown dwarf.  Only modest sodium
and potassium lines appear in the EGP spectrum, compared with very
strong lines for the brown dwarf.  This fact is not due to a dominance
of Rayleigh scattering cross sections; the alkali cross sections
are actually much larger.  Rather, the equilibrium abundances of the
alkalis are very low in the outermost atmospheres of both the EGP
and the brown dwarf.  The alkali absorption in a brown dwarf of this
effective temperature occurs at pressures of $\sgreat$ 1 bar,  but
the strong impinging radiation on the EGP in the visible region
is reflected by Rayleigh scattering higher in the atmosphere, producing
emergent flux in the visible,  before this
pressure region is well
probed.  So although alkali absorption occurs in both objects at moderate
pressures, strong alkali absorption is seen only in the emergent
spectrum of the brown dwarf.  This is not the case if the outer
atmosphere is at higher temperature, as for Class IV
objects.

In the near-infrared, the peak-to-trough variations in the molecular
absorption are greater in the brown dwarf than in the Class III EGP, which is
expected due to the more isothermal profile of the EGP.  In the case
of the brown dwarf, more flux escapes in the $Z$, $J$, and to a lesser
degree, the $H$ band.  But beyond the $K$ band, the EGP infrared
flux is somewhat higher.

\subsection{Close-In Class IV EGPs}
\label{sec_class4}

The close-in Class IV EGPs are those orbiting their stars well within the
distance at which Mercury orbits the Sun.  Many are roughly 
0.1-0.2 AU from their central stars, or even closer for late-type
stars.  Giant planets at such small orbital
distances are quite exotic by the standards of our Solar System, and
their spectra are as well.  With atmospheric temperatures in the
1000 K range, the alkali metal abundances increase substantially
and CO takes up much of the carbon in the low-pressure outer
atmosphere.

Figure \ref{fig_class4} shows a Class IV EGP spectrum.  Perhaps
most striking is the strong absorption due to the sodium (0.59 $\mu$m)
and potassium (0.77 $\mu$m) resonance lines.  These lines are
strongly pressure-broadened and clearly dominate the visible
spectral region.  Also present are the lithium ($\sim$ 0.67 $\mu$m),
rubidium ($\sim$ 0.78, 0.795 $\mu$m; not labeled), and cesium
doublets ($\sim$ 0.85, 0.895 $\mu$m; not labeled).

CO absorption is very strong in the 4.4-5.0 micron region, as well
as at $\sim$ 2.3 $\mu$m.  This comes at the expense of methane,
which competes for carbon.  However, methane absorption
is still apparent, particularly the 3.3-$\mu$m $\nu_3$ fundamental feature.  Although
methane absorption is weaker, water absorption remains very strong,
and, as for Class III objects, H$_2$ CIA absorption affects the
near-infrared.  At these high temperatures, N$_2$ takes up most
of the nitrogen, so ammonia abundances are very low.

Silicate and iron clouds do form, but too deeply ($\sgreat$ 10 bars)
to affect the emergent spectrum significantly.  Figure \ref{fig_cloudclass4}
shows the effect of a silicate cloud on the T-P profile.  Despite
the changes in the T-P structure at depth, the emergent
spectrum is nearly identical to the cloud-free model.  Both spectra
are shown in Figure \ref{fig_cloudclass4spec}.

The very strong alkali absorption shown here is likely an upper
limit since the neutral alkali abundances could be reduced somewhat
by ionizing ultraviolet flux
or scattering by low-abundance, nonequilibrium photochemical products. 
In any event, strong alkali lines are expected to be prominent
features of close-in Class IV EGPs.

\subsection{Class V Roasters}
\label{sec_classV}

The hottest of all EGPs are the extremely close-in Class V
EGPs, or ``roasters.''  These objects orbit their stars around
0.05 AU.  The first EGP discovered, 51 Pegasi b, is a good
example of a roaster.   Many of these objects have equilibrium
temperatures that rival hot brown dwarfs ($\sgreat$ 1400 K).
Silicate and iron clouds can form quite high in the atmospheres
of these EGPs,
especially for low-mass, low-gravity roasters, and they have
significant effects on the emergent spectra.

Figure \ref{fig_class5} shows the emergent spectrum of a Class V roaster.
Like Class IV EGPs, Class V objects have strong alkali metal absorption
lines, although their strengths can be reduced somewhat due to reflection
by condensates at altitude, and/or if the outer atmosphere becomes isothermal.
Water and CO absorption remain strong, but unlike Class IV objects,
where some methane absorption is seen, methane features in the visible
and near infrared are either very weak or nonexistent, even at the
3.3-$\mu$m fundamental band.

The presence of high clouds in Class V EGPs results in smaller
wavelength-dependent flux variations relative to Class IV EGPs.
This is mainly a result of cloud opacity blocking the flux windows
between the molecular absorption features, thereby reducing the
flux peaks.  As a result, the absorption troughs are not as deep,
since some of the blocked flux escapes
in these spectral regions.  Additionally,
clouds reflect some of the incident stellar radiation,
increasing the incident flux where the scattering opacity is high,
a phenomenon that tends to be more noticeable in the vicinity of the gaseous
absorption troughs.

\section{Generic Model Sequence from 5 AU to 0.05 AU and Parameter Studies}
\label{sec_genericseq}

A generic sequence of EGP models at various orbital distances from a
given primary can be very instructive.  For simplicity,
we produce a cloud-free sequence, and the inner
boundary flux is set equal to that of an isolated object with
T$_{\textrm{eff}}$ = 125 K.  The resulting EGP models are compared
with a model of an isolated EGP/brown dwarf with the 
same inner boundary flux (i.e., effective temperature) and
gravity ($3\times 10^3 $ cm s$^{-2}$).

Figure \ref{fig_PTG0AU} illustrates the effects of irradiation on
the temperature-pressure structure of an EGP at various distances,
from 5 AU to 0.05 AU from a G0V primary.  Even at a distance of
5 AU, the stellar irradiation significantly alters the T-P profile
of the planet relative to the isolated object, and this effect becomes
quite extreme for objects closer than $\sim$1 AU.

Figure \ref{fig_spectrumG0AU} shows the emergent spectra of
these models, including that for the isolated object.  From high to
low atmospheric temperature, one can see certain atomic and molecular features
appear and disappear.  For the objects orbiting most closely to
the primary, the alkali absorption lines in the visible are strong,
as are the CO features at 2.3 $\mu$m and $\sim$4.4 to 5 $\mu$m.  But
as the EGP is moved away from the primary, the alkali lines weaken,
giving way to Rayleigh scattering by the gaseous species high in the
atmosphere, where the alkali abundances are negligible.
The CO features quickly weaken as well, 
while methane absorption becomes quite strong, particularly at
3.3 $\mu$m.  Water absorption remains strong for the full model set,
which is not surprising considering its high abundance throughout
a very broad range of temperatures and pressures.

It is clear from Figure \ref{fig_spectrumG0AU} that for wavelengths
greater than $\sim$ 0.8 $\mu$m and larger orbital distances 
the emergent spectrum of an EGP approaches
that of an isolated object with the same inner boundary flux. 
However, this is not true in
the visible region of the spectrum.  Here, reflection of the incident
stellar radiation keeps the emergent flux from matching
the very low flux of the isolated object.  

Figure \ref{fig_spectrumG0AU500} shows spectra for the same run
of distances, but with an inner boundary flux associated with an
effective temperature of not 125 K, but 500 K.  This model sequence
mimics the expected behavior for an irradiated low-mass ($\sim$3 \mj), young ($\sim 10^8$ yr) brown dwarf,
and is very similar to that of a higher-mass, older brown dwarf (e.g., 10$^9$ yr/15 \mj or 10$^{10}$ yr/35 \mj).
As in Fig. \ref{fig_spectrumG0AU}, Rayleigh scattering is responsible
for the elevated flux in the optical, but exterior
to $\sim$0.5 AU in the near- and mid-infrared the flux
becomes independent of orbital distance.  The comparison between Figs.
\ref{fig_spectrumG0AU500} and \ref{fig_spectrumG0AU} is dramatic.
Again, in general the infrared flux becomes
independent of the EGP/brown dwarf's orbital distance once
this distance is great enough that the object's internal luminosity 
begins to dominate the emergent spectrum.

The planet-to-star flux contrast as a function of wavelength is of
central importance in any campaign to detect EGPs directly.  We plot
in Fig. \ref{fig_spectrumG0AU_contrast} these planetary phase-averaged, planet-to-star
flux ratios for the models of Figure \ref{fig_spectrumG0AU}.  It is
important to stress that these are
generic, cloud-free models, which assume a central star of type
G0V and an inner flux that corresponds to an effective temperature of 125 K. 
Nevertheless, Fig. \ref{fig_spectrumG0AU_contrast} illustrates 
clearly the generic dependence of the planet-to-star contrast on orbital distance.
For a very close-in object, the flux ratio varies
from $\sless 10^{-5}$ in the visible to $\sless 10^{-3}$ in the
3.8-4.0 $\mu$m region.  In almost all cases,
this ratio becomes smaller at all wavelengths with increasing
orbital distance (as expected from the $a^{-2}$ diminution of stellar flux
at the planet).   However, for orbital distances between $\sim$0.1 and 0.2 AU 
and in the 4.4 to 4.8 $\mu$m wavelength range, CO's absorption strength wanes
with increasing distance and the associated decrease in
atmospheric temperature.  This temporarily buoys the flux in the $M$ band
until the CO has been replaced by CH$_4$.   Nevertheless, the $M$ band between
4.0 and 5.0 $\mu$m is always a relatively bright region of the spectrum in
which to probe for EGP's, irradiated or otherwise.

For a low inner boundary flux and cloud-free atmospheres, the EGP contrast ratio in the near infrared
plummets and can reach values of 10$^{-10}$ to 10$^{-15}$.  The corresponding
ratio in the visible approaches 10$^{-9}$ to $10^{-10}$ (again, for cloud-free atmospheres).   
However, since the flux in the Rayleigh-Jeans tail
is proportional to T$_{\textrm{eq}}$, not T$_{\textrm{eq}}^4$,
and T$_{\textrm{eq}}$ $\propto a^{-1/2}$, the planetary flux in the mid- and
far-infrared does not decrease as fast.  This makes the planet-to-star contrasts 
in the mid-IR wavelengths longward of $\sim$10 $\mu$m quite high and suggests
that mid-IR searches could be quite profitable.

\subsection{Effects of Inner Boundary Flux}

Above we studied the effect of varying the
outer boundary flux, while holding the inner boundary flux
constant.  In this section, we'll explore just the opposite, holding
the outer boundary flux constant while varying the inner boundary
flux.

Without evolutionary models, it is not possible to know the correct
inner boundary flux when modeling the atmospheres of EGPs.  This
inner boundary flux fixes the $net$ flux at the surface (not to
be confused with the emergent flux, the outward flux at the surface),
so long as there are no sources or sinks throughout the atmosphere.
That is, integrated over all frequencies, the net flux at every depth
zone is constant.  For an isolated object with no incident
radiation, the emergent flux is equivalent to this net flux. (This,
of course, is not true for irradiated objects such as EGPs.)

When modeling atmospheres of EGPs without evolutionary models,
one must make reasonable assumptions about the inner boundary.
One may choose an inner boundary flux based on the evolutionary
models of isolated objects \citep{Burrows97}.  Such a choice would
provide the correct intrinsic luminosity of the EGP in isolation,
assuming its mass and age are known.  While this may provide a good
approximation in some cases,
such an approach ignores the fact that the interior and atmosphere
are inexorably coupled, and so an incident flux will affect the
interior and the evolution of the object as a whole
\citep{Burrows00b}.  For heavily
irradiated objects, strong horizontal winds are expected, and it
has been suggested recently that even a small vertical component can
transport kinetic energy to the interior, acting as an additional
heat source \citep{GuillotShowman02}.

Since deciding upon the inner boundary condition for an atmosphere
model is obviously not a trivial task, it is important to explore the
variation in the emergent spectrum and T-P structure as a function
of the inner boundary flux.  Using a Class IV EGP model (a cloud-free
version of 55 Cancri b), we investigate the effects of varying this
flux by up to 5 orders of magnitude.  Figure \ref{fig_PTboundary}
shows the EGP T-P structures for inner boundary fluxes
(= $\sigma$\teff$^4$) corresponding
to \teff  = 1000 K (top curve), 750 K, 500 K, 300 K, 150 K, and 50 K
(bottom curve).  The dashed portions of the curves indicate the
convective regions.  At pressures greater than $\sim$ 1 bar, each
of these profiles is substantially different.  Clearly, the T-P
structure in the deepest regions is determined by the chosen inner
flux.  However,
at pressures of less than a bar, only the 1000 K and 750 K models
deviate significantly from those with lower temperature inner
boundary fluxes.

The emergent spectra associated with these models are depicted
in Figure \ref{fig_spectrumboundary}.  The lowest curves (300 K, 150 K,
and 50 K; overlapping at the bottom) are nearly identical,
indicating that below
300 K, the inner boundary condition ceases to have any noticeable effect
on the emergent spectrum of an object as hot as Class IV.  The 500 K boundary
model is also very close to these lower temperature models at most, but not
all wavelengths, with the most significant deviations appearing in the
flux peaks between $\sim$ 0.8 and 1.4 $\mu$m.  The 750 K and 1000 K models
have higher fluxes than the other models at most wavelengths.  Still,
the $\sim$ 5 dex increase in the inner boundary flux between the 50 K model and
the 1000 K model translates to well under a 1 dex difference in the
emergent spectrum at most wavelengths.  

\subsection{Effects of Surface Gravity}

Since only the minimum masses of most EGPs are known, and their radii
are given only by theoretical models, the surface gravities of
EGPs are problematic.  In
this subsection, we explore the dependence of the T-P structure and
emergent spectrum on surface gravity, which is an input to the
atmosphere models. 

An estimate of the minimum surface gravity
for most objects can be made by assuming a radius close to that of
Jupiter, or more accurately, obtaining the radius from evolutionary
models of isolated objects.  Statistically speaking, upper limits can be estimated crudely
by noting that the actual mass of an EGP should rarely be more than
a factor of 3 or 4 greater than its M$_p$sin($i$).  Otherwise a clearly
disproportionate number of detected EGPs would be nearly face-on.
But no degree of certainty for a given EGP can be claimed without
accurate astrometry or the fortuitous transit of a primary by the EGP. 
The only certain case presently known is HD 209458b.
As a low-mass roaster, HD 209458b has a particularly low surface
gravity ($\sim$1 $g$). 
For the majority of EGPs, regardless of their proximity to their
central star, surface gravities are rather poorly constrained.  Hence,
a parameter study on gravity is useful.

Using a Class III EGP model, we vary the surface gravity from 10$^{3}$
to $3\times 10^5$ cm s$^{-2}$ (large) in steps of 1/2 dex.  Figure
\ref{fig_PTgravity} depicts the resulting T-P profiles.  (The
top curve is the lowest gravity model.)  Increasing the gravity
by 1/2 dex increases the pressure at a given atmospheric
temperature by roughly 1/2 dex as well, as one could estimate
from hydrostatic equilibrium.  Of course, since compositions
and opacities vary with pressure and temperature, this problem
is not merely one of hydrostatic equilibrium, but such
simple arguments hold reasonably well.  Since the surface
gravity varies roughly in concert with the mass (the radius does
not vary too strongly with mass), a 1/2 dex uncertainty in the
mass of an EGP might correspond to the differences between two
adjacent T-P profiles in Figure \ref{fig_PTgravity}.

The associated emergent spectra for
these models are shown in Figure \ref{fig_gravity}.
For these Class III models, an increase in gravity results in
a general decrease in flux shortward of
$\sim$2.2 $\mu$m, and an increase longward.  Additionally, a larger
gravity tends to reduce the peak-to-trough variations throughout
the spectrum.  While there certainly are some
large differences between the $10^3$
and $3\times 10^5$ cm s$^{-2}$ (a massive brown dwarf) models, a 1/2-dex difference in surface
gravity produces a relatively small difference in the emergent spectrum.

\subsection{Effects of Condensate Particle Size}

The scattering and absorption of radiation by condensates is a
wavelength-dependent function of particle size.  
Here, we vary the median particle size for a water cloud size distribution
to investigate the effects on the emergent spectrum of the Class II
EGP, $\epsilon$ Eridani b.

Our fiducial median particle size
for $\epsilon$ Eridani is 5 $\mu$m.  For two other models, we use
median sizes of 0.5 $\mu$m and 50 $\mu$m.  The effects on the visible and
near-infrared emergent spectrum are shown in Figure \ref{fig_partsize}
along with a cloud-free spectrum of the planet (bottom curve).  The
0.5-$\mu$m model (top curve) produces the most reflection, followed by
the 5-$\mu$m model, and then the 50-$\mu$m model.  The cloud-free model
visible and near-infrared fluxes are substantially lower because Rayleigh
scattering alone provides limited reflection with increasing wavelength.
In the visible, the three cloudy models differ by less
than a factor of 2 at any given wavelength.  However, in the infrared
$J$ and $H$ bands the variation is significantly larger, with the 
0.5-$\mu$m and 50-$\mu$m models differing by roughly an order of magnitude. 
The effects out to 30 $\mu$m are depicted in Figure \ref{fig_partsize2}.
The cloudy models simply do not differ much beyond the near-infrared.
However, each of the cloudy models does differ from the cloud-free model in the
4-5 $\mu$m opacity window by up to two orders of magnitude.  The clouds
act to close off the opacity window, blocking radiation from the deeper
atmosphere, which can escape through this window in the cloud-free case.
This is reminiscent of Jupiter, in which ammonia clouds block the $M$
band flux, which emerges only through the clear gaps between the cloud bands.  

In general, different condensate particle sizes will translate into significant
differences at some wavelengths.  However,
such differences are unlikely to rival the differences between
cloud-free and cloudy EGP models.

\section{Specific Theoretical Models for Known EGP Systems}
\label{ch_specific}

Models of specific EGP systems will prove to be invaluable both to
inform observational surveys and in the interpretation of the
resulting data.  Like any subfield of
astrophysics, the accuracy and improvement of EGP models will in part depend
upon the availability of observational data to constrain
the rich variety of consistent theoretical models that are possible.
At this early stage, assumptions must be made with respect to
metallicities, gravities, inner boundary fluxes, cloud scale heights,
cloud particle sizes, etc.  When modeling a specific EGP system with
some well-measured parameters (semimajor axis, stellar spectral type, etc.),
we choose one set of values based on the available data.  (We explored
the effects on the emergent spectra and T-P structures of variations
in some of these parameters in \S \ref{ch_classes}.)
Some aspects, such as cloud patchiness and atmospheric winds, are not
tackled in the current modeling effort.  Furthermore, we do 
not attempt to model the complex photochemistry of
nonequilibrium species that are likely produced by the ultraviolet
irradiation of molecules in the outermost
atmospheres of EGPs.  Such stratospheric processes can alter
the T-P profile at low pressures, producing significant
temperature inversions that can reduce the strengths of molecular
absorption troughs or result in emission features,
such as the 7.7-$\mu$m feature of methane.  Additionally,
photochemical products likely would reduce the flux in the
ultraviolet and blue (Sudarsky \etal 2000).

To date, there are roughly 100 known EGPs.  We model a representative
sample of these that encompasses a full range of EGP atmospheric
temperatures and angular separations.  A broad set 
of Kurucz model stellar spectra \citep{Kurucz94}, each spanning
a wavelength range of 0.4 to 300 $\mu$m,
is utilized to simulate the central stellar spectrum for each planetary
system modeled.  These spectra are scaled and normalized using the measured
effective temperatures and gravities of the actual stars
\citep{Santos01, Ryan01}.

For the purpose of modeling the atmospheres and emergent spectra of
these objects, the mass and age determine the intrinsic luminosity
and, hence, the inner boundary flux.  Unfortunately, we know neither 
the actual planetary masses nor the ages of most of the EGPs discovered
to date.  We do know the minimum planetary masses, as given by the
radial velocity method (M$_p$sin($i$)), but we do not know the orbital 
inclination ($i$) for most systems.  Only with a transit ($e.g.$ HD 209458b)
or an astrometric measurement ($e.g.$, $\epsilon$ Eridani, Gatewood 2000) can we
discern the inclination and mass.  The ages are difficult as well.
Detailed spectroscopy of the central star may result in an estimated
age, but such work has yet to be completed for most systems.
The inner boundary is of
little consequence to the emergent spectra of closely orbiting
EGPs so long as the inner flux corresponds to \teff  $\sless$500 K, but it
can be quite important in the modeling of planets with
larger orbital distances.  We use M$_p$sin($i$) and, where
available, estimates of the primary's age along with evolutionary
models \citep{Burrows97} as guides to our chosen inner
boundary fluxes.  In all cases, we'll state what inner boundary
condition is used.

\subsection{The Upsilon Andromeda System}
\label{sec_UpsAnd}

The Upsilon Andromeda system \citep{Butler99} is currently known
to be comprised of three EGPs in orbit around an F8 main sequence star
of distance $\sim$ 13.5 parsecs.
The innermost planet ($\upsilon$ And b) has an M$_p$sin($i$) of 0.71\mj and
is in a roughly circular orbit ($e$ $\sim$ 0.03) about its primary
at only 0.059 AU.  $\upsilon$ And c is somewhat more massive
(M$_p$sin($i$) = 2.11 \mj) and has an elliptical orbit ($e$ $\sim$ 0.2) with
a semimajor axis of 0.83 AU.  With M$_p$sin($i$) = 4.61\mj, $\upsilon$ And d
is even more massive and has a semimajor axis of 2.50 AU.  It has a
significant eccentricity of $\sim$ 0.4, so its periastron is actually
1.5 AU and its apastron is 3.5 AU.  Since we do not currently know the
inclinations of their orbits, we know only the minimum masses of these
planets.  However, if we assume that the orbits are roughly coplanar,
as indicated by our own solar system and the theory of planet formation,
the ratios of the M$_p$sin($i$) are also the ratios of the planetary masses.  

The emergent model spectra for the planets in the Upsilon Andromeda
system, along with an estimated spectrum of the primary (a scaled
Kurucz model stellar spectrum), are shown in the upper panel of Figure
\ref{fig_spectrumUpsAnd}.  The T-P profiles of these planets are depicted
in Figure \ref{fig_TPUpsAnd}.

Upsilon And b is a Class V ``roaster'' with silicate and iron
clouds high in the atmosphere.  We choose a cloud particle size distribution that is
peaked at 10 microns for both forsterite and iron.  Each of these
clouds has a vertical extent of one pressure scale height.  The iron cloud
base resides at $\sim$20 mbar, while the forsterite condenses
at slightly lower pressures ($\sim$10 mbar).  Given the prescribed
vertical extents of these clouds, they overlap spatially.  At pressures
where the condensation curves of forsterite and iron are relatively
close, as in this case, one might reasonably argue that a
number of species from the olivine sequence, in addition to forsterite,
could form (as discussed in \S \ref{sec_condandrainout}).  For this
reference model of $\upsilon$ And b, we do not include such species.
The mixing ratio of the condensed forsterite is set equal to the
elemental mixing ratio of magnesium ($\sim 3.2\times 10^{-5}$).
Similarly for the iron cloud, the condensate is assumed to consume
all the iron (mixing ratio $\sim 5.4\times 10^{-5}$).

The lower panel of Figure \ref{fig_UpsAndb} is a comparison of the
T-P profile of a cloud-free version of
$\upsilon$ And b (thin curve) with that of our fiducial model (thick
curve).  Both models have a lower boundary flux equal to that of
an isolated object with \teff = 500 K and a surface gravity of
$3\times 10^3$ cm s$^{-2}$. The primary effect of
removing the clouds is a somewhat
lower temperature in the outer atmosphere (P $\sless$ 0.1 bar)
and a slightly hotter atmosphere deeper (P $\sgreat$ 0.1 bar).  At
large depth, in the convective region (P $\sgreat$ 50 bars),
these high clouds appear to have essentially no effect.  A comparison
of the fiducial spectrum (thick curve) and cloud-free spectrum (thin
curve) is depicted in upper panel of Figure \ref{fig_UpsAndb}.
The removal of the clouds results in a wider variation from peak to
trough throughout most of the spectrum.  This is expected to be the
case because condensates produce relatively gray extinction in comparison
with strongly wavelength-dependent gaseous absorption.
In this same vein, the removal of the clouds somewhat deepens the sodium resonance
line.

While the presence of clouds in $\upsilon$ And b (and other roasters)
has significant effects on its T-P structure and emergent spectrum,
these effects are not so drastic as those predicted in hot EGPs by some researchers, 
such as Barman, Hauschildt, \& Allard (2001, BHA).
The main differences
are the vertical extent of the clouds and the cloud particle sizes.
Our clouds have a finite scale height like those observed in Earth's
atmosphere and in the atmospheres of other planets in our Solar
System.  Such finite scale heights are also supported by the latest
theoretical cloud models for EGPs and brown dwarfs
\citep{AckermannMarley01, Cooper02}.
BHA use a ``dusty'' model in which the condensates
extend to the lowest pressure of their atmosphere model, an assumption 
that was also made by SWS.  Furthermore,
BHA assume that a large number of condensates form at these very low
pressures, including (in addition to forsterite and iron) enstatite,
spinel, diopside, akermanite, and others.  As
detailed in \S \ref{sec_condandrainout},
some of these species should not form because one or more of their
elements will be consumed by a condensate that forms at higher
temperature.  Furthermore, gravitational
settling should prevent any such species from remaining suspended at
such low pressures ($\sim$ 1-10 microbars).  BHA suggest
that the nature of their treatment of clouds provides an upper limit
to the effect of clouds in EGP atmospheres.

The other difference between our cloud layers and those
of BHA is approximately two
orders of magnitude in the cloud particle sizes.  BHA assume
an interstellar size distribution.  We have been guided in our
choice of particle size (10 $\mu$m) by the work of Cooper \etal
(2002) and Ackerman and Marley (2001). 

Upsilon And c is a Class III EGP.  That is, no principal condensates
are expected to appear in its outer atmosphere.  Since it is more massive
than $\upsilon$ And b, we have chosen a higher surface gravity of
$10^4$ cm s$^{-2}$.  Using evolutionary models \citep{Burrows97}
as a guide, we have chosen an inner boundary flux to be equivalent to that
of an isolated of object with \teff  = 200 K.  Its much larger orbital
distance and lack of condensates to reflect incident radiation result in
a flux that is a few-to-several orders of magnitude lower than
that of $\upsilon$ And b.  However, in the 4-5 $\mu$m region, a lack
of opacity reduces this difference to 1-2 dex. 
In actuality, this 4-5 micron opacity window could be partially
closed due to a nonequilibrium process, namely the rapid vertical transport
of CO (Saumon et al. 2002), an effect that we do not include here.

Upsilon And d contains condensed H$_2$O in its outer atmosphere, and
so it is a Class II EGP.  It's M$_p$sin($i$) is roughly twice that of $\upsilon$
And c, so we have chosen a surface gravity of $2\times 10^4$ cm s$^{-2}$
and a lower boundary flux of \teff  = 250 K.  Its semimajor axis is
three times that of $\upsilon$ And c, yet its flux at Earth is actually
larger at some wavelengths shortward of 2 microns.  This is due
to the fact that scattering by
the condensate reduces the depths of the molecular absorption features.
In the windows between these features, $\upsilon$ And c still has the
higher flux.  The rather large eccentricity ($e$ $\sim$ 0.4)
of $\upsilon$ And d means that its atmospheric temperature may vary
from periastron to apastron, depending upon the radiative time constants.
At periastron, its outer atmospheric temperature may be too high for
water to condense, rendering $\upsilon$ And d a Class III object,
like $\upsilon$ And c.  At apastron, the water clouds of our fiducial
model may reside somewhat deeper with greater column and optical depth.

The lower panel of Figure \ref{fig_spectrumUpsAnd} 
shows the visible and near-infrared phase-averaged
contrast between the planetary and stellar fluxes for $\upsilon$ And b, c,
and d.  Again, this quantity is an average
of the planetary flux emerging into all solid angles
divided by the stellar flux (uniform in solid angle) at the same
distance from the system.  In each case, a planetary radius equal to that
of Jupiter is assumed.  For $\upsilon$ And b, the planet-to-star flux ratio is greater
near 4 $\mu$m, where it approaches $3\times 10^{-4}$.  For comparison,
it is worse than $10^{-5}$ throughout the visible.
For $\upsilon$ And c and d, the planet-to-star flux ratio
is greatest in the $\sim$ 4.1 to 4.8 $\mu$m range,
where it is $\sim$ 10$^{-5}$ and $3\times 10^{-6}$, respectively.
Elsewhere, it is below 10$^{-6}$ (below $10^{-8}$ throughout most
of the visible).  Since these numbers are based on a phase-averaged
ratio, they are a bit pessimistic compared with what one would expect
to observe at full phase, which could improve the ratios by over a factor
of two, depending upon the details of the planetary phase function
\citep{SWS}.  Note that due to the large M$_p$sin($i$) of $\upsilon$ And d
and the correspondingly higher inner boundary flux, its spectrum in the near-IR $Z$, $J$, $H$, and $K$
bands is brighter than that of an old, lower-mass EGP at the same distance (see \S\ref{sec_genericseq}
and Figs. \ref{fig_spectrumG0AU} and \ref{fig_spectrumG0AU500}).

\subsection{The Transiting Planet, HD 209458b}
\label{sec_HD209458b}

HD 209458b was discovered by the usual radial velocity
method \citep{Henry00,Mazeh00} to be orbiting with a semi-major axis of only
0.045 AU ($e$ $\approx$ 0) around a G0V star, but
subsequent dramatic photometry \citep{Charbonneau00, Henry00}
revealed that it transits its
primary.  At the time of this writing, HD 209458b is still the
only transiting EGP certain to exist (there are several candidates
currently under analysis).  However, transits should not be
very rare for Class V roasters.  A transit
will be observed if the inclination of a planet's orbit ($i$) is
greater than
$\cos^{-1}[(R_*+R_p)/a]$, where $R_*$ is the stellar
radius, $R_p$ is the planetary radius, and $a$ is the
orbital distance of the planet.  For roasters like HD 209458b,
this corresponds to an inclination greater than $\sim$84 degrees.
HD 209458b-like systems with    
a random distribution of inclinations on the sky should transit 
approximately 10\% of the time
\footnote{The integrated probability
for a system to be between 0 and $i$ degrees (P[$0^o$,$i$]) is just
1-cos($i$), while P[$i$,$90^o$] = cos($i$).}.
Of course, such
a fortunate orientation immediately reveals the mass of the planet.
A good estimate of the radius of the planet can be obtained from
the photometric transit depth, since the percentage
decrease in the stellar flux
is approximately given by $(R_p/R_*)^2$. 
With a mass of 0.69\mj \citep{Mazeh00, CodySasselov02} and a radius of 1.35R$_{\textrm{J}}$,
HD 209458b has a surface gravity of $\sim$ 980 cm s$^{-2}$.

A Class V roaster, the atmosphere of HD 209458b is expected to 
have high silicate and iron clouds with bases at $\sim$ 5-10 mbar.
Our fiducial model has a surface gravity
of 980 cm s$^{-2}$ and an inner boundary
flux corresponding to \teff  = 500 K.  Figure \ref{fig_spectrumHD209458b}
depicts our HD 209458b model spectrum and an assumed spectrum for 
the primary (a scaled Kurucz model; upper panel).  Additionally, the
phase-averaged planet-to-star flux ratio is shown in the lower panel
of this figure.  The T-P profile of HD 209458b is depicted in Figure
\ref{fig_TPGJ876}. As with
$\upsilon$ And b, the planet-to-star flux ratio is large near 4 $\mu$m, where for HD
209458b it approaches 10$^{-3}$.  The sodium and potassium absorption
keeps the ratio low in the visible, but between the resonance lines
($\sim$ 0.7 $\mu$m) the ratio is approximately $3\times 10^{-5}$, as
it is blueward of the sodium line near 0.4 $\mu$m.  One caveat is that
nonequilibrium photochemical products (not included in these calculations) might absorb strongly in the
ultraviolet and blue regions of the spectrum (cf. Jupiter), so the
contrast at such short wavelengths is quite uncertain for HD 209458b
and other roasters.

Recent landmark observations of the HD 209458b transit by Charbonneau
\etal  (2002) in the vicinity of the Na-D doublet reveal
a photometric dimming in this region of the spectrum relative to 
adjacent bands.  This has been interpreted as the first detection of
sodium in the atmosphere of an EGP (and, importantly, the first detection
of an EGP atmosphere) and it is in general accord with theoretical
predictions \citep{Hubbard01,SeagerSasselov00}.  However, the level of
absorption inferred is less than what 
one would expect from basic theoretical models that assume a solar abundance
of sodium in neutral form and no clouds.  It should be pointed out that the transit spectrum
and the emergent phase-averaged spectrum are in general different.  The transit spectrum
samples to a greater degree the characteristics of the planetary limb and terminator.
The emergent spectrum is a whole-body quantity.  As a result, differences
between transit spectra for different models do not translate into corresponding
differences in the phase-averaged emergent spectrum.  

In their model for HD 209458b, BHA assume that heavy metals such 
as Mg, Ca, and Al exist in gaseous form in its outermost
atmosphere.  Their justification for the presence of these 
metal gases at low pressures is their particularly hot
T-P profile, which is above the condensation temperatures of forsterite, enstatite,
gehlenite, and other condensates that would sequester and rain out these
metals if the outer atmosphere were somewhat cooler.  In fact, our
fiducial model for HD 209458b is several hundred degrees Kelvin cooler
in the outermost atmosphere, leading us to conclude that 
when account is taken of the condensation and rainout of metal-rich refractories,
the outer atmosphere should indeed be cooler and clear of these metals. 
In this regard, our general model results jibe with the work 
of others on close-in EGPs \citep{SWS, Goukenleuque00}.

\subsection{Orbiting a Cool Star: The GJ 876 System}

The GJ 876 system (a.k.a. Gliese 876)\citep{Marcy98, Delfosse98}
is comprised of two planets orbiting at 0.13 AU and 0.21 AU about
a late main sequence star of spectral type M4.  The more massive
and first of the two planets discovered, GJ 876b (M$_p$sin($i$) = 1.89\mj),
has the wider orbit (with $e$ $\approx 0.10$), while GJ 876 c
(M$_p$sin($i$) = 0.56\mj) has the more eccentric orbit ($e$ $\approx$ 0.27).
A recent astrometric study using the Fine Guidance Sensor on HST
determined the mass of GJ 876b to be 1.89$\pm$0.34\mj (Benedict \etal  2002).

GJ 876b and c are both Class III planets because their temperatures
are too cool for a silicate layer to appear in the troposphere, but 
too hot for H$_2$O to condense.  The result
of such a ``clear'' atmosphere is a strongly wavelength-dependent emergent
spectrum, which is governed largely by the gaseous molecular absorption
bands of water, methane, and ammonia.  Given their similarities in composition,
GJ 876b and c are expected to have very similar spectra.  The factor
of $\sim$ 3 difference in their M$_p$sin($i$) implies somewhat different
surface gravities and lower boundary fluxes, but both objects are close
enough to their star that the T-P structure in their outer atmospheres
is determined more strongly by the stellar irradiation than by
their intrinsic luminosities.  

For GJ 876b, we use an inner boundary flux corresponding to
\teff  = 150 K and a surface gravity of $6\times 10^{3}$ cm s$^{-2}$.
For GJ 876c, we use an inner boundary flux corresponding to \teff  = 100 K and a
surface gravity 1/3 that of GJ 876b, or $2\times 10^{3}$ cm s$^{-2}$.
The emergent spectra of the two planets, along with an assumed spectrum
of the primary (a scaled Kurucz model stellar spectrum), are shown in Figure
\ref{fig_contrastGJ876} (upper panel).  Figure
\ref{fig_TPGJ876} depicts the planetary T-P profiles.  Given their
similarities in composition, GJ 876b and c are expected to have
very similar emergent spectra.  One caveat is that, while the T-P
profile of GJ 876b never crosses the water condensation curve, it
does come within $\sim$ 10 K of it near about 1 mbar.  Given 
somewhat lower incident irradiation than that of our scaled Kurucz
model for GJ 876, or given an observation of GJ 876b at apastron,
some water condensation may occur in its outermost atmosphere, rendering
it a Class II EGP.  However, we retain the Class-III designation for
our fiducial model of GJ 876 b for its given semimajor axis and any
reasonable range in surface gravity or lower boundary flux.
The lower panel of Figure \ref{fig_contrastGJ876} depicts the phase-averaged
planet-to-star flux ratios
for GJ 876b and c.  For both planets, this ratio peaks near 4.4 $\mu$m,
where it reaches a value of $\sim$ 10$^{-5}$.  In contrast, in the visible
region, the M-dwarf primary flux overpowers that of either planet by a
factor of at least 10$^8$.

\subsection{51 Pegasi b and Tau Bootes b}
\label{sec_PegBoo}

In 1995, Mayor \& Queloz discovered the first EGP, 51 Pegasi b
(M$_p$sin($i$) = 0.45 \mj), via
the radial velocity method.  The discovery of such a planet around
another star would alone cause a great deal of excitement in the
astronomical community, but the fact that its orbital period was only
4.23 days at a minuscule 0.05 AU was nothing less than astonishing.   
Perhaps the fact that this planet orbited a solar-type star made the
observation even more dubious to some.  After all, this is a much
different beast than the giant planets of the Solar System we know.  
The 51 Peg b observation is not due merely to non-radial stellar pulsations
\citep{Gray97, GrayHatzes97},
nor is the radial velocity variation due to a nearly face-on orbit
of two identical stars \citep{BlackStepinski01, StepinskiBlack01}.
And although 51 Peg b
remains to be detected {\it directly}, almost everyone in the astronomical
community now accepts it and the other roasters as bona fide planets.

The $\tau$ Boo system \citep{Butler97} gained popularity
when Cameron \etal  (1999) claimed
a detection in reflected light near 0.48 $\mu$m.  This detection was
not confirmed by Charbonneau \etal  (2000), who constrained
the albedo to be low in this region ($\sless$ 0.3 assuming a phase
function and a highly inclined orbit).  If Cameron's detection were
true, $\tau$ Boo b would have had an albedo as high as that of Jupiter
in this region and a troubling radius of $\sim$ 1.8\mj---troubling because
such a large radius would be well out of bounds relative to theoretical
models for a high gravity planet \citep{Guillot96, Burrows00b}.  This detection was
later retracted by the authors, but $\tau$ Boo b remains a particularly
interesting object because, with M$_p$sin($i$) = 4.09 \mj, it is more
massive than most of the other roasters known.

Since they orbit their primaries so closely,
51 Peg b and $\tau$ Boo b are both Class V roasters,
but these systems do differ.
51 Peg b orbits a G2-2.5V Sun-like star at 0.05 AU, while
$\tau$ Boo b orbits a hotter F7V star at 0.046 AU.  Additionally, their
projected masses differ by about a factor of 10.

As the prototype EGP, 51 Peg b has been modeled by a number
of researchers (Seager \& Sasselov 1998; Goukenleuque \etal  2000;
BHA)
and the model T-P structures and emergent spectra vary considerably.
In all cases, a self-consistent planar atmosphere code is used.  Perhaps the
greatest difference is between the models of Goukenleuque \etal  and
BHA.  Specifically, BHA's atmospheric temperature at P = 1 mbar is
above 1650 K (``AMES-Cond'' model; cloud-free), while that of
the most analogous Goukenleuque \etal  model
is under 900 K.  At P = 1 bar, the temperature of the BHA model is
$\sim$ 2000 K, while that of Goukenleuque is $\sim$ 1400 K.  To first
order, this difference can be explained by the fact that Goukenleuque
\etal  redistribute the incident power over the surface
of the full sphere (weighting the incident flux by a factor of 1/4,
as described in \S \ref{ch_numerical}), while BHA treat only
the substellar point of the planetary atmosphere (no weighting of the
incident flux).

In order to investigate the dependence of the T-P structure using our
own models, we ran a cloud-free 51 Peg b model with no flux weighting,
one with a weighting of 1/4, and one with a weighting of 1/2.  The
flux weighting of 1/2 corresponds to the irradiation of the
day-side hemisphere only, the default weighting used in the present
work.  The resulting T-P profiles are depicted in Figure
\ref{fig_TP51Pegbweight}.  For our 1/4-weighted model, the atmospheric
temperature at 1 mbar is 965 K, and at 1 bar it is 1830 K.  For our
unweighted model, the temperature at 1 mbar is 1390 K, and at 1 bar it
is 2370 K.  So for our models, this difference in the flux weighting
results in a 35\% difference in the temperature at 1 mbar and a 25\%
difference at 1 bar.  The difference between the Goukenleuque \etal and
BHA models at 1 mbar is 60\%, while at 1 bar this difference
is a more modest 35\%.  Hence, while differences in the incident flux weighting
can in part explain the differences in the T-P structures of published models, 
such differences are not a complete explanation. 
Other less obvious differences must
exist as well, such as atmospheric composition, opacities, incident flux
(beyond the weighting issue), or the inner boundary condition implemented.   

Our fiducial model of 51 Peg b has a silicate (forsterite) cloud base
at 20 mbar and an iron cloud base at 30 mbar.  we assume a cloud height
of one pressure scale height, an inner boundary flux
of \teff  = 500 K, and a low surface gravity of 10$^3$ cm s$^{-2}$, due
to the small M$_p$sin($i$) (0.44 \mj) of 51 Peg b.  In order to achieve
numerical convergence of the 51 Peg b model, we found it
necessary to attenuate the clouds to 10\% of the elemental iron and
magnesium abundances.  This should be taken into account when comparing
the T-P structure and emergent spectrum of 51 Peg b with those of
other roasters, for which such an adjustment to the condensate concentration
has not been made. 

Tau Boo b \citep{Butler97} orbits at only 0.046 AU
from a hot F7 main sequence star.
For our fiducial $\tau$ Boo b model, we use a surface gravity of 10$^4$
cm s$^{-2}$ (an order of magnitude larger than that used for 51 Peg b)
and an inner boundary flux of \teff  = 500 K.  As can be seen in Figure
\ref{fig_TPPegBoo}, the T-P profiles of $\tau$ Boo b (both with and without
clouds) and 51 Peg b are
fairly similar, especially considering their very different surface gravities
and primary spectral types. 
This can be explained quite simply:  the same
object orbiting a hotter star naturally would have a higher temperature at a
given pressure, but on the other hand, hydrostatic equilibrium dictates
that pressure increases with surface gravity.
So in this case, the two effects work against each other, resulting
in T-P profiles that are actually closer to each other in some regions
than they would be if either the surface gravities of the two planets
were the same, or if the planets orbited similar stars.  Gravity and
temperature also have
implications for the positions of cloud decks in these and other atmospheres.
A cloud deck will reside deeper in an atmosphere as
surface gravity increases, but higher in an atmosphere with increasing temperature. 
For 51 Peg b and $\tau$ Boo b, the profiles are so close that
they are coincident with the condensation curves at essentially the
same pressures.  As a result, these planets should condense clouds at very
similar atmospheric pressures.

The atmosphere model of $\tau$ Boo b is particularly interesting because it has
a convection zone at low pressures, toward the top of the forsterite cloud
around 10 mbar.  This is the only model for which such a low-pressure
convective region has resulted, and it persists with either fine or coarse
zoning within the cloud region.  The strong incident flux
from the primary (F7V at 0.046 AU) in combination with the opacity of the
forsterite cloud results in a steep temperature gradient and, hence,
the onset of convective energy transport, as depicted by the dashed portion
of the T-P profile of $\tau$ Boo in Figure \ref{fig_TPPegBoo}.
The model spectra of $\tau$ Boo b and 51 Peg b, and their planet-to-star
flux contrasts, are shown in Figure
\ref{fig_spectrumPegBoo}.  These spectra are characteristic of those in
Class V described in \S\ref{sec_classV}.

\subsection{The 55 Cancri System and its Long-Period Planet}
\label{sec_55Cancri}

The 55 Cancri system is made up of at least two EGPs orbiting
a G8 main sequence star.  The inner planet \citep{Butler97} has a
nearly circular orbit ($e$ $\approx$ 0.03) at 0.11 AU, with
M$_p$sin($i$) = 0.84\mj.  The recently discovered outer planet \citep{Marcy02}
is the first EGP found to orbit its primary at a distance greater than
Jupiter's distance from the Sun.  This object has a semimajor axis
of  $\sim$ 5.5 AU, an eccentricity of 0.16, and M$_p$sin($i$) of 4.0\mj.
The outer object has been designated, 55 Cancri d (rather than c), because
judging by the radial velocity residuals, a third planet of sub-Jupiter
mass may exist at 0.24 AU.  Currently, Marcy \etal consider the existence
of this third planet to be a good possibility, but not a firm detection
because the additional periodicity may be due to rotating inhomogeneities
on the stellar surface.

55 Cancri b is a Class IV planet.  Our fiducial model of this EGP
assumes a surface gravity of $3\times 10^3$ cm s$^{-2}$ and an
inner boundary flux of \teff  = 500 K.  A silicate cloud resides
at depth, with a base at 20 bars and a vertical extent of one pressure
scale height.  As detailed in \S \ref{sec_class4}, this deep cloud
does not have a significant effect on the emergent visible and near-infrared
fluxes of 55 Cnc b.

55 Cancri d is a Class II
object.  One might estimate that, with an orbital distance greater than
that of Jupiter and a primary cooler than our Sun, 55 Cnc d would be
a Jovian-like Class I object with ammonia clouds.  However, because
of its large minimum mass, such a planet will be intrinsically hotter
than Jupiter, even for a relatively late age \citep{Burrows97}.
With a conservative inner boundary flux corresponding to
\teff  = 150 K, the atmosphere of 55 Cnc d is too warm for ammonia
to condense.

Figure \ref{fig_spectrum55Cnc} depicts the spectra of 55 Cnc b
and d and their planet-to-star flux contrasts, while Figure \ref{fig_PT55Cnc} shows the T-P profiles
of these planets.
The Class IV designation of 55 Cnc b is manifest, due to its strong
sodium and potassium resonance features and the methane
absorption at 3.3 $\mu$m.  The spectrum of 55 Cnc d exhibits strong
molecular absorption features in the visible and near-infrared, but
a water cloud between $\sim$ 0.6 and 1.6 bars (one pressure scale height)
reduces the strengths of these features considerably.  Note that this water cloud is
deeper and more optically thick than that in the atmosphere of$\upsilon$ And d
(\S \ref{sec_UpsAnd}) simply because the T-P profile of 55 Cnc d intersects
the water condensation curve at significantly higher pressure.
However, whether a water cloud is
high and optically thin or relatively deep and optical thick, the basic
result of an elevated visible and near-infrared spectrum holds.

Figure \ref{fig_contrast55Cnc30} shows the phase-averaged
planet-to-star flux ratios
for 55 Cnc b and d out to 30 microns.
55 Cnc d has an unusually large angular separation from its
primary of $\sim$0.45$^{\prime\prime}$, making it an enticing candidate for
direct detection.  A good bet may be between 4.2 and 4.8 $\mu$m,
where there is little opacity and the planet-to-star flux ratio is
better than 10$^{-6}$.  Between $\sim$5 and 6 $\mu$m, the methane,
water, and ammonia opacities strengthen, but beyond a strong methane
band between $\sim$7 and 8 $\mu$m, the planet-to-flux ratio
improves substantially.  There are a number of striking differences
between 55 Cnc b and d.  In the visible region, the strong sodium
and potassium lines of 55 Cnc b contrast with the methane-dominated
spectrum of 55 Cnc d, whose continuum is provided by both Rayleigh
scattering and scattering off condensed H$_2$O.  In the near-infrared,
the troughs and peaks of 55 Cnc d (due to gaseous H$_2$O and methane)
vary more widely than those of 55 Cnc b.  This effect can be explained
largely by H$_2$O opacity differences at high and low temperatures.
Additionally, the methane opacity is stronger
in the lower-temperature 55 Cnc d due to a higher equilibrium abundance
of this molecule.  Between $\sim$ 4 and 5 microns, CO opacity
in 55 Cnc b closes the gaping opacity window seen in the
55 Cnc d spectrum.  Toward longer wavelengths, the 55 Cnc d
planet-to-star flux ratio continues to increase through 30 $\mu$m, while that of 55 Cnc b
is fairly flat.

\subsection{Some Other EGPs of Interest}
\label{sec_otherEGPs}

In addition to the systems chosen above, many more
deserve specific attention due to their interesting orbital
parameters, masses, primaries, and/or proximity.  We model a few of these in
the present work, including HD 114762b, $\epsilon$ Eridani b, and HD 83443b.

The massive HD 114762b \citep{Latham89} is often referred to as a brown
dwarf, not a planet.  It has a projected mass of 11\mj  \citep{Marcy98},
and its mass has been
estimated to be as high as 145\mj \citep{Han01}, which of course would 
make this object a low-mass star, but this result is quite controversial.
Whatever its mass, HD 114762b orbits its F9V primary at a distance of
$\sim$ 0.3 AU with an eccentricity of 0.335.
The spectrum out to 5 $\mu$m of our fiducial HD 114762b model (inner
boundary flux of 500 K, surface gravity of $10^5$ cm s$^{-2}$) is shown
in Figure \ref{fig_remainder}.

Epsilon Eridani b is a controversial, but potentially very important,
EGP because of its wide orbit around a star that is only 3.2 parsecs
away from Earth.  If it is real, its separation from its K2V primary is
$\sim$ 1 arcsecond, significantly larger than that of any other known system.
The radial velocity detection of $\epsilon$ Eri b \citep{Hatzes00} is
considered marginal by many, but its orbital parameters are M$_p$sin($i$) = 0.86
\mj, $a$ = 3.3 AU, and $e$ $\sim$ 0.61, a very eccentric orbit.
This system is probably under 1 Gyr of age \citep{Greaves98}, indicating
that $\epsilon$ Eri b is quite a bit warmer than Jupiter if its mass is
$\sim$ 1\mj.  Indeed, an astrometric study claimed a mass of
1.2 $\pm$0.33 \mj \citep{Gatewood00}.  Using a fiducial lower boundary flux
corresponding to \teff  = 170 K and a gravity of $3\times 10^3$ cm s$^{-2}$,
$\epsilon$ Eri b is a Class II EGP.  The resulting spectrum and
planet to star flux ratio from 0.4 to
30 $\mu$m are shown in Figure \ref{fig_spectrumEpsErib}.   A water
cloud resides at $\sim$1 bar pressure in $\epsilon$ Eri b's atmosphere,
but given the large eccentricity of this EGP, the cloud
may be deeper in the atmosphere at apastron and higher at periastron.
Even at apastron, it is unlikely that the atmospheric temperature will
be low enough for ammonia to condense.  In fact, an atmosphere model for
$\epsilon$ Eri b at 5.3 AU indicates that the T-P profile falls
just short of the ammonia condensation curve.

HD 83443b \citep{Mayor00} is a Saturn-mass (M$_p$sin($i$) = 0.35\mj) planet with
the smallest orbital distance of any EGP known to date.  However, unlike
most known roasters, HD 83443b orbits a later-type
main sequence star (K0V).  Therefore, despite its status as the most
closely-orbiting roaster, it is a little cooler than its brethren. 
Because it is cooler, its silicate and iron clouds are somewhat deeper
than in most roasters.  In our fiducial model, the silicate (forsterite)
cloud resides at $\sim$ 30 mbar and an iron cloud is located
near 100 mbar.  We use a lower boundary flux corresponding to \teff  = 500 K and
a gravity of 900 cm s$^{-2}$.  Initially, due to radial velocity residuals,
this system was believed to have another planet orbiting at 0.174 AU,
with M$_p$sin($i$) = 0.17\mj.  However, follow-up measurements
failed to reveal such a planet \citep{Butler02}.

Figure \ref{fig_remainder} depicts the visible and near-infrared
spectra of HD 83443b, HD 114762b, and $\epsilon$ Eri b, and their
planet-to-star flux contrasts.
The T-P profiles for each of these objects are given in Figure
\ref{fig_TPremainder}. 

\section{Detection and Imaging of EGPs in Light of Theoretical Models}
\label{ch_future}

A number of very innovative ground-based
and space-based methods are now under development, or under serious
consideration, to detect EGPs directly.
Theoretical models have had and will continue to play an important
role in guiding observational campaigns.  This
section reviews some of the many observational methods and
their prospects for success in light of our theoretical
models.  Where available, instrumental sensitivities are used in
combination with spectral models to reveal which of the currently
known EGPs are likely to be detectable in the near term.

The central problem in direct imaging, photometric phase detection,
and spectral separation methods is the huge contrast between the
central star and its companion(s).  The tiny planet-to-star
flux ratios and minuscule angular separations require 
observers and instrumentalists to develop very clever detection methods.
Space-based instrumentation has a clear advantage due to the
absence atmospheric turbulence, but these missions are costly and
will not be in operation for several years.  Hence, a plethora of
ground-based techniques are already in use or in the advanced stages
of development.  Many of these techniques are limited by the angular 
separation of the planet from the primary, whose maximum value is given
by $a/D$, where $D$ is the distance from the Earth.  Table 1 lists the
maximum angular separations at the Earth of the EGPs studied in this paper, along
with those of others with the greatest separations.  The objects on this
list will be some of the most important targets during the coming decade 
of direct EGP research.

\subsection{Spectral Separation Methods}

The close-in EGPs, such as 51 Peg b or $\tau$ Boo b, are separated from
their primaries by only 0.8--4 milliarcseconds.  While direct imaging of such 
systems from the ground at any wavelength would be a great challenge,
a radial-velocity-based spectral separation method is being
used by some researchers \citep{Cameron99, Charbonneau00}.
With knowledge of the radial velocity variation of the stellar absorption
lines due the primary's orbit about the system's center of mass, one can also
expect a time-dependent opposite shift of the component of the stellar spectrum
that is reflected by the planet.  By modeling and subtracting the
radial velocity-dependent stellar spectrum, the reflected component
of the stellar spectrum should be left in the noisy residuals.  The
great difficulty is the large contrast between the actual stellar
spectrum and the reflected component.

The intensity of the reflected stellar component depends upon the
albedo of the EGP, which for a Class V EGP tends to be highest in the visible region of the
spectrum.  At short enough wavelength (i.e., in the blue region of
the spectrum), the thermal emission of even a Class V roaster is
negligible.  Hence, to a good approximation, all the flux in this region
is reflected flux, so the wavelength dependence of the reflected
stellar component can be derived by combining the stellar spectrum with
the planetary albedo.  According to our models, the phase-averaged
planet-to-star flux ratio in this region is below $2\times 10^{-5}$ in all
cases, and worse than $10^{-5}$ in some.  When the planet is near full
phase, these values may improve by a factor of 2-3, bringing the ratio
into the vicinity of the detection limit
of $\sim$ $0.5-1\times 10^{-4}$ \citep{Charbonneau00, Cameron99}, although the velocity shift in
the reflected component relative to the stellar spectrum approaches
zero as the planet approaches opposition.

The difficulty of this method is illustrated by the false detection of $\tau$ Boo b,
as discussed in \S \ref{sec_PegBoo}.  Yet, this technique may be our
best hope for detecting a roaster in reflected light from the
ground.  Even if no detections are made, improvements in the accuracy 
of this method can provide upper limits to planetary albedos as
a function of wavelength and orbital inclination, thereby putting soft
contraints on the outer atmospheric compositions.  However, the motivation and
rationale for spectral separation methods will continue to be
the detection of reflected light.

\subsection{Ground-Based Differential Direct Imaging}

Differential imaging entails the simultaneous imaging of a
candidate system in two or more adjacent narrow band filters.
The bandpasses are chosen so that at least one resides in
a spectral region where a deep planetary absorption feature
is expected, and another is situated in an adjacent opacity
window, where the planet flux is predicted to be relatively
large.  However, the central star is bright in both
bandpasses, so by imaging the candidate system and then differencing
the two bandpass fluxes, one can subtract out the
offending starlight, hopefully revealing the dim planet.
Just how dim a planet can be detected depends on a number
of parameters, including the angular separation, the telescope size,
instrument design, filter bandpasses, and CCD efficiency.

One instrument design makes use of dual Wollaston prisms for
image splitting using the Arizona Infrared Imager and
Echelle Spectrograph (ARIES), the MMT 6.5-meter telescope, and an
adaptive optics (AO) secondary \citep{Freed02}.  With this set-up,
on schedule for use in mid-2003, a 3$\sigma$ detection in the H-band
region is expected for
an EGP-primary separation of
0.2$^{\prime\prime}$ and planet-to-star flux ratio of $\sim$ $3\times 10^{-6}$.
For larger separations, the detectable flux ratio improves significantly.
For example, at a 1$^{\prime\prime}$ separation, the limit is $\sim$ $10^{-7}$ for
a 2-hour exposure.  These estimations are based on the placement
of three narrow-band filters between 1.56 and 1.68 $\mu$m, toward
the short-wavelength edge of a strong methane absorption feature.
According to our models, none of the currently known EGPs will be detectable
with this technique (given these sensitivities), since there are only a few with separations
on the order of $\sgreat$ 0.2$^{\prime\prime}$ and their theoretical plant-to-star flux ratios
are too low by 2-3 orders of magnitude.  However, this method
may succeed in discovering new, young Class III EGPs at 5-10$^+$ AU
from their primaries.

Another similar instrument, TRIDENT, is optimized for the
detection of EGPs with orbital distances greater than 5 AU
in a narrow spectral region between 1.57 and 1.68 $\mu$m \citep{Marois02}.
Using the CFHT 3.6-meter telescope with an AO system, a
test observation
was made of a faint companion 0.5$^{\prime\prime}$ from its primary.  In this case,
the contrast ratio was $\sim$ $2\times 10^{-4}$, a very easy
detection relative to that of an EGP.  However, this detection confirms that
the basic technique does work, and plans to use this
instrument on the Gemini 8.2-meter telescope are under way.

\subsection{Transit Searches and Photometric Reflected Light Detection}

EGPs that have highly inclined orbits from our vantage point
have a high probability of transiting their stars.  In particular,
this is true of Class V roasters, for which the overall probability
of a transit is approximately 10\%, assuming a random distribution
of orbital inclinations
(\S \ref{sec_HD209458b}).  The resulting stellar photometric dimming of
$\sim$1.5\% is easily observable from the ground.  This fact has led to
an abundance of inexpensive ground-based transit searches far too numerous
to discuss here.  One such search, STARE \citep{BrownCharbonneau99},
proved its worth with the
detection of HD 209458b (see also Henry \etal\ 2002), the first and only
verified transiting EGP to date.
Like STARE, most other programs use small telescopes and 
monitor stars in an automated fashion.  One search collaboration is OGLE
\citep{Udalski02}.
Originally developed for the
detection of optical gravitational lensing events in the
galactic disk and bulge, OGLE has amassed an impressive collection of stellar light curves.
Over 50,000 stars have been analyzed for variability due to a possible transiting planet (a
light curve showing a flat-bottomed eclipse).  For 42 of the stars,
multiple transits were observed, and two of these may be
close-in transiting EGPs given their short periods and photometric
dimmings similar to that of HD 209458b \citep{Udalski02}.  Follow-up work, including a
radial velocity determination of M$_p$sin($i$) for these putative companions will
be necessary to confirm their planetary nature.

In principle, one should be able to detect photometric variations for
EGPs that do $not$ transit their stars, due to differences in reflected
light as a planet runs through its various orbital phases.
The difficulty is that such variations are likely to be
only a micromagnitude ($\mu$mag) to tens of micromagnitudes, depending upon
the inclination of the orbit, the existence of high clouds, and their
constituent condensates \citep{Green02, SWS}.  Such variations
are too small to be observed from the ground, but several space-based
missions are in the works, including MOST \citep{Matthews01},
COROT \citep{AntonelloRuiz02}, MONS \citep{Christensen00},
and $Kepler$ \citep{Koch98}.

MOST is essentially complete and due for launch in April of 2003.  This
small, but very sensitive instrument, will be used for transit observations
and for reflected light detection.  MOST contains a 15-cm mirror and a
single broad-band filter from 0.35 to 0.70 $\mu$m.  An array of microlenses
projects a large, stable image onto a CCD.  Sensitivities are on
the order of a few $\mu$mag, easily sufficient according to our models for the detection
of Class V roasters with high enstatite clouds, which are quite reflective
in the visible.  However, if the clouds are buried and/or are composed
of more highly absorbing species (such as forsterite), detection will
be much more difficult.  At the very least, MOST will constrain the outer
atmospheric compositions of a number of close-in EGPs, but hopefully a
few ground-breaking detections will be made.

COROT, scheduled for launch in 2005, will be capable of multicolor
photometry.  Its wide-field-of-view, 27.4-cm telescope is suitable
for the simultaneous monitoring of many stars.  Throughout its mission,
6,000-12,000 stars will be searched for transits by terrestrial and gas
giant planets.  Additionally, detections of reflected
light from close-in nontransiting planets through blue and red bandpasses may
be possible.  Such observations could provide rudimentary color data
for some EGP atmospheres. 

$Kepler$ has been selected as a NASA Discovery mission for the detection of
transits by Earth-like planets, and it is scheduled for launch in 2006.  
A much larger instrument than MOST or COROT, Kepler has a 1.6-meter primary mirror and an array
of 42 CCD detectors.  It
will monitor $\sim$100,000 stars in visible light for planetary transits.
Hundreds of EGP transits are
expected, and follow-up radial-velocity measurements will be possible for
some of the closer systems.  A determination of the planetary albedos
of transiting planets should be
possible by inferring a planetary radius from the transit depth, a
semimajor axis from the period of the orbit, and the
reflected light modulation between transits, because the fraction of
reflected stellar light as a function of planetary phase can yield the
albedo.

\subsection{Interferometric Imaging}

The precision of interferometry provides promising methods for the
detection of EGPs.  These include both astrometric and
direct imaging techniques, which are being developed for both ground-based
and space-based programs.

One of the most innovative methods is nulling interferometry, whereby
the light from a star is strongly suppressed while that from
a companion is not.
A nulling interferometer is similar to a Michelson interferometer,
except that the path lengths from a pair of collectors are purposely
maintained at a difference of half a wavelength so that the light from
the star interferes destructively.  However, at a minuscule separation
angle from the central star (perhaps at the position of a planetary companion),
the phase relation for the incoming radiation is such that the light
will interfere constructively.  Moreover, the instrument can be
tuned so that the peak of the constructive interference occurs
at various angles from the central star.

The nulling of a star was tested successfully on the multiple-mirrored
MMT telescope before its decommissioning and upgrade to a 6.5-meter telescope
\citep{Hinz00}.  However, the successful detection of EGPs likely will be left to
the Large Binocular Telescope Interferometer (LBTI), a twin 8.4-meter
mirror common-mount telescope due to be operational by late 2005.
Using the telescope's AO system along with a nulling beam combiner, which
includes a dielectric material to correct for the color-dependence
of light interference, at least a few of the currently known EGPs
should be detectable according to our model spectra.  An $image$ of the field around a nulled star,
not fringes, will be observable because the point spread function is
broader than the transmission function due to the interferometer
\citep{Hinz01}.  In the infrared L$^{\prime}$ Band ($\sim$ 3.6 $\mu$m), the
LBTI sensitivity is expected to be 2.1 $\mu$Jy for a 1-hour integration, and the
M-band (4.5-5 $\mu$m) sensitivity should be 21 $\mu$Jy for a 1-hour
integration.
A planet-star separation of $\sgreat$$0.03^{\prime\prime}$ will
be required \citep{Hinz01}.  Optical wavelengths are not currently being
considered for ground-based nulling due to the enormous difficulty of wavefront
corrections at these shorter wavelengths.

Among the systems we have modeled, $\upsilon$ And c and d,
55 Cnc d, and $\epsilon$ Eri b all have sufficient angular separations from
their primaries to be detected with the LBTI.  In the L$^{\prime}$ band, $\upsilon$ And c's average flux
density is near 10 $\mu$Jy, roughly a factor of five greater than
the LBTI sensitivity for a 1-hour integration.  Upsilon And d will
be too dim in the L$^{\prime}$ band by about an order of magnitude,
but its M-band flux density of $\sim$ 20 $\mu$Jy is comparable to
the LBTI sensitivity in that wavelength region.
$\upsilon$ And c's M-band flux density
is approximately a factor of 5 greater.  Although their angular separations
are large, our $\epsilon$ Eridani b and 55 Cnc d models indicate that these
objects will be too dim for imaging via nulling interferometry
with the LBTI in either the L$^{\prime}$ or M bands. 

A nulling interferometer is one of two proposed designs for the
Terrestrial Planet Finder (TPF), part of a very ambitious plan by NASA
for the detection and spectroscopic measurement of Earth-sized planets.  The
LBTI will provide a ground-based test of the techniques that in
principal can work even better outside of Earth's turbulent
atmosphere.  In addition to the characterization of terrestrial
planets, TPF should provided unprecedented spectral data
for EGP atmospheres.  Launch is anticipated between 2012 and 2015.

Another innovative interferometric development is the differential phase
method \citep{AkesonSwain00}.  Using the Keck Interferometer in the
infrared, a faint companion might be detected by simultaneously measuring
the fringe phase at two (or more) wavelengths at which the
planet-to-star flux ratios are different.  Because the primary 
and secondary sources provide different fractions of the total light
received at each of two wavelengths, and because the fractional fringe
separation depends on wavelength, a small phase difference will
occur.  With the twin 10-meter apertures of the Keck, a phase difference
as small as 0.1 milliradian might be detectable \citep{AkesonSwain00}.  Due
to their higher planet-to-star flux ratios, hot EGPs
with temperatures of $\sgreat$1000 K are favored by this
technique.  Of the currently known EGPs,
most of the close-in Class V objects meet this criterion.

The differential phase signal in radians is roughly given by
twice the difference of the planet-to-star flux ratios (Akeson 2002,
private communication).  This estimate holds for separations
as small as $\sim$2 milliarcseconds.  Using the $H$ and $K$ bands
\citep{AkesonSwain00}, the Class V roasters that we
have modeled, such as 51 Peg b, $\tau$ Boo b, and $\upsilon$ And b,
fall into the 0.1-0.2 milliradian phase-difference range using the H and K bands, the
two bands emphasized by Akeson \& Swain (2000), indicating that these roasters
may be detectable by this method.  In contrast, Class IV EGPs translate to well
below 0.1 milliradians.

A recent test run on the Palomar Testbed Interferometer (PTI) revealed that
the differential phase method may have some difficulties in ultimately
reaching 0.1 milliradian accuracy, due to water vapor turbulence
in Earth's atmosphere \citep{Akeson00}.  Some ideas for overcoming
these difficulties are currently in the works.  A refinement of the
differential phase method is highly desirable, because if it does prove
to be successful, not only detections, but low-resolution spectra of roasters
may be obtained by applying it at various wavelengths.
 
\subsection{Interferometric Astrometry}
Another class of interferometric techniques for the discovery of new EGPs is
high-accuracy, infrared astrometry.  This approach would complement the
radial-velocity method used
for those EGPs currently known,
providing orbital inclinations from which companion masses can be
derived.  The two major ground-based efforts now in development are
the VLTI (Very Large
Telescope Interferometer) and the Keck Interferometer (KIA).
Astrometric interferometry is an indirect detection method in which
the position of the central star is monitored as it orbits
the system's center of mass.  The reflex amplitude of the star
is proportional to the mass and distance of the secondary ($a_* =
a_pM_p/M_*$).  Therefore, with a system such as $\tau$ Boo b, for which the
planet-star angular separation is $\sim$2.7 milliarcseconds, the
reflex motion of the F7V primary will be $\sim$11 microarcseconds,
assuming a planetary mass of 5 \mj (the most massive of the Class V EGPs
to date).  
Astrometric accuracies
of $\sim$10 microarcseconds are expected for KIA and the VLTI.  Hence, such precision will enable
the measurement of planet masses and inclinations for the majority of
Class I through Class IV
EGP systems known today.

In the case
of the VLTI, its auxiliary 1.8-meter telescopes will be dedicated to
full-time interferometry, while the 8.2-meter telescopes will be used
for interferometry only intermittently.  Due to a baseline of up to 202 meters,
very high accuracy will be possible.  At 10 parsecs, a Jupiter could be detected at
an orbital distance as small as 0.1 AU, and a 0.1
\mj  object at 1 AU
should be measurable as well \citep{Paresce01}.  The Keck Interferometer,
with a somewhat shorter baseline,
is expected to have similar, though perhaps slightly inferior, resolving
power \citep{vanBelleVasisht98}.

The Space Interferometry Mission (SIM), scheduled for launch in
2009, is expected to achieve 1-microarcsecond astrometry, an accuracy that
will enable the measurement of the orbits of all the EGP systems discovered to date.  This
10-meter-baseline instrument could also act as an ideal space-based
testbed for one of two possible designs for the larger TPF mission.
GAIA, a European equivalent of SIM in terms of astrometric accuracy,
is scheduled for launch between 2010 and 2012. 

\subsection{Coronagraphic Imaging}
Coronagraphic imaging is a direct imaging method in which
starlight is blocked
by a specially-designed phase mask, or coronagraph.
Without such a mask,
the extremely faint image of an off-axis EGP would be washed out.
Due to the difficulties of visible and near-infrared
imaging from the ground, coronagraphic instruments suitable
for EGP detection are space-based.

$Eclipse$ is a 1.8-meter visible and near-infrared coronagraphic space telescope currently in
development at Jet Propulsion Laboratory \citep{Trauger00, Trauger01}.
Specifically designed for direct imaging of nearby ($\sim$15 pc) EGPs,
this ambitious instrument will contain precision
wavefront control technology to correct for even minute imperfections
in the optical systems.  $Eclipse$ is expected to be 3 orders of magnitude
better at
reducing the scattered and diffracted starlight between 0.3 and a
few arcseconds than any HST instrument
\citep{Trauger00}.  Within 1$^{\prime\prime}$ of an F or G star,
$Eclipse$ is being designed to achieve high-contrast imaging of 10$^9$
in a wide R-band filter (0.55-0.85 $\mu$m.); for
angular separations of only 0.1--0.3$^{\prime\prime}$, imaging contrasts
of 10$^8$ may be
possible.  $\epsilon$ Eri b, has a separation of $\sim$1$^{\prime\prime}$
and a modeled R-band plant-to-star flux ratio of $\sim$$3\times 10^{-9}$, making it
a very likely candidate for direct imaging by $Eclipse$.  With a
separation of $\sim$0.45$^{\prime\prime}$ and a planet-to-star ratio of 
$\sim$$3\times 10^{-10}$, 55 Cnc d will be close to the detection limit, but it
may be slightly too dim for direct R-band imaging.

The 6.5-meter Next Generation Space Telescope (NGST), scheduled for launch
in 2010, will house the Near Infrared Camera (NIRCam), including a
coronagraphic module.  From 2 to 5 $\mu$m, the NIRCam coronagraph will be capable
of 10$^8$--10$^9$ high-contrast imaging for planet-star separations of
$\sgreat$$0.1^{\prime\prime}$ \citep{Rieke02}.  Tunable narrow-band
filter modules from $\sim$2.5 $\mu$m to 4.5 $\mu$m will allow for the low-resolution reconstruction of
infrared EGP spectra (M. Rieke 2002, private communication).  Of the
detected EGPs that we have modeled, $\upsilon$ And d, 55 Cnc d,
and $\epsilon$ Eri b all have sufficient angular separations and
planet-to-star flux ratios to be imaged by NGST/NIRCam.  According
to our theoretical models, for $\upsilon$
And d, the most advantageous wavelength regions for imaging are the $K$ band
(2.2 $\mu$m) and the broad wavelength region between 3.6 and 5 $\mu$m.  55 Cnc d
and $\epsilon$ Eri b have
sufficient planet-to-star flux ratios for coronagraphic imaging within
only the 4-5 $\mu$m opacity window.
     
The Extra-solar Planet Imager (ESPI) is a space telescope with
a 1.5-meter square aperture and
a custom mask \citep{Melnick01}.
The resulting diffraction pattern is cross-shaped rather than circular,
and the telescope can be rotated so that a faint companion EGP
appears between two edges of the pattern.  The planet-star separation
limit is expected to be in the 0.3$^{\prime\prime}$ range.  Hence, for many systems,
Jupiter-like EGPs orbiting at a few AU or more may be detectable.
Another space-based coronagraphic instrument
of similar size is the Jovian Planet
Finder (JPF), a 1.46-meter coronagraphic optical telescope that would
be located on the International Space Station.  The JPF would survey
50 of the closest stars for EGPs orbiting between 2 and 20 AU and
would survey 2000 of the closest stars for more massive brown dwarf companions.

A visible coronagraphic instrument is also a design concept for TPF.  If
this path is chosen instead of the interferometric design, extremely
precise optics will be required to image terrestrial planets. 
Even minute imperfections scatter light and degrade contrast,
so an AO system will be used to correct the wavefront for such
imperfections (as with $Eclipse$).  Whichever concept is adopted, it is assured that
any instrument designed to reveal information about terrestrial planets
and their atmospheres will provide even more
detailed information about the gas giants known today and the 
hundreds more that will be discovered in the coming years.

\section{Summary and Conclusions}
\label{sec_conclusion}

In this paper we have conducted an extensive exploration of spectral
and atmospheric models of irradiated extrasolar giant planets. Our central
purpose has been to provide a map to observers, as well as a comprehensive
view of the theoretical possibilities.  We have investigated the dependence
of the phase-averaged emergent spectra on orbital distance, stellar type,
the presence or absence of clouds, cloud particle sizes, surface gravity, and
the inner flux boundary condition.  Included are calculations for irradiated
brown dwarfs, specific known EGP systems, a generic sequence around a G0V star, 
and our composition Classes I to V.  In addition, we have reviewed 
representative direct detection techniques in the context of our 
theoretical models.  

There are many overall, as well as specific, conclusions represented in
the previous sections and a full summary of them here would be heavily redundant.
Nevertheless, a small subset of the most salient points, in no particular order,
is useful.  They are that:

\begin{itemize}

\item the planet-to-star flux ratio is a very sensitive function of wavelength
and orbital distance.

\item EGP band fluxes are not strictly monotonic functions of orbital distance,
nor are the Bond and geometric albedos.

\item EGPs fall naturally into classes due
to qualitative similarities in the compositions and spectra of
objects within several broad atmospheric temperature ranges.

\item the mid-infrared region of the spectrum from 10 to 30 \mic has a favorable planet-to-star
flux ratio, even for distant EGPs, and does not decay as fast as $1/a^2$.

\item due to Rayleigh and/or grain scattering, the optical spectrum of an irradiated
brown dwarf can be very much brighter than that of a brown dwarf in isolation,
even when its near- and mid-infrared spectra remain relatively unaffected.

\item the fluxes in the $Z$, $J$, $H$, and $K$ bands of an EGP in a long-period orbit
can be enhanced above baseline levels (normally determined by stellar irradiation and
scattering alone), if the planet is either young or massive.

\item there is a $relatively$ bright feature within the 3.8 to 5 \mic
wavelength region for all irradiated EGPs, and in particular for the more distant
EGPs (such as 55 Cnc d, $\epsilon$ Eridani, 47 UMa c, Gliese 777A, and $\upsilon$ And d).  

\item as a result of the progressive decrease in atmospheric CO abundance,
the center of the 3.8 to 5.0 \mic feature shifts systematically from shorter to
longer wavelengths with increasing orbital distance (or decreasing stellar luminosity).

\item Rayleigh and grain scattering elevate the optical and near-IR fluxes,
but grain absorption depresses the mid-infrared fluxes (in particular in the ``5-\mic" band).

\item increasing surface gravity slightly decreases the flux shortward of $\sim$2.2 \mic,
but also slightly increases it longward of $\sim$2.2 \mic.   
The larger the gravity the smaller the peak-to-trough variations throughout
the spectrum.

\item the Na-D and K I resonance doublets are prominent features of the hottest, 
close-in EGPs (such as 51 Peg b, $\tau$ Boo b, HD209458b),
but quickly wane in importance with increasing orbital distance.

\end{itemize}

Table 1 lists some of the most prominent near-term targets for direct detection, as well
as all of the known EGPs for which we have provided theoretical spectra in this paper.
Our models collectively span the variety of discovered systems with their wide range
of orbital distances, stellar types, and M$_p$sin($i$)s.  However, much remains to be
done, including calculating phase functions (SWS), exploring metallicity and abundance
dependences, predicting the character of EGPs irradiated by stars of other spectral types
and luminosity classes, incorporating 3D meteorology, and investigating the effects of NLTE 
and non-equilibrium chemistry.  As a result, the theorist will be profitably engaged
for the foreseeable future.  However, the next major phase of EGP research should see
the direct measurement of EGP photons, first at low spectral resolution, then higher.
The many instruments, techniques, and telescopes being employed
to detect EGPs directly, summarized in \S\ref{ch_future}, 
suggest that we do not have long to wait before the theory we have presented 
here will be tested.

\acknowledgments
 
The authors are happy to thank Bill Hubbard, Jonathan Lunine,
Jim Liebert, Aigen Li, Christopher Sharp, Drew Milsom,
Maxim Volobuyev, Curtis Cooper, and Jonathan Fortney
for fruitful conversations and help during the
course of this work, as well as
NASA for its financial support via grants NAG5-10760
and NAG5-10629.

{}

\newpage

\begin{deluxetable}{cccccccc}
\tablewidth{16.5cm}
\tablenum{1}
\tablecaption{Interesting EGPs Listed by Angular Separation\label{EGPwide_tab}}
\tablehead{
\colhead{EGP} & \colhead{separation ($^{\prime\prime})$} & \colhead{star}
& \colhead{a (AU)} & \colhead{d (pc)}
& \colhead{P} & \colhead{Msin$i$ (\mj)} & \colhead{$e$}}
\startdata

$\epsilon$ Eri b & 1.0 & K2V & 3.3 & 3.2  & 6.85 yrs. & 0.86 & 0.61 \\
55 Cnc d          & 0.44 & G8V & 5.9 & 13.4 & 14.7 & 4.05 & 0.16  \\
47 UMa c          & 0.28 & G0V& 3.73& 13.3 & 7.10 & 0.76 & 0.1 \\
Gl 777A b         & 0.23& G6V& 3.65& 15.9 & 7.15 & 1.15 & $\sim$0 \\
$\upsilon$ And d  & 0.19 & F8V& 2.50& 13.5 & 3.47 & 4.61 & 0.41 \\
HD 39091b         & 0.16& G1IV& 3.34& 20.6 & 5.70 & 10.3 & 0.62 \\
47 UMa b          & 0.16 & G0V& 2.09& 13.3 & 2.98 & 2.54 & 0.06 \\
$\gamma$ Cephei b & 0.15 & K2V& 1.8& 11.8 & 2.5 & 1.25 & $\sim$0 \\
14 Her b          & 0.15 & K0V& 2.5 & 17   & 4.51 & 3.3  & 0.33 \\
HD 216437b        & 0.10 & G4V & 2.7 & 26.5 & 3.54 & 2.1  & 0.34 \\
HD 147513b        & 0.098 & G3V& 1.26& 12.9 & 1.48 & 1.0  & 0.52 \\
HD 33636b         & 0.094 & G0V& 2.7 & 28.7 & 4.43 & 7.71 & 0.41 \\
HD 13507b         & 0.091 & G0V& 2.39& 26.2 & 3.61 & 3.45 & 0.13 \\
HD 168443c        & 0.087 & G5V& 2.87& 33   & 4.76 & 17.1 & 0.23 \\
HD 50554b         & 0.077 & F8V& 2.38& 31.03& 3.50 & 4.9  & 0.42 \\
HD 106252b        & 0.070 & G0V& 2.61& 37.44& 4.11 & 6.81 & 0.54 \\
HD 10697b         & 0.067& G5IV& 2.0 & 30   & 2.99 & 6.59 & 0.12 \\
\\
$\upsilon$ And c  & 0.061 & F8V & 0.83 & 13.5 & 241 days & 2.11 & 0.18 \\
GJ 876b           &0.045 & M4V & 0.21 & 4.72 & 61.0 & 1.89 & 0.1 \\
GJ 876c           &0.028 & M4V & 0.13 & 4.72 & 30.1 & 0.56 & 0.27 \\
HD 114762b        &0.013& F9V&0.35 & 28 & 84.0 & 11.0 & 0.34 \\
55 Cnc b          &$8.2\times 10^{-3}$&G8V& 0.12 & 13.4 & 14.7 & 0.84 & 0.02 \\
$\upsilon$ And b  &$4.4\times 10^{-3}$&F8V& 0.059 & 13.5 & 4.62 & 0.71 & 0.034 \\
51 Peg b          &$3.4\times 10^{-3}$&G2V& 0.05 & 14.7 & 4.23 & 0.44 & 0.01 \\
$\tau$ Boo b      &$3.3\times 10^{-3}$&F7V& 0.05 & 15 & 3.31 & 4.09 & $\sim$0 \\
HD 209458b        &$9.6\times 10^{-4}$&G0V& 0.045& 47 & 3.52 & 0.69 & $\sim$0 \\
HD 83443b         &$8.7\times 10^{-4}$& K0V&0.038& 43.5 & 2.99 & 0.35 & 0.08 \\

\enddata
\end{deluxetable}

\newpage

\include{plots1}

\include{plots2}

\include{plots3}

\end{document}

%% file: plots1.tex

\begin{figure} 
\vspace*{6.0in}
\hbox to\hsize{\hfill\includegraphics{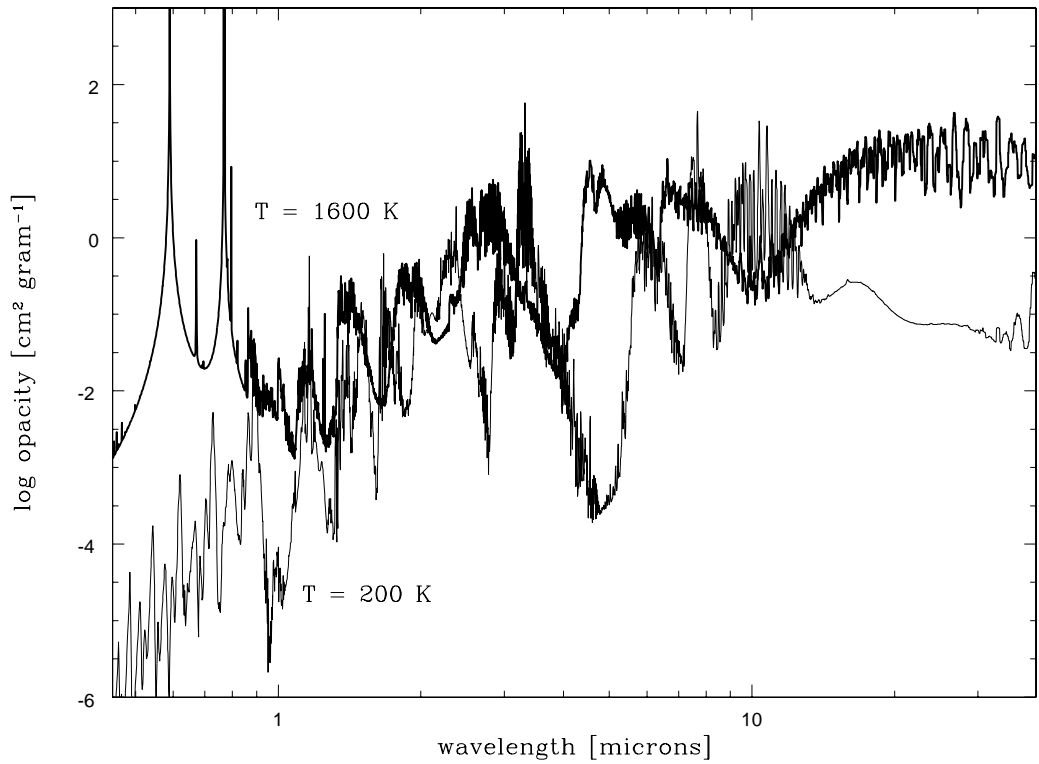}\kern+6in\hfill}
\caption{Depiction of the differences in total, abundance-weighted
gaseous opacities in hot and cold EGP atmospheres at P = 1 bar.
The opacity in the visible region of a hot atmosphere is up to
several orders of magnitude larger than that of a relatively
cold atmosphere.  In contrast, there are regions in the infrared where
the total opacity of the cold atmosphere is greater.}
\label{fig_comboopacs}
\end{figure}

\newpage

\begin{figure} 
\vspace*{6.0in}
\hbox to\hsize{\hfill\includegraphics{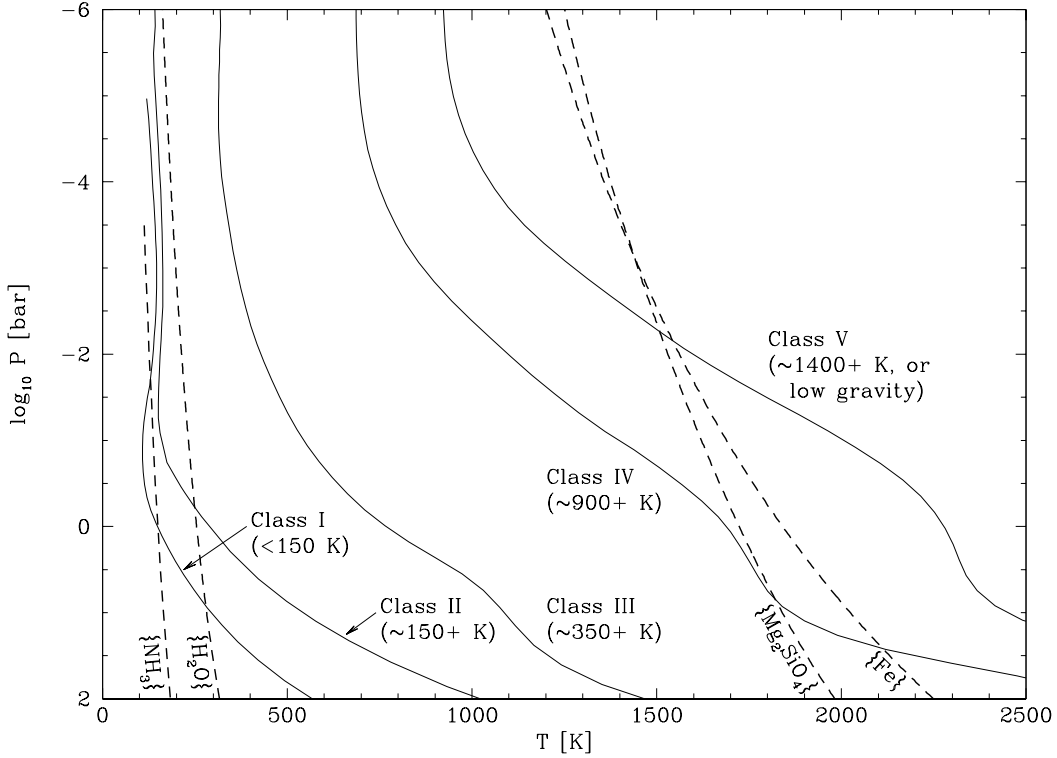}\kern+6in\hfill}
\caption{Temperature-Pressure (T-P) profiles of EGPs over the full range
in effective temperature.  The profiles shown represent the different
EGP classes.  The corresponding emergent spectra are depicted in
Figures \ref{fig_class1}, \ref{fig_class2}, \ref{fig_class3},
\ref{fig_class4}, and \ref{fig_class5}.  Condensation curves for four
high-abundance condensates (ammonia, water, a silicate, and iron)
are also shown.  The intersection of these curves with the T-P
profiles indicates the positions of cloud bases.  (For clarity,
cloud-free T-P profiles are shown here.)}
\label{fig_classcurves}
\end{figure}

\newpage

\begin{figure} 
\vspace*{6.0in}
\hbox to\hsize{\hfill\includegraphics{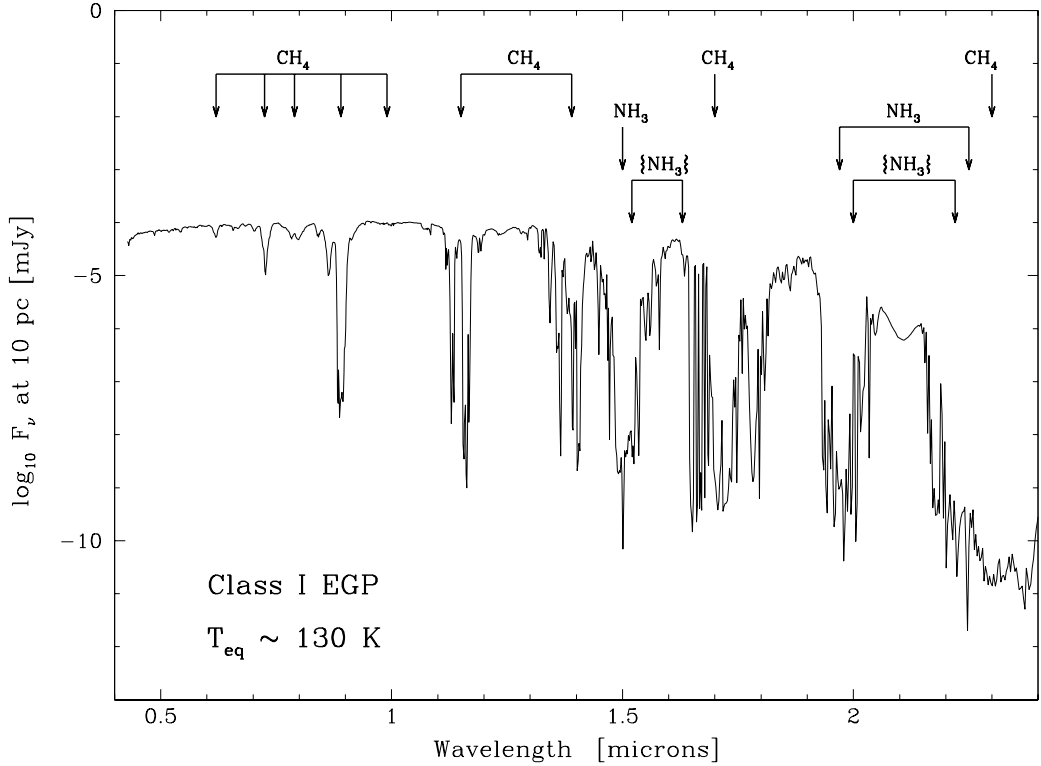}\kern+6in\hfill}
\caption{Emergent spectrum of a Class I (``Jovian''; $\sim$ 5 AU) EGP.  The
temperature in the outer atmosphere is cold enough ($\sless$ 150 K)
that ammonia condenses into ice, which provides significant reflection.
The upper troposphere is depleted of water, which is condensed into
a cloud layer at a pressure of a few-to-several bars.  Methane and
ammonia absorption dominate the visible and near-infrared spectrum.
Note that `$\{$NH$_3$$\}$' indicates a feature of condensed ammonia,
while `NH$_3$' indicates that of gaseous ammonia.
\label{fig_class1}}
\end{figure}

\newpage

\begin{figure} 
\vspace*{6.0in}
\hbox to\hsize{\hfill\includegraphics{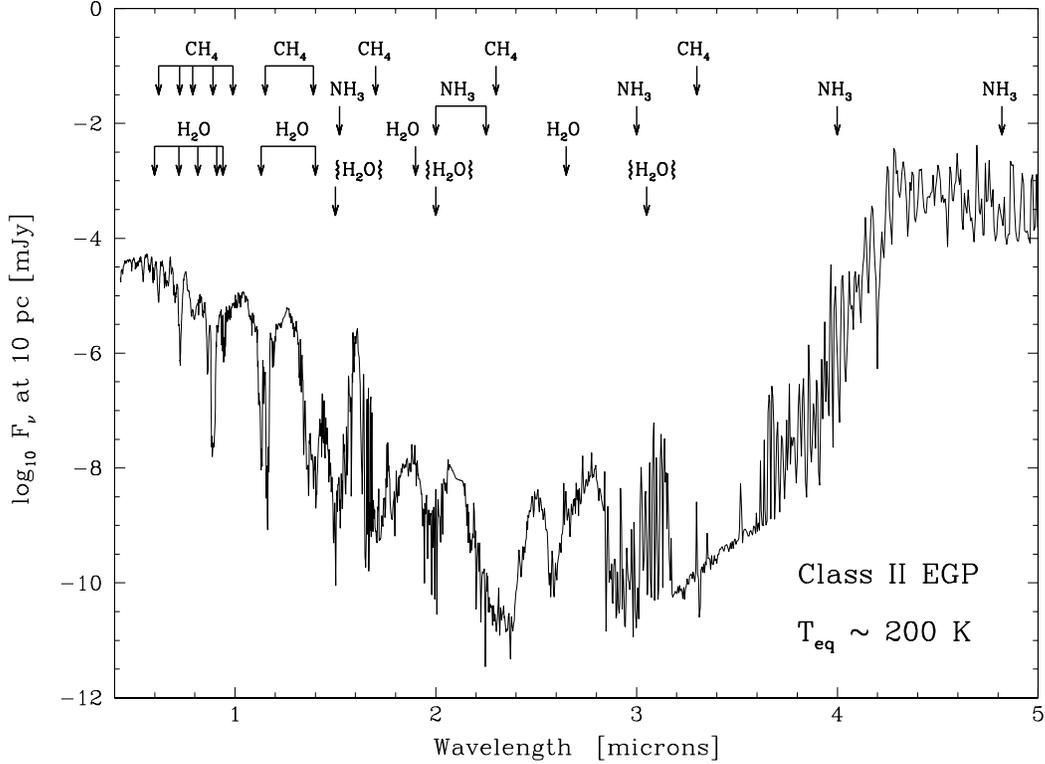}\kern+6in\hfill}
\caption{Emergent spectrum of a Class II (``water class''; $\sim$ 1-2 AU) EGP.
A water cloud resides in the troposphere, but the temperature
at every pressure is hot enough that ammonia remains in gaseous form.
Reflection by water keeps the emergent flux higher than it would
otherwise be in the visible and near-infrared.  Gaseous absorption
by methane, water, and ammonia produce strong spectral features.  Note
that `$\{$H$_2$O$\}$' indicates a feature of condensed water, while
`H$_2$O' indicates that of gaseous water.
\label{fig_class2}}
\end{figure}

\newpage

\begin{figure} 
\vspace*{6.0in}
\hbox to\hsize{\hfill\includegraphics{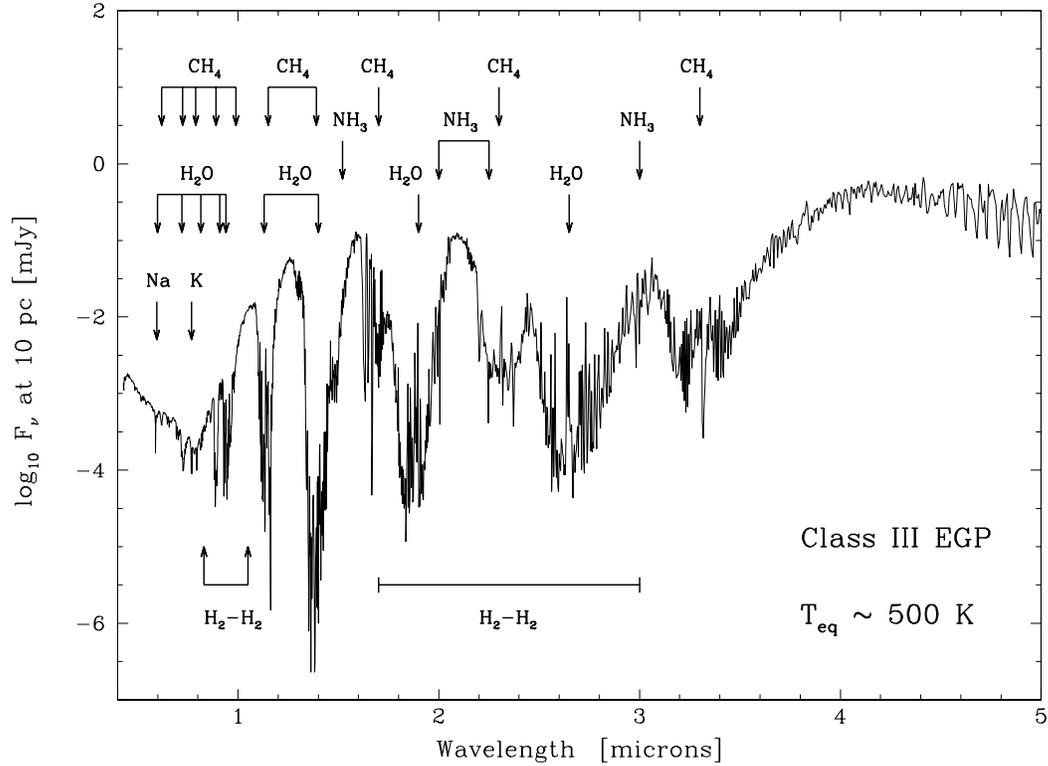}\kern+6in\hfill}
\caption{Emergent spectrum of a Class III (``clear''; $\sim$ 0.5 AU) EGP.
These objects are too warm for water to condense, but too cold for
a silicate layer to appear in the upper troposphere.  In the absence
of any high-abundance, principal condensates, the ro-vibrational
molecular absorption features are very strong.  Of particular
importance are absorption by gaseous water, methane, and molecular
hydrogen (collision-induced).  Additionally, alkali metal lines
appear in the visible.
\label{fig_class3}}
\end{figure}

\newpage

\begin{figure} 
\vspace*{6.0in}
\hbox to\hsize{\hfill\includegraphics{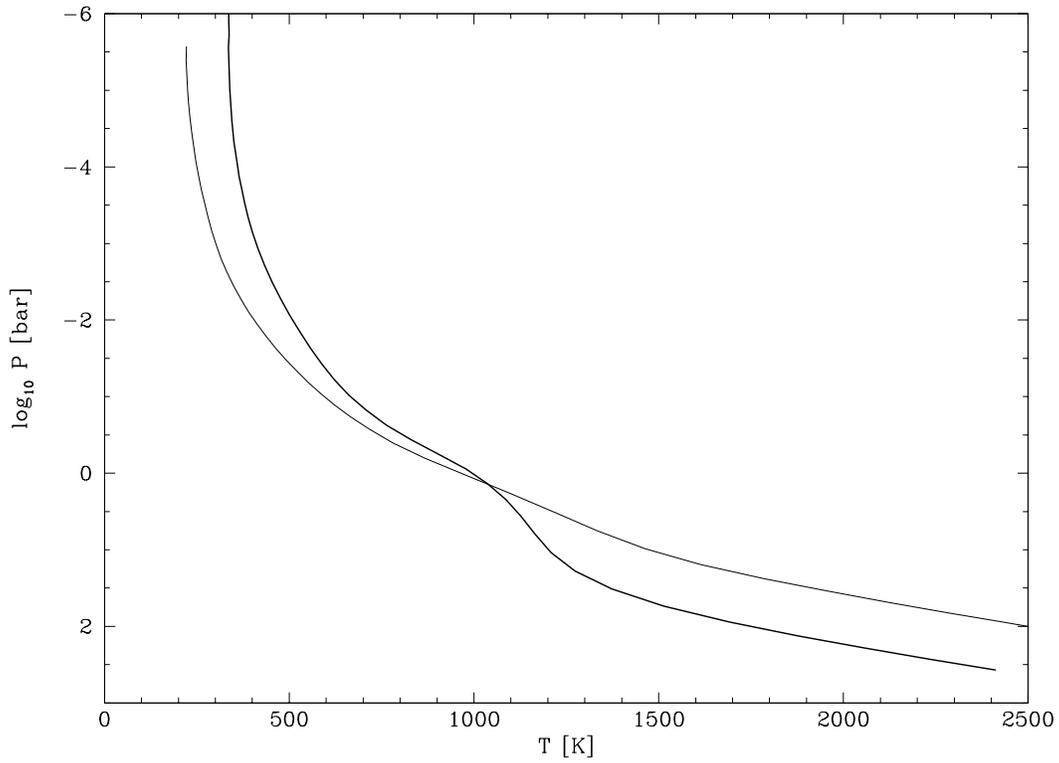}\kern+6in\hfill}
\caption{Comparison of the T-P structure of a Class III EGP (thick curve) with that of
a brown dwarf (thin curve) with the same surface gravity and total integrated
emergent flux.  The
outer atmosphere of the EGP is more isothermal due to irradiation by
the primary.}
\label{fig_bd3TP}
\end{figure}

\newpage

\begin{figure} 
\vspace*{6.0in}
\hbox to\hsize{\hfill\includegraphics{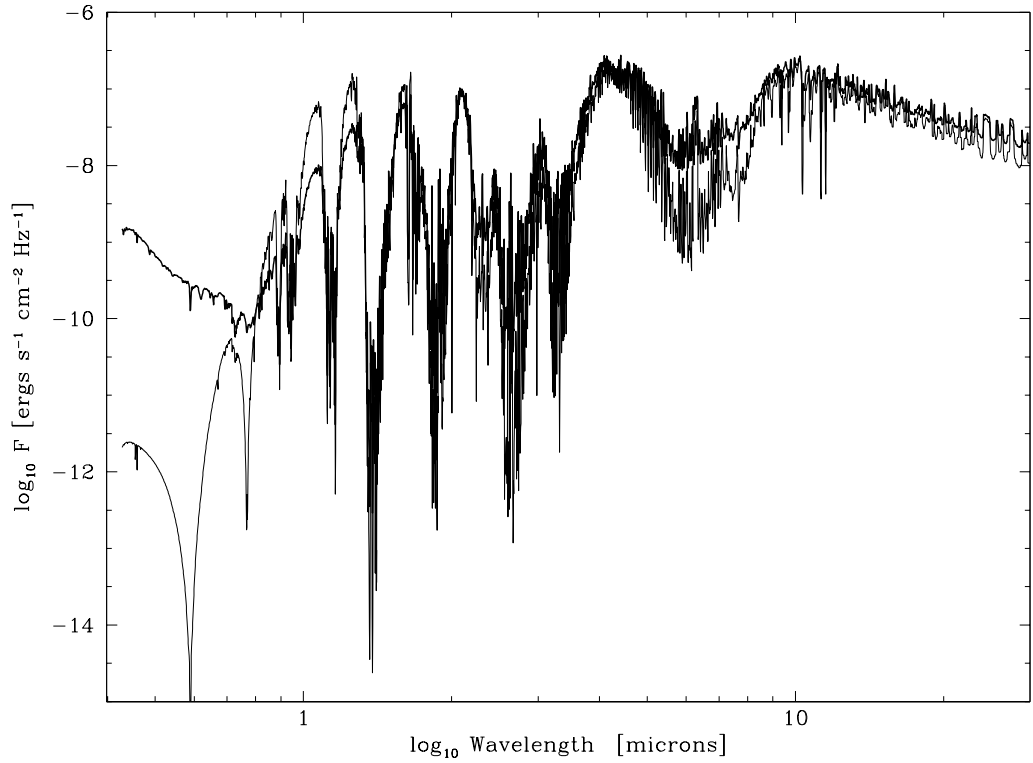}\kern+6in\hfill}
\caption{Comparison of the emergent spectrum of a Class III EGP (thick curve) with that of
a cloud-free brown dwarf (thin curve) of the same surface gravity and
integrated emergent flux.  By construction, the fluxes
are identical, but the spectral energy distributions differ
significantly.  (Fluxes shown are at the objects' surfaces.)}
\label{fig_bd3spec}
\end{figure}

\newpage

\begin{figure} 
\vspace*{6.0in}
\hbox to\hsize{\hfill\includegraphics{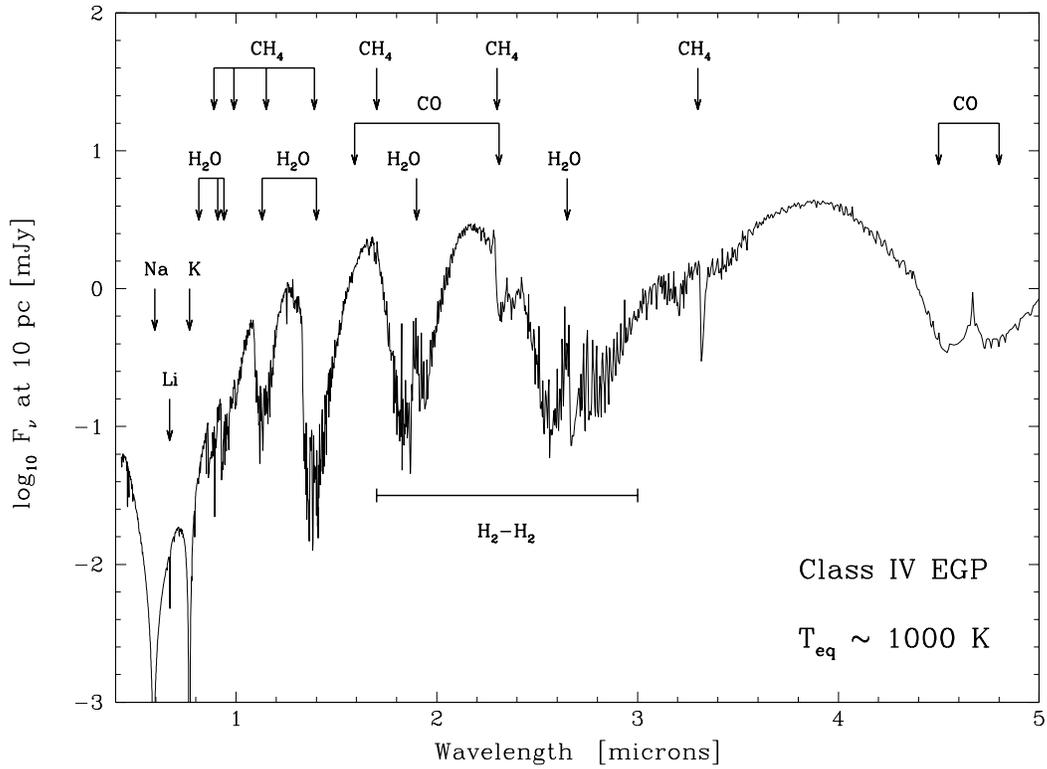}\kern+6in\hfill}
\caption{Emergent spectrum of a Class IV EGP ($\sim$0.1 AU).  The visible
spectrum is dominated by strong pressure-broadened sodium and
potassium resonance lines, while gaseous water, methane, and carbon
monoxide absorption are strong in the infrared.  Silicate and iron
clouds exist at depth, but they have no significant effects on the visible
and near-infrared emergent spectra.}
\label{fig_class4}
\end{figure}

\newpage

\begin{figure} 
\vspace*{6.0in}
\hbox to\hsize{\hfill\includegraphics{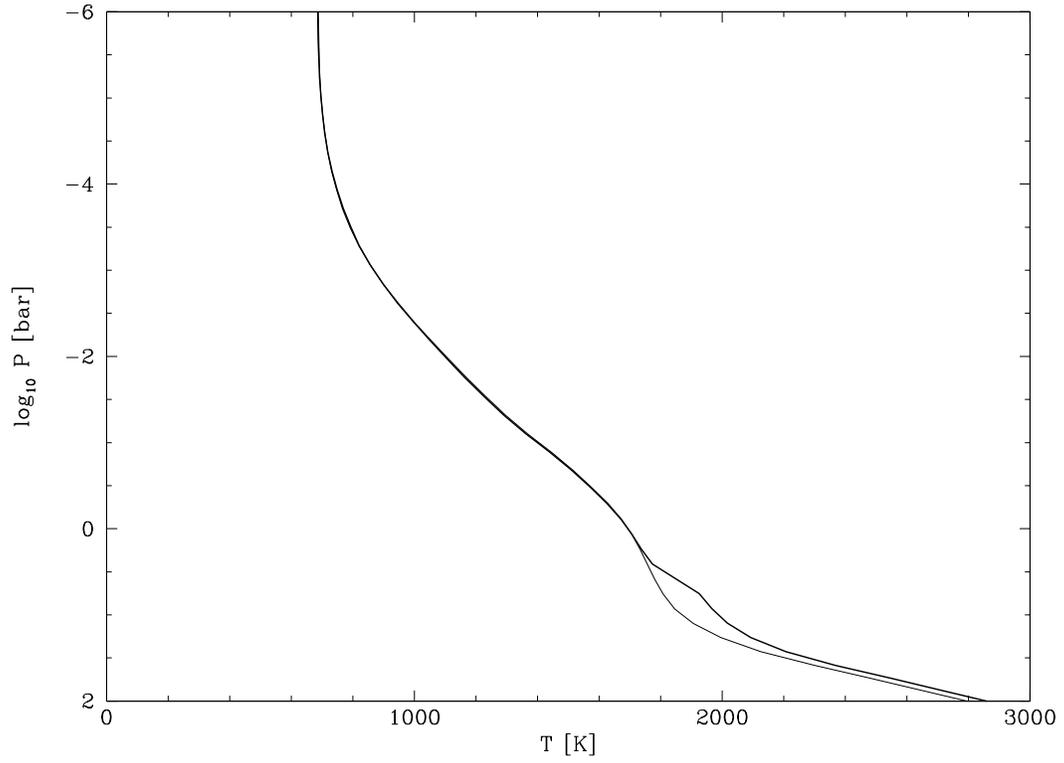}\kern+6in\hfill}
\caption{Effect of a deep silicate cloud on the T-P structure of a Class IV
EGP.  The resulting T-P profile (thick curve) differs from that of 
a cloud-free model (thin curve), but only at depth.}
\label{fig_cloudclass4}
\end{figure}

\newpage

\begin{figure} 
\vspace*{6.0in}
\hbox to\hsize{\hfill\includegraphics{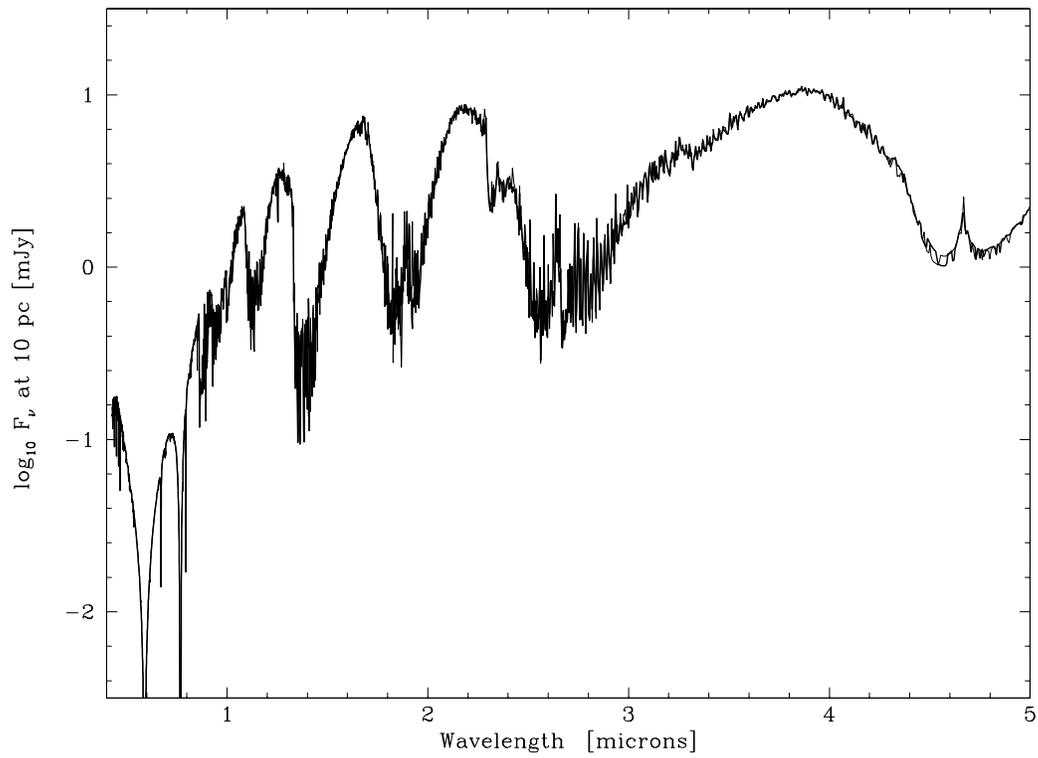}\kern+6in\hfill}
\caption{Emergent spectrum of a Class IV EGP with its deep silicate
cloud (thick curve) relative to that of a cloud-free model (thin
curve).  Silicate clouds in Class IV EGPs form too deeply to have
significant effects on their emergent spectra.}
\label{fig_cloudclass4spec}
\end{figure}

\newpage

\begin{figure} 
\vspace*{6.0in}
\hbox to\hsize{\hfill\includegraphics{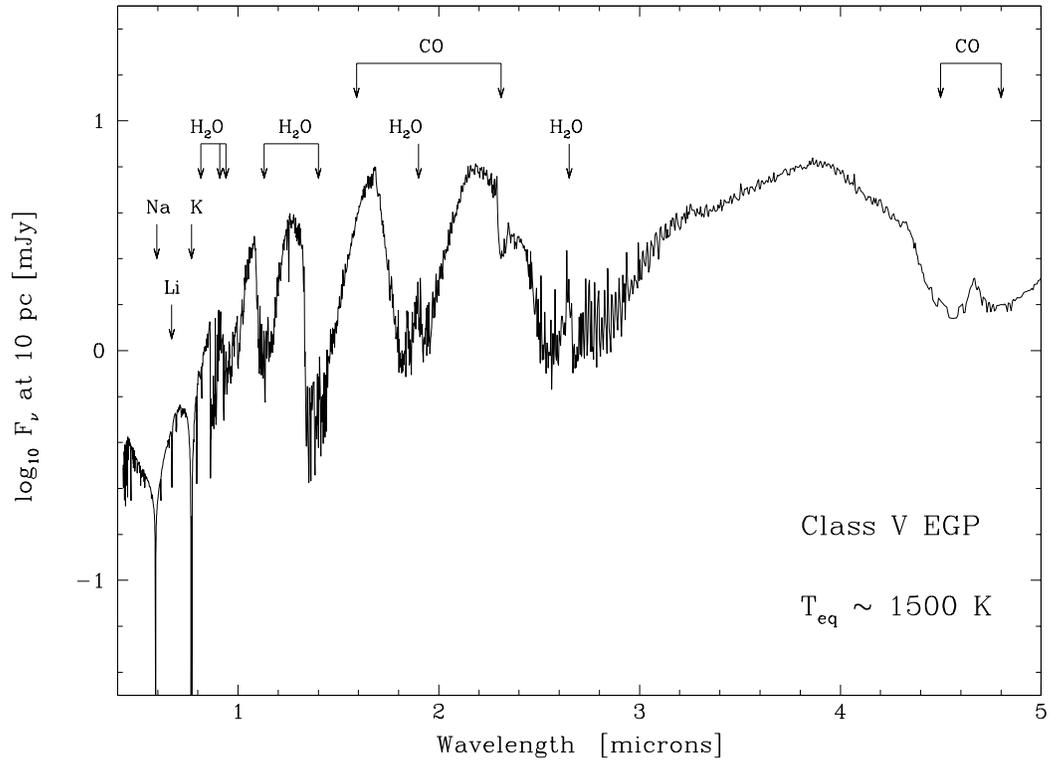}\kern+6in\hfill}
\caption{Emergent spectrum of a Class V EGP (a ``roaster''; $\sim$ 0.05 AU).
Like Class IV objects, these EGPs have strong alkali lines, but
scattering and absorption by silicate and iron grains high in the
atmosphere alter the character of the spectrum.
Furthermore, nearly all of the carbon resides in carbon monoxide, so
methane features are very weak or nonexistent.}
\label{fig_class5}
\end{figure}

\newpage

\begin{figure} 
\vspace*{6.0in}
\hbox to\hsize{\hfill\includegraphics{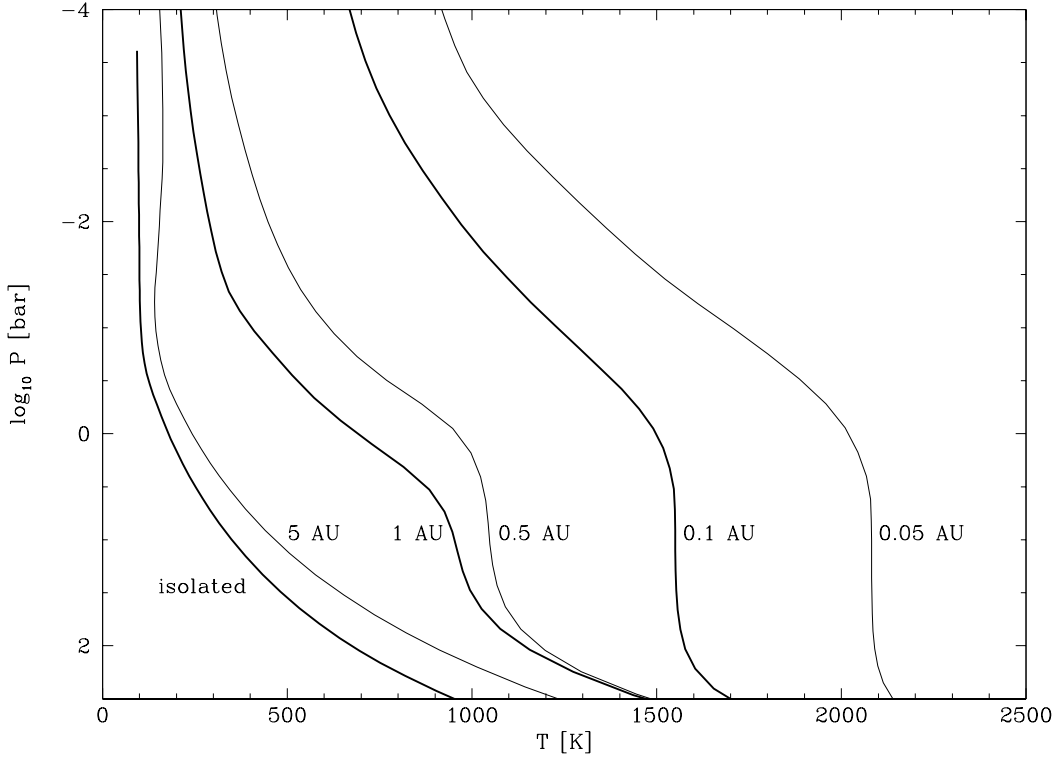}\kern+6in\hfill}
\caption{Cloud-free EGP T-P profiles as a function of orbital distance from a G0V
primary.  From right to left, the orbital distance is 0.05 AU, 0.1 AU,
0.5 AU, 1 AU, and 5 AU.  Additionally, the leftmost profile
(thick curve) is that of an isolated EGP/brown dwarf.  In all cases, the gravity is
$3\times 10^3$ cm s$^{-2}$ and the inner boundary
flux is set equal to that of an isolated object with T$_{\textrm{eff}}$
= 125 K.
(Note that we have alternated between thin and thick line
types in order to facilitate a correspondence with the spectra in Figures
\ref{fig_spectrumG0AU} and \ref{fig_spectrumG0AU_contrast}.)}
\label{fig_PTG0AU}
\end{figure}

\newpage

\begin{figure} 
\vspace*{6.0in}
\hbox to\hsize{\hfill\includegraphics{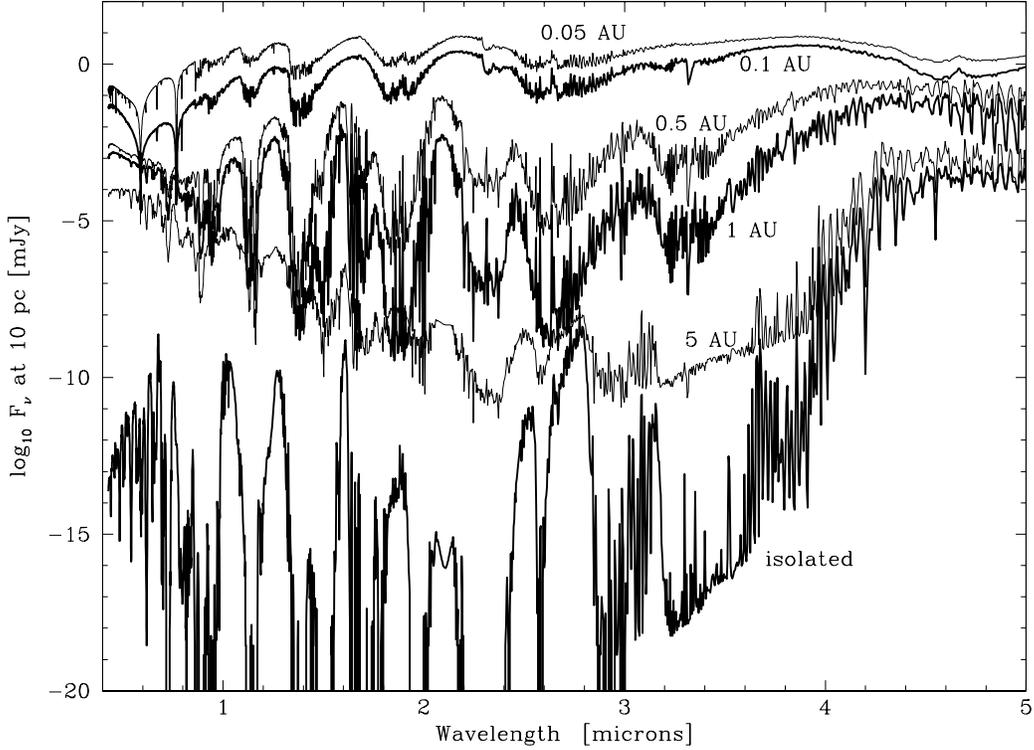}\kern+6in\hfill}
\caption{Cloud-free EGP emergent spectra as a function of orbital distance from
a G0V primary.  From top to bottom, the orbital distance is 0.05 AU, 0.1 AU,
0.5 AU, 1 AU, and 5 AU.  Additionally, the bottom curve is that
of an isolated EGP/brown dwarf.  In all cases, the gravity is
$3\times 10^3$ cm s$^{-2}$ and the lower boundary
flux is set equal to that of an isolated object with T$_{\textrm{eff}}$
= 125 K.  (Note that we have alternated between thin and thick line types
in order to facilitate a correspondence with the profiles
in Figure \ref{fig_PTG0AU}.)}
\label{fig_spectrumG0AU}
\end{figure}

\newpage

\begin{figure} 
\vspace*{6.0in}
\hbox to\hsize{\hfill\includegraphics{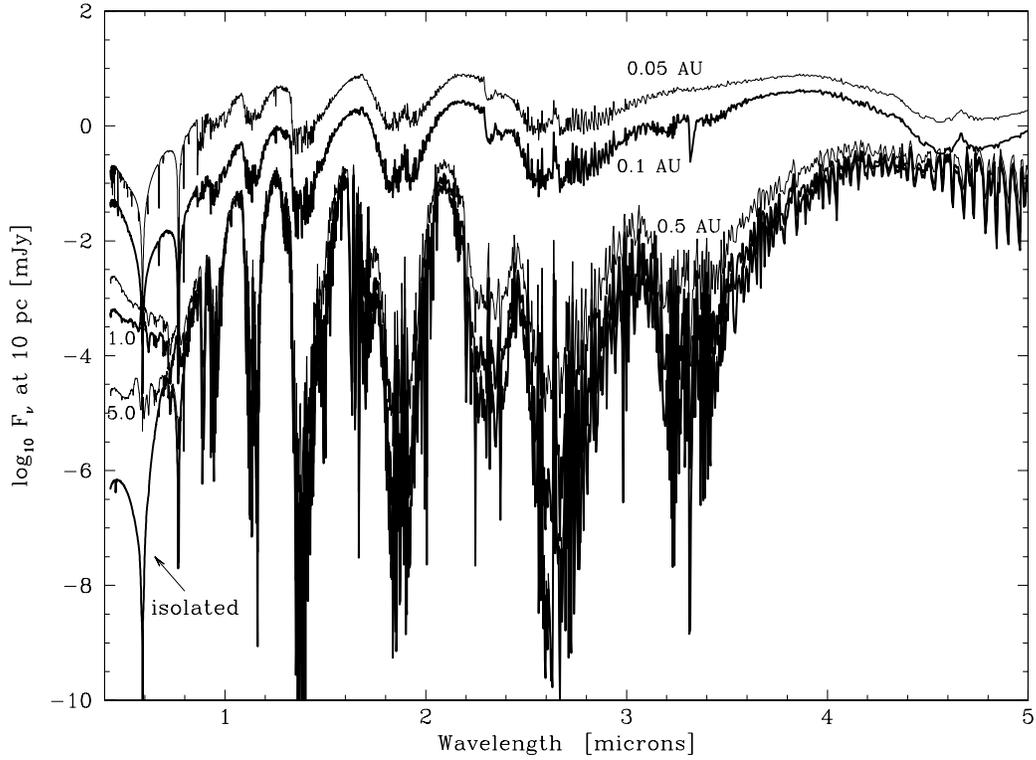}\kern+6in\hfill}
\caption{Same as Figure \ref{fig_spectrumG0AU}, but for an inner boundary
flux characteristic of an effective temperature of 500 K.  
From top to bottom, the orbital distance is 0.05 AU, 0.1 AU,
0.5 AU, 1 AU, and 5 AU.  The bottom curve is that
for an isolated 500 K EGP/brown dwarf.  In all cases, the gravity is
$3\times 10^3$ cm s$^{-2}$.  
Rayleigh scattering of the incident stellar
radiation keeps the EGP emergent flux in the visible significantly above that
of the isolated EGP/brown dwarf, but the high inner boundary flux causes the spectrum
in the near and mid-infrared to be roughly independent of distance 
exterior to $\sim$0.5 AU. This sequence shows our approximate
expectations for an irradiated brown dwarf and stands in marked 
contrast with Fig. \ref{fig_spectrumG0AU}.}
\label{fig_spectrumG0AU500}
\end{figure}

\newpage

\begin{figure} 
\vspace*{6.0in}
\hbox to\hsize{\hfill\includegraphics{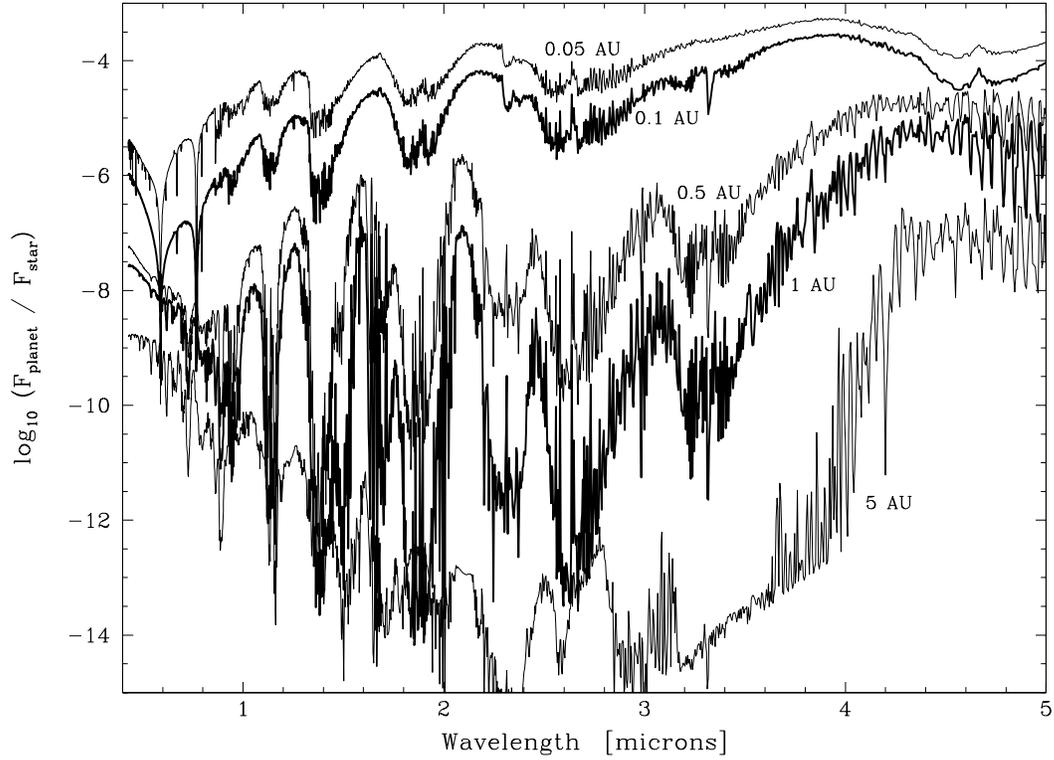}\kern+6in\hfill}
\caption{The phase-averaged planet-to-star flux contrast 
as a function of wavelength and orbital distance for the models portrayed 
in Figure \ref{fig_spectrumG0AU}.  The inner boundary
flux is characteristic of an effective temperature of 125 K,
the planet models are cloud-free, and the primary star is of subtype G0V.  
From top to bottom, the orbital distance is 0.05 AU, 0.1 AU,
0.5 AU, 1 AU, and 5 AU.  In all cases, the gravity is
$3\times 10^3$ cm s$^{-2}$. See text for details.} 
\label{fig_spectrumG0AU_contrast}
\end{figure}

\newpage

\begin{figure} 
\vspace*{6.0in}
\hbox to\hsize{\hfill\includegraphics{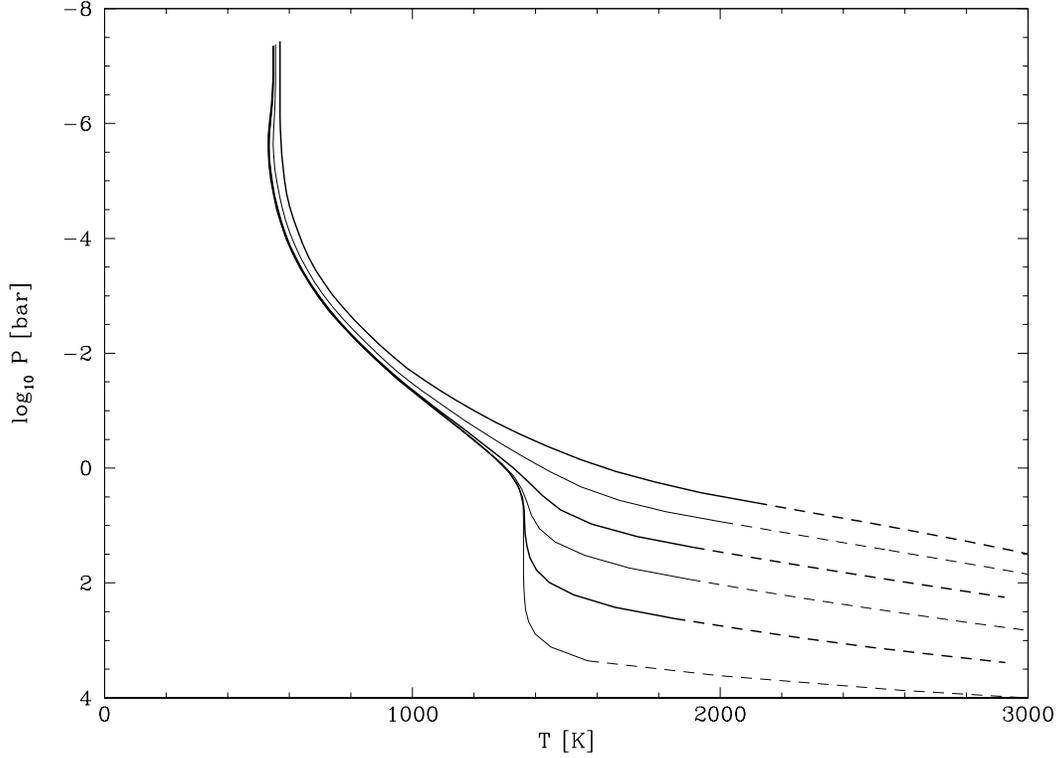}\kern+6in\hfill}
\caption{EGP T-P structure as a function of inner boundary flux.  The inner
boundary flux of a cloud-free Class IV model is varied.  From the
top profile to the bottom profile, models with inner boundary fluxes
(= $\sigma$\teff$^4$) corresponding to \teff = 1000 K, 750 K, 500 K,
300 K, 150 K, and 50 K are shown.
(Note that we have alternated between thin and thick line
types in order to facilitate a correspondence with the spectra in Figure
\ref{fig_spectrumboundary}.)}
\label{fig_PTboundary}
\end{figure}

\newpage

\begin{figure} 
\vspace*{6.0in}
\hbox to\hsize{\hfill\includegraphics{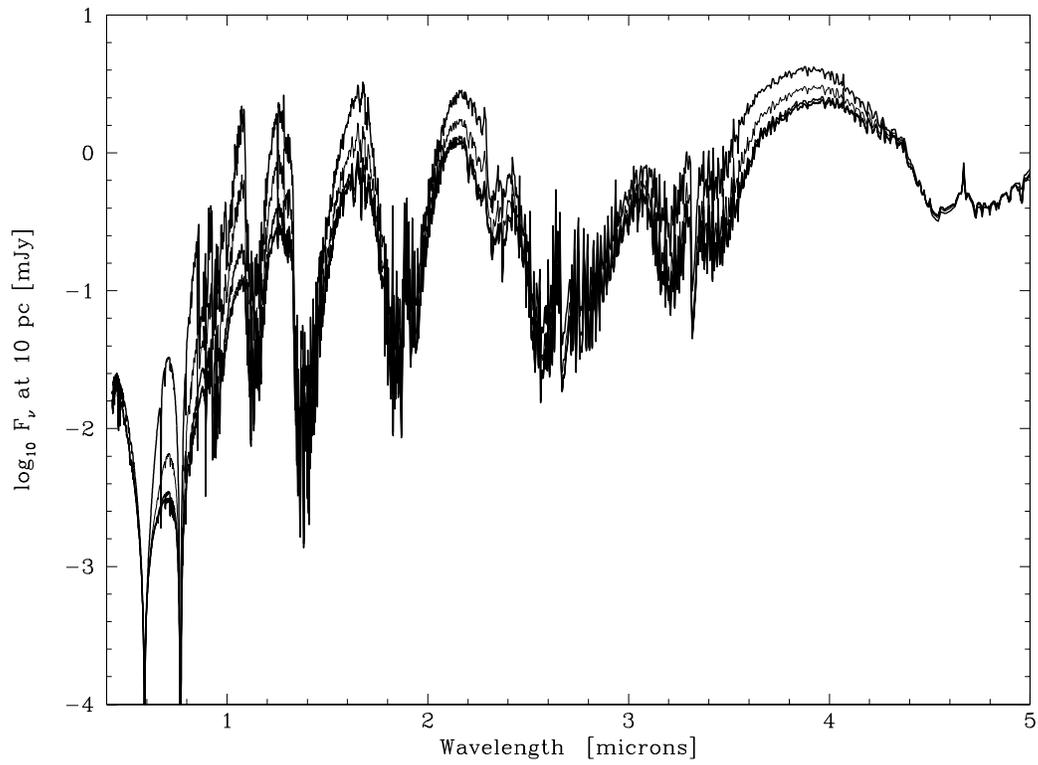}\kern+6in\hfill}
\caption{Class IV EGP emergent spectra as a function of inner boundary flux.  Each
spectral curve corresponds to each of the models depicted in
Figure \ref{fig_PTboundary}, from 1000 K (top curve) to 50 K (bottom
curve, overlapping with that of 150 K and 300 K).}
\label{fig_spectrumboundary}
\end{figure}

%% file: plots2.tex
\newpage

\begin{figure} 
\vspace*{6.0in}
\hbox to\hsize{\hfill\includegraphics{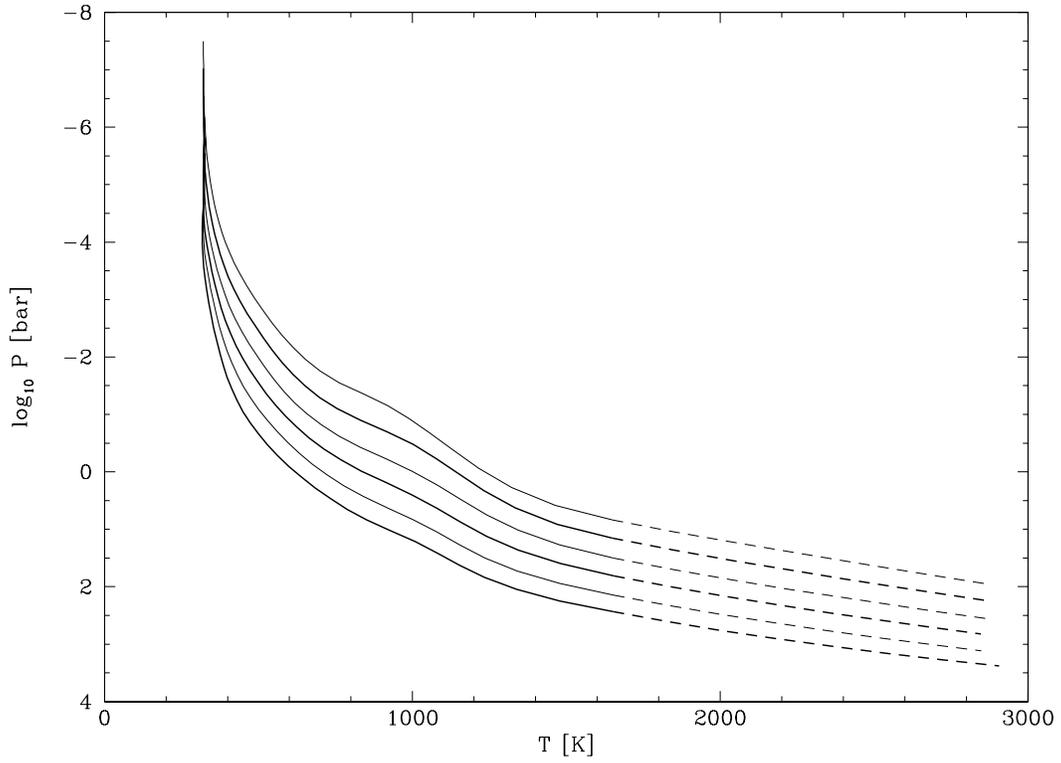}\kern+6in\hfill}
\caption{EGP T-P structure as a function of surface gravity.  The surface
gravity of a Class III model is varied from 10$^3$ cm s$^{-2}$
(top curve) to $3\times 10^5$ cm s$^{-2}$ (a massive brown dwarf;
bottom curve) in intervals
of 1/2 dex.
(Note that we have alternated between thin and thick line
types in order to facilitate a correspondence with the spectra in Figure
\ref{fig_gravity}.)}
\label{fig_PTgravity}
\end{figure}

\newpage

\begin{figure} 
\vspace*{6.0in}
\hbox to\hsize{\hfill\includegraphics{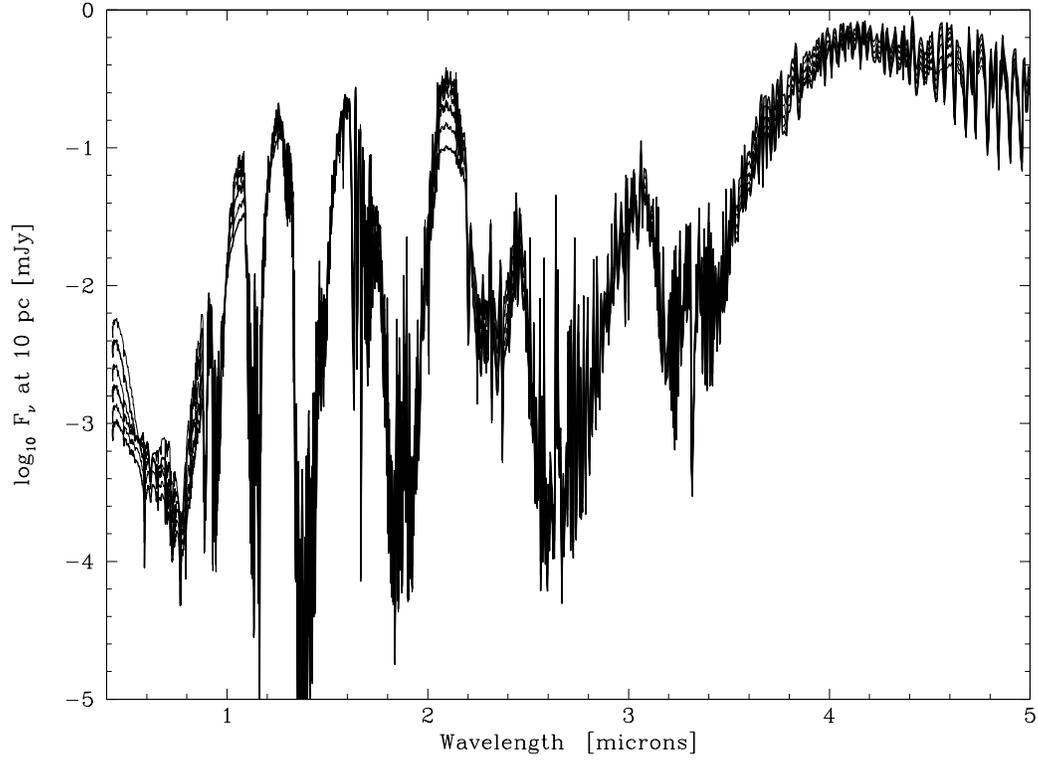}\kern+6in\hfill}
\caption{Class III EGP emergent spectra as a function of surface gravity.  Each
spectrum corresponds to each of the models depicted in
Figure \ref{fig_PTgravity}.
For these models, an increase in gravity results in
a general decrease in flux shortward of 
$\sim$2.2 $\mu$m, and an increase longward.  Additionally, a larger
gravity tends to reduce the peak-to-trough variations throughout
the spectrum.}
\label{fig_gravity}
\end{figure}

\newpage

\begin{figure} 
\vspace*{6.0in}
\hbox to\hsize{\hfill\includegraphics{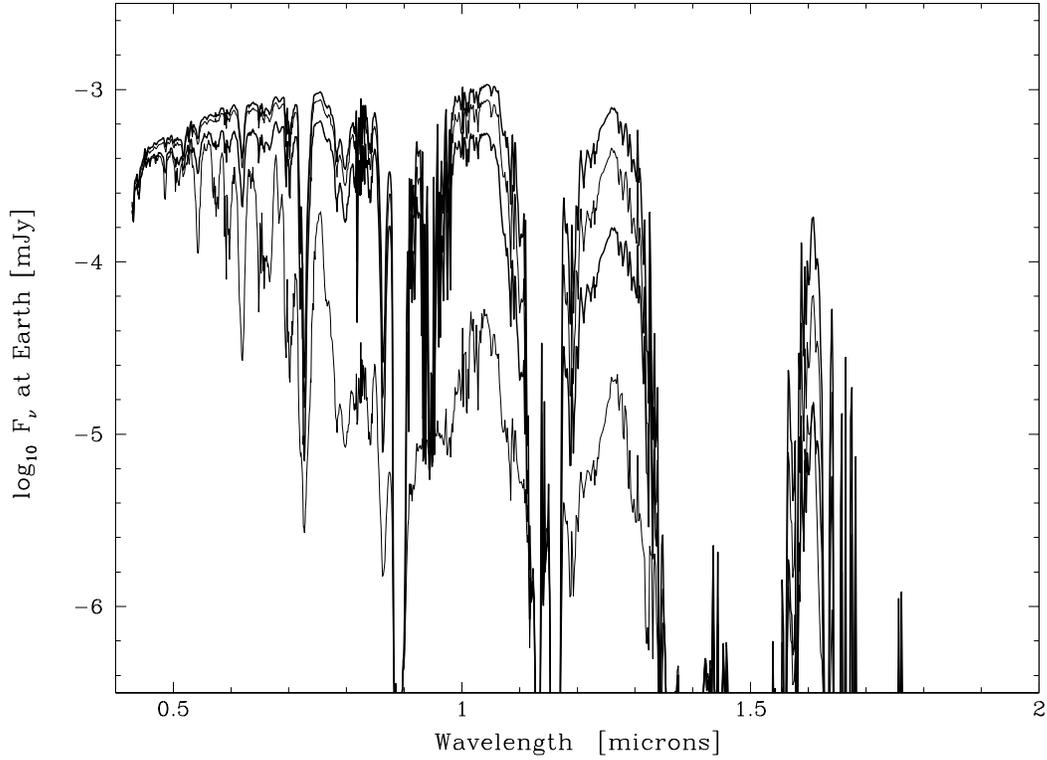}\kern+6in\hfill}
\caption{Effects of cloud particle size variation on the emergent spectrum
of a Class II EGP, such as $\epsilon$ Eridani b.  The resulting emergent spectra are shown for
EGPs with three different median ice particle sizes, along with
a cloud-free model of the same object.  From
top to bottom, the curves correspond to a size distribution peaked
at 0.5 $\mu$m, one peaked at 5 $\mu$m, and another at 50 $\mu$m.
In each case, 10\% of the available H$_2$O was assumed to condense.
The bottom curve is that of the cloud-free model.}
\label{fig_partsize}
\end{figure}

\newpage

\begin{figure} 
\vspace*{6.0in}
\hbox to\hsize{\hfill\includegraphics{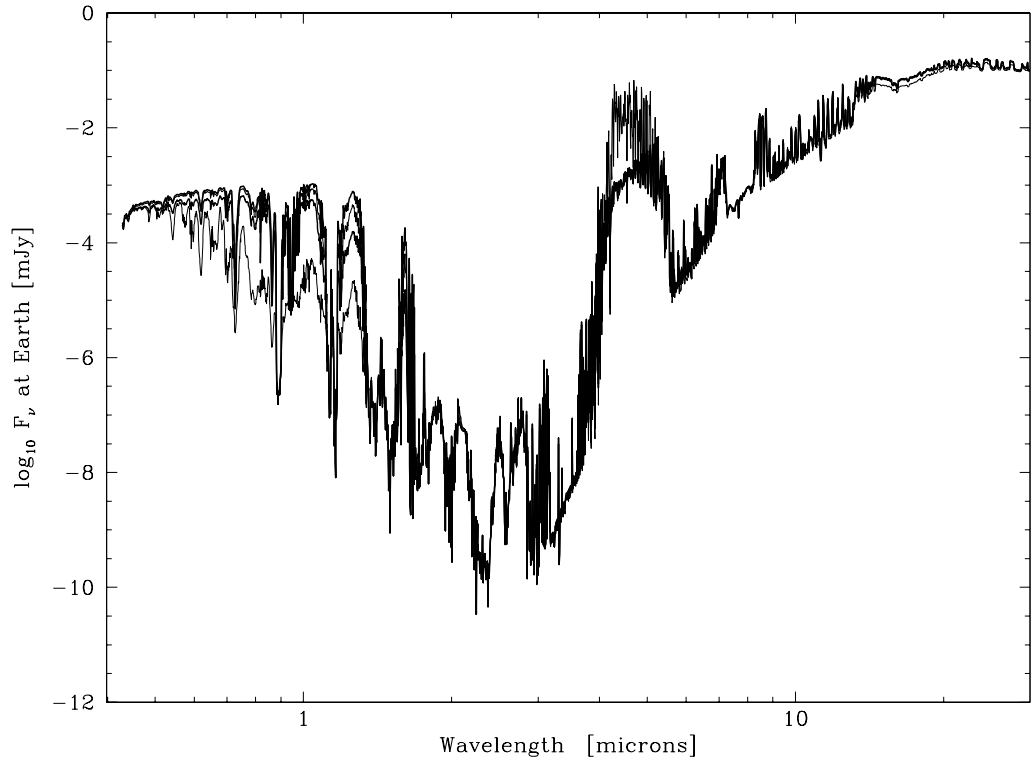}\kern+6in\hfill}
\caption{Effects of cloud particle size variation on the emergent spectrum
of a Class II EGP, such as $\epsilon$ Eridani b, out to 30 $\mu$m.  Models are the same as depicted
in Figure \ref{fig_partsize}.  In the 4-5 micron opacity window,
the cloud-free model flux is up to 2 orders of magnitude greater than
those that include clouds of any particle size.}
\label{fig_partsize2}
\end{figure}

\newpage

\begin{figure} 
\vspace*{6.0in}
\hbox to\hsize{\hfill\includegraphics{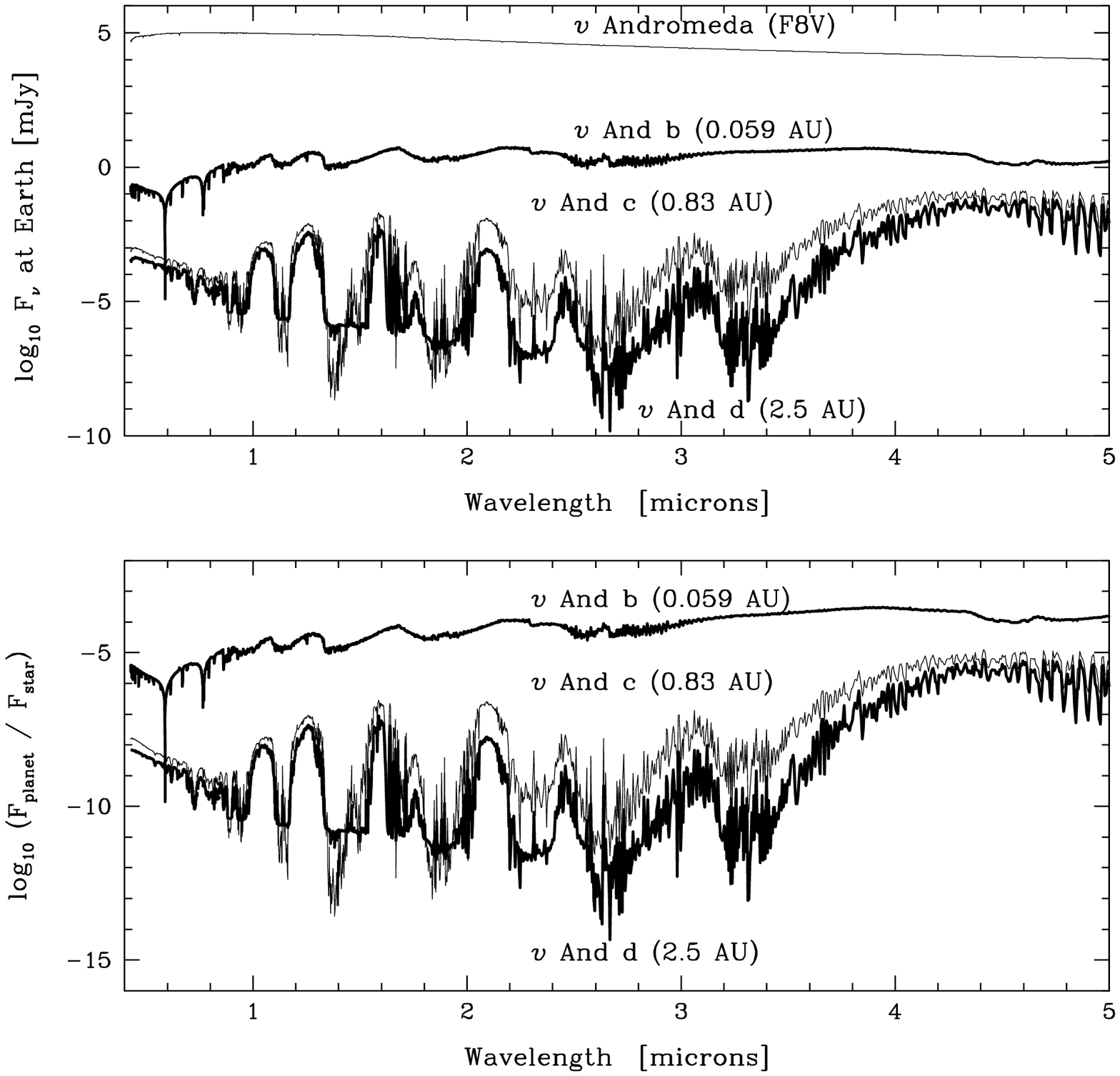}\kern+5.9in\hfill}
\caption{$Upper$ $panel$: Visible and near-infrared spectra of $\upsilon$ And b, c, and d.  An
assumed spectrum of the primary (a scaled Kurucz model) is also
depicted.  $\upsilon$ And b, c, and d are Class V, III, and II EGPs, respectively.
$Lower$ $panel$: Wavelength-dependent, phase-averaged planet-to-star flux ratios for $\upsilon$ And b,
c, and d.}
\label{fig_spectrumUpsAnd}
\end{figure}

\newpage

\begin{figure} 
\vspace*{6.0in}
\hbox to\hsize{\hfill\includegraphics{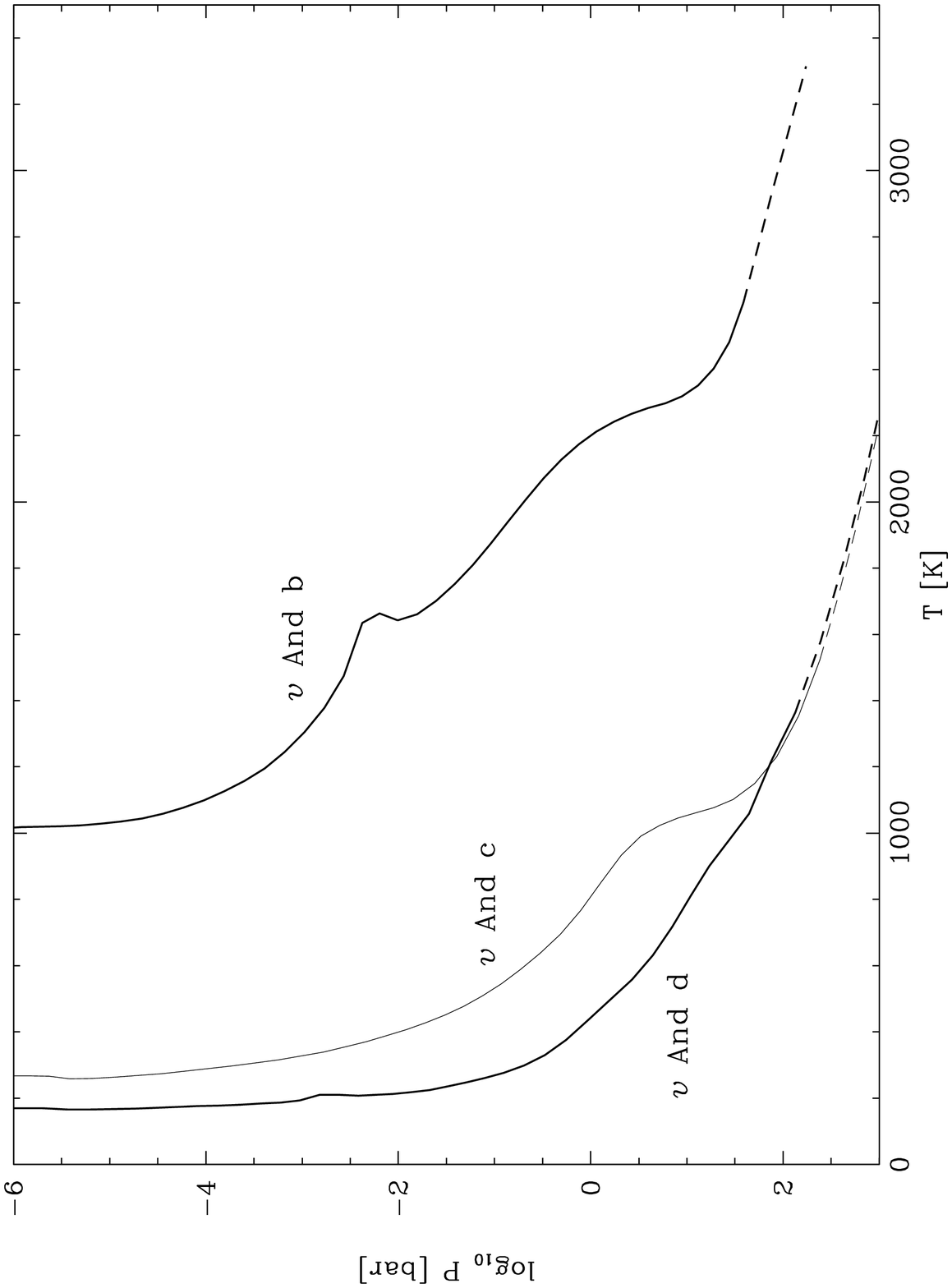}\kern+5.9in\hfill}
\caption{T-P profiles of $\upsilon$ And b, c, and d.  The dashed portions
of the profiles indicate convective regions.}
\label{fig_TPUpsAnd}
\end{figure}

\newpage

\begin{figure} 
\vspace*{6.0in}
\hbox to\hsize{\hfill\includegraphics{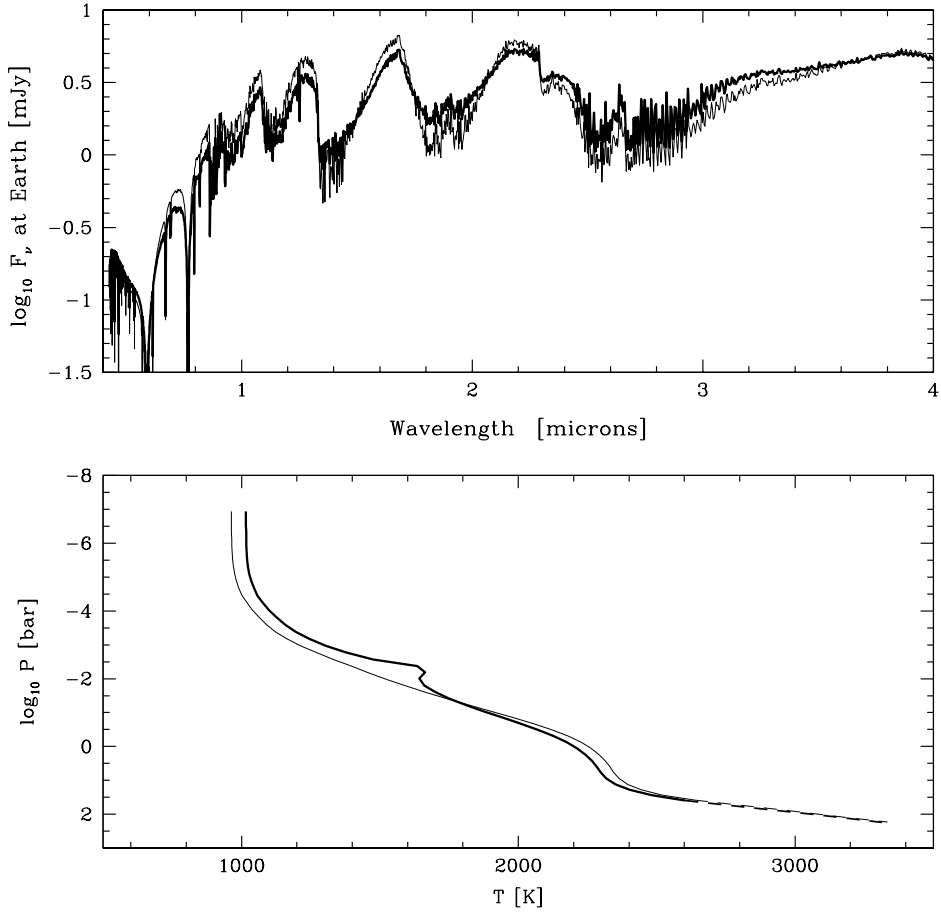}\kern+5.9in\hfill}
\caption{
$Upper$ $Panel$: Fiducial model spectrum of $\upsilon$ And b (thick curve) versus that
of a cloud-free model (thin curve).  The removal of the clouds results
in a wider variation from peak to trough throughout most of the spectrum.
$Lower$ $panel$: Fiducial T-P profile of $\upsilon$ And b (thick curve)
versus that of a cloud-free
model (thin curve).  Removing the iron and silicate clouds results in a
cooler outer atmosphere, but a hotter deeper atmosphere.  At large pressures
($\sgreat$ 50 bars), the high clouds appear to have essentially no effect
on the T-P profile.}
\label{fig_UpsAndb}
\end{figure}

\newpage

\begin{figure} 
\vspace*{6.0in}
\hbox to\hsize{\hfill\includegraphics{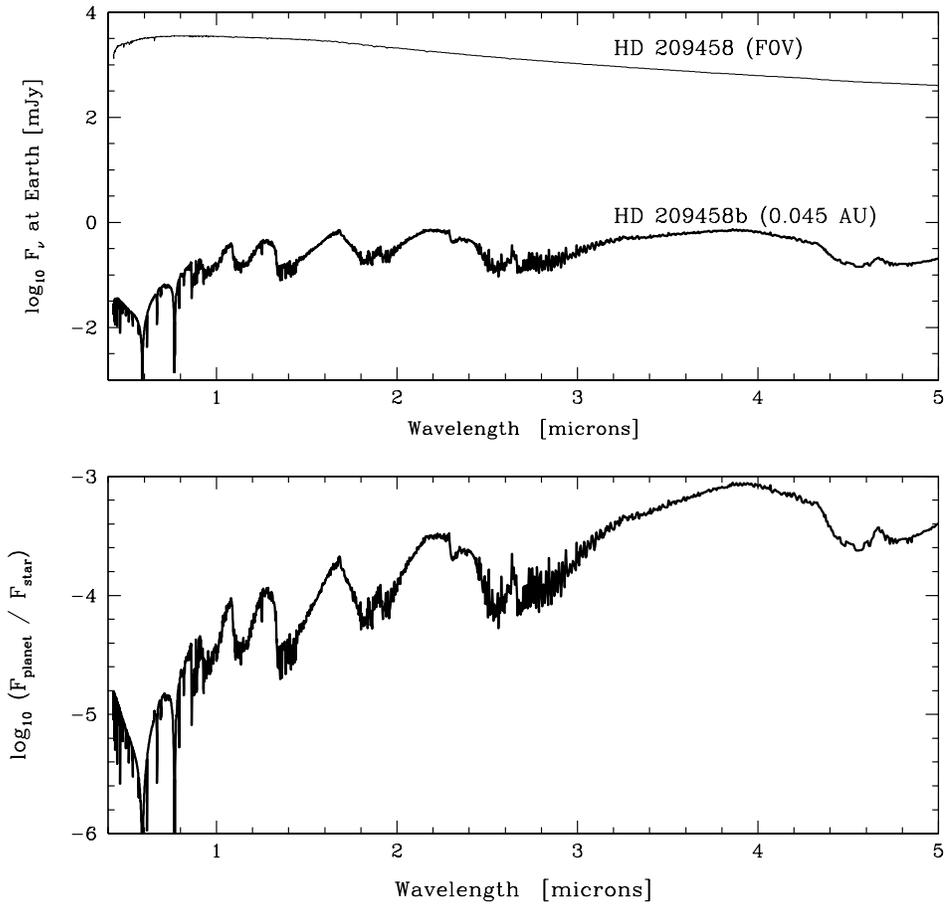}\kern+5.9in\hfill}
\caption{$Upper$ $panel$: Visible and near-infrared spectrum of HD 209458b along with an assumed
spectrum of its primary (a scaled Kurucz model).
These fluxes were calculated using the planetary radius of 1.35 R$_{\textrm{J}}$.
$Lower$ $panel$: Wavelength-dependent, phase-averaged planet-to-star
flux ratio of HD 209458b.}
\label{fig_spectrumHD209458b}
\end{figure}

\newpage

\begin{figure} 
\vspace*{6.0in}
\hbox to\hsize{\hfill\includegraphics{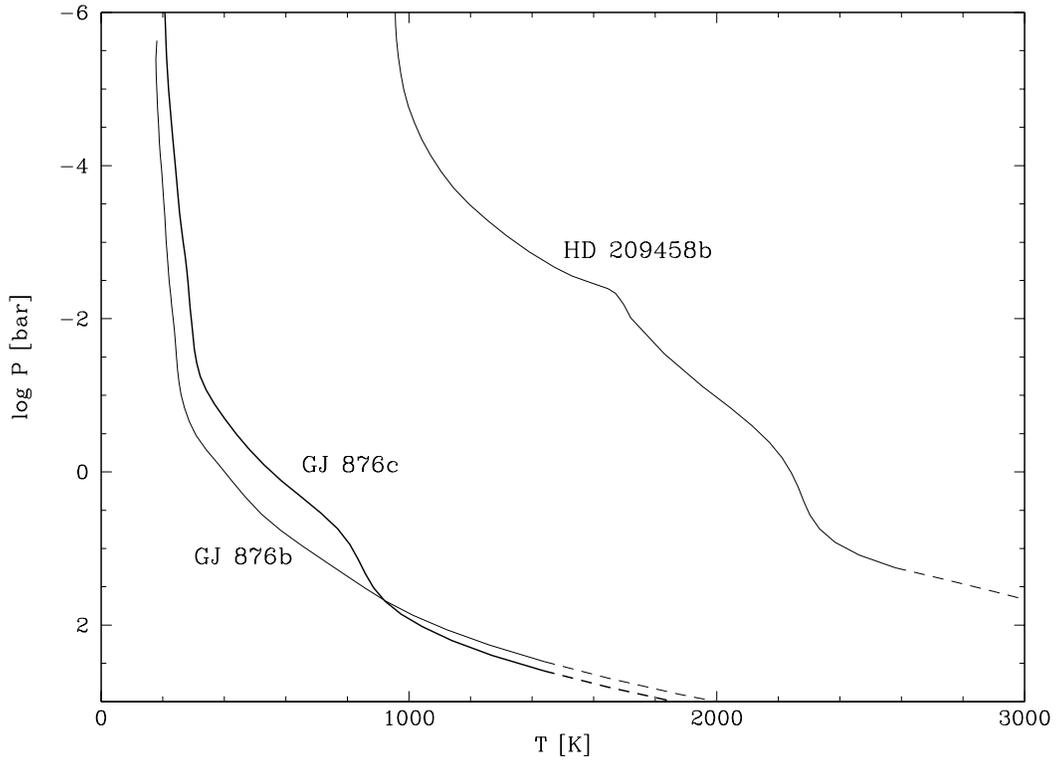}\kern+5.9in\hfill}
\caption{T-P profiles of HD 209458b, GJ 876b and GJ 876c.  The dashed portions
of the profiles indicate convective regions.}
\label{fig_TPGJ876}
\end{figure}

\newpage

\begin{figure} 
\vspace*{6.0in}
\hbox to\hsize{\hfill\includegraphics{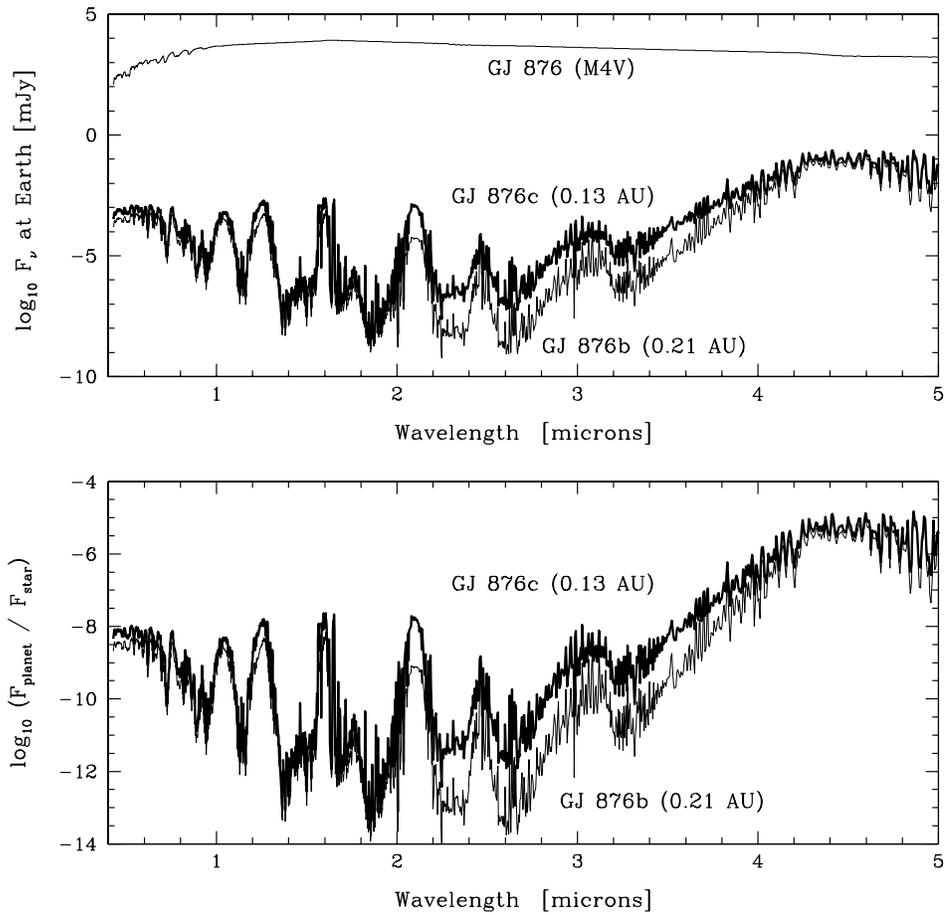}\kern+5.9in\hfill}
\caption{$Upper$ $panel$: Visible and near-infrared spectra of GJ 876b and c.  An
assumed spectrum of the primary (a scaled Kurucz model) is also
depicted.  Both GJ 876b and c are Class III EGPs.
$Lower$ $panel$: Wavelength-dependent, phase-averaged planet-to-star flux ratios for GJ 876 b and c.}
\label{fig_contrastGJ876}
\end{figure}

\newpage

\begin{figure} 
\vspace*{6.0in}
\hbox to\hsize{\hfill\includegraphics{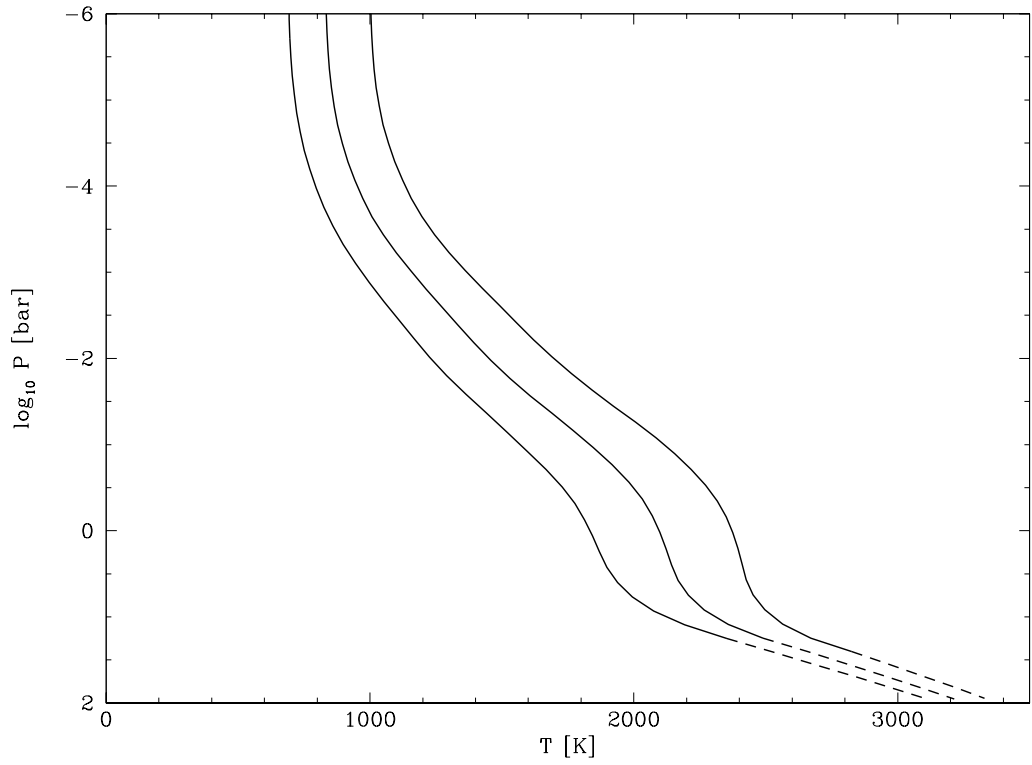}\kern+5.9in\hfill}
\caption{Cloud-free 51 Peg b T-P profiles with incident flux weighting of 1/4, 1/2,
and 1.  The coolest model shown has the flux weighting of 1/4.  See text for
details.}
\label{fig_TP51Pegbweight}
\end{figure}

\newpage

\begin{figure} 
\vspace*{6.0in}
\hbox to\hsize{\hfill\includegraphics{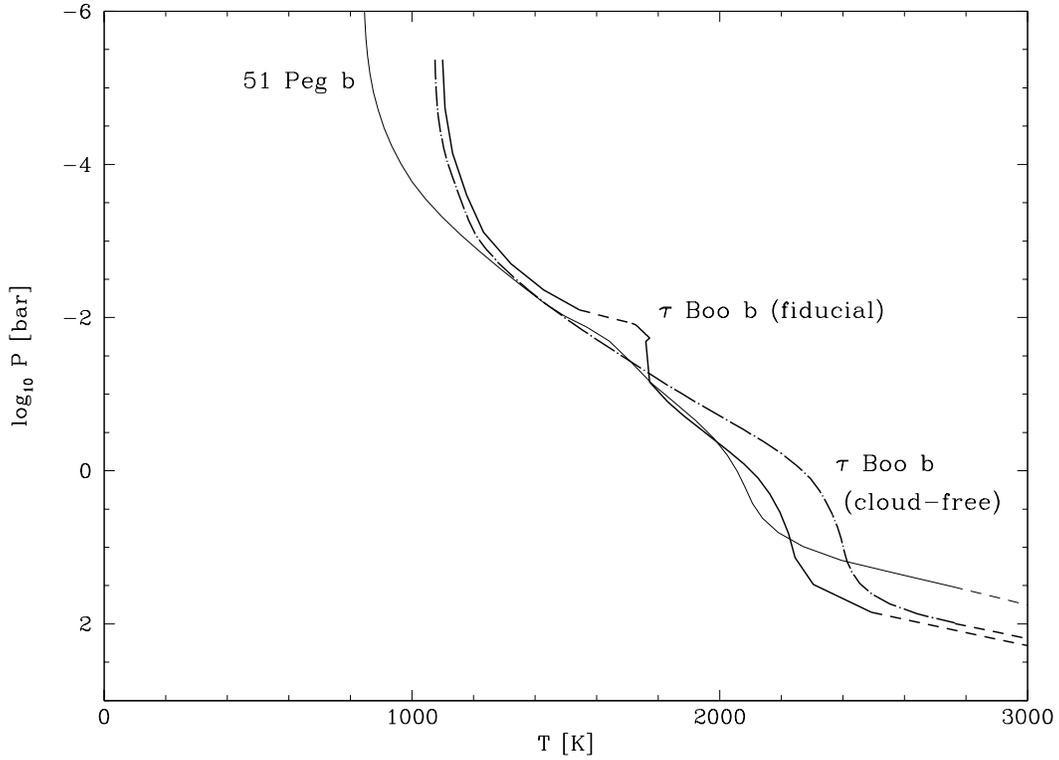}\kern+5.9in\hfill}
\caption{T-P profiles of $\tau$ Boo b and 51 Peg b.  The dashed portions
of the profiles indicate convective regions, and the fiducial model
of $\tau$ Boo b is convective at low pressures, toward the top of
the silicate cloud layer.  Note that in order to achieve numerical
convergence, the silicate and iron cloud layers of 51 Peg b have
been attenuated to 10\% of the elemental abundances of magnesium and
iron, respectively.}
\label{fig_TPPegBoo}
\end{figure}

\newpage

\begin{figure} 
\vspace*{6.0in}
\hbox to\hsize{\hfill\includegraphics{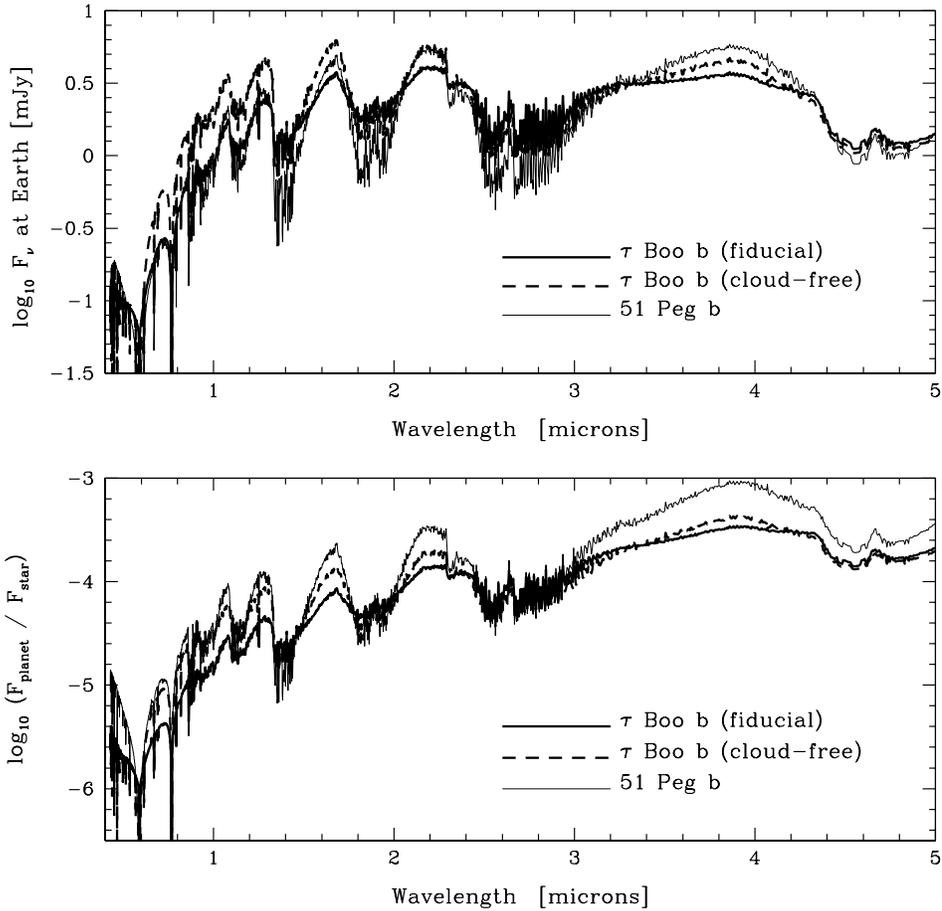}\kern+5.9in\hfill}
\caption{$Upper$ $panel$: Visible and near-infrared spectra of $\tau$ Boo b and 51 Peg b.
(Note that in order to achieve numerical
convergence, the silicate and iron cloud layers of 51 Peg b have
been attenuated to 10\% of the elemental abundances of magnesium and
iron, respectively, which may yield higher than actual fluxes in some
spectral regions.)
$Lower$ $panel$: Wavelength-dependent planet-to-star flux ratios for $\tau$ Boo b and
51 Peg b. } 
\label{fig_spectrumPegBoo}
\end{figure}

\newpage

\begin{figure} 
\vspace*{6.0in}
\hbox to\hsize{\hfill\includegraphics{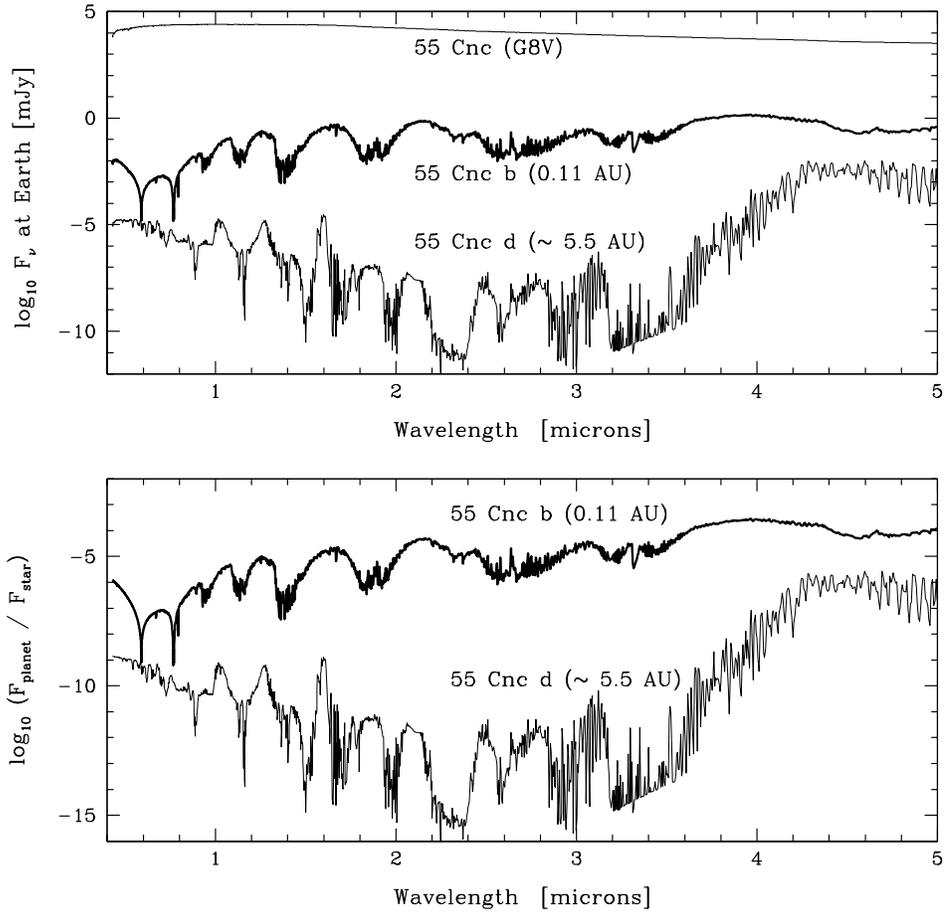}\kern+5.9in\hfill}
\caption{$Upper$ $panel$: Visible and near-infrared spectra of 55 Cancri b and d from
0.4 to 5 $\mu$m.  An assumed spectrum of the primary (a scaled Kurucz model) is also
depicted.  55 Cnc b and d are Class IV and II EGPs, respectively.
$Lower$ $panel$: Wavelength-dependent, phase-averaged planet-to-star flux ratios for 55 Cnc b and d.}
\label{fig_spectrum55Cnc}
\end{figure}

\newpage

\begin{figure} 
\vspace*{6.0in}
\hbox to\hsize{\hfill\includegraphics{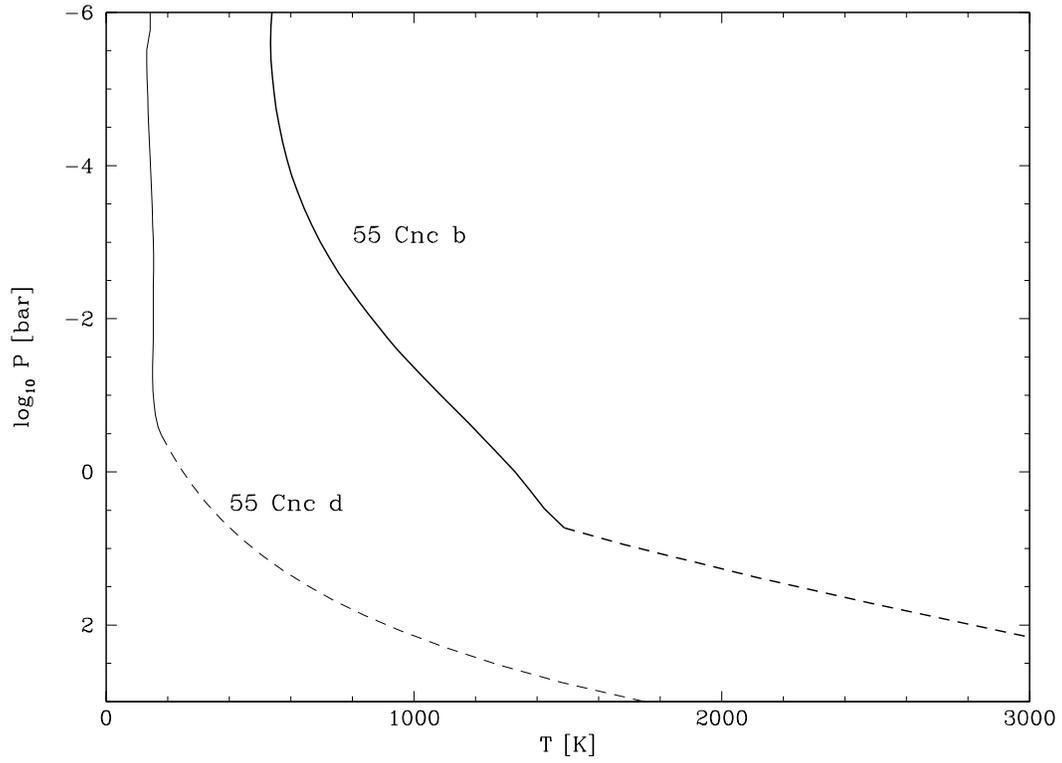}\kern+5.9in\hfill}
\caption{T-P profiles of 55 Cancri And b and d.  The dashed portions
of the profiles indicate convective regions.}
\label{fig_PT55Cnc}
\end{figure}

\newpage

\begin{figure} 
\vspace*{6.0in}
\hbox to\hsize{\hfill\includegraphics{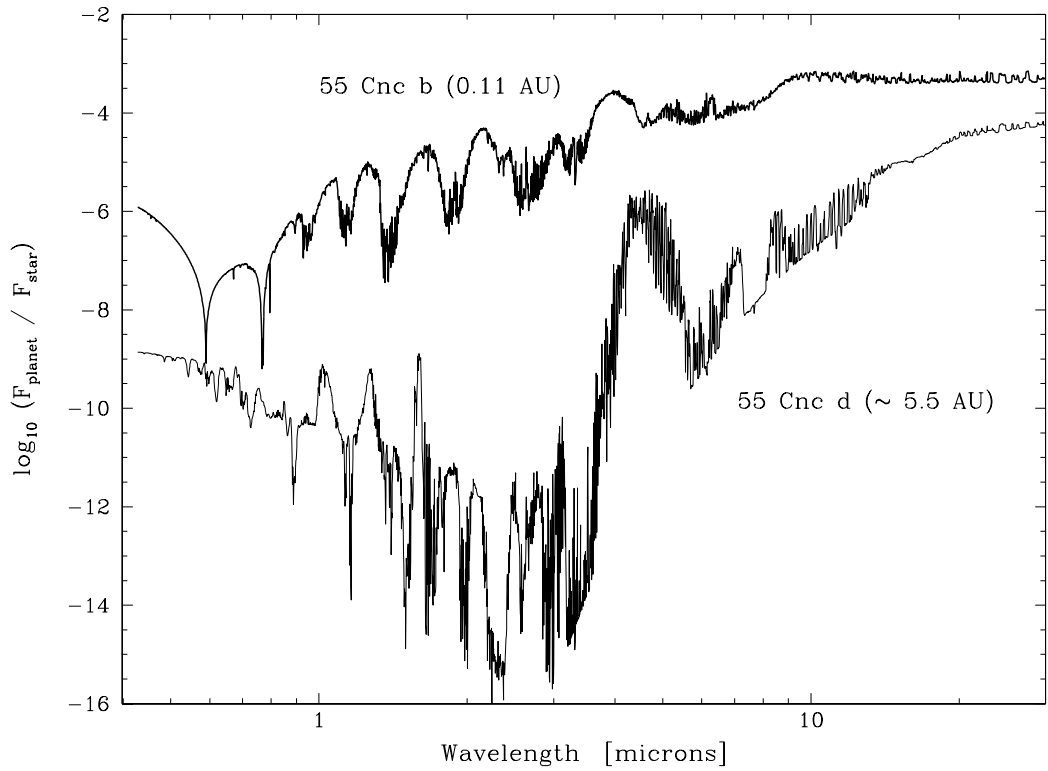}\kern+5.9in\hfill}
\caption{Wavelength-dependent, phase-averaged planet-to-star flux ratios for 55 Cnc b and d from 0.4 to 30 $\mu$m.}
\label{fig_contrast55Cnc30}
\end{figure}

%% file: plots3.tex
\newpage

\begin{figure}
\vspace*{6.0in}
\hbox to\hsize{\hfill\includegraphics{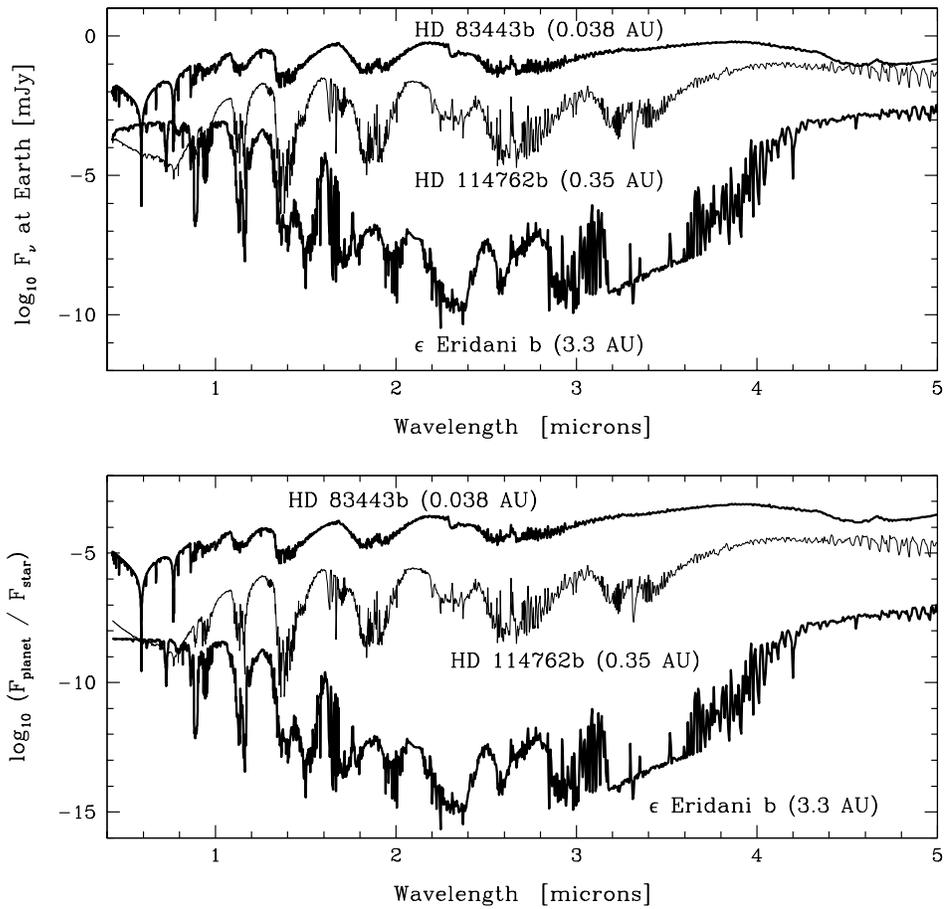}\kern+5.9in\hfill}
\caption{$Upper$ $panel$: Visible and near-infrared spectra of HD 83443b, HD 114762b, and
$\epsilon$ Eri b.  These planets are Class V, III, and II EGPs,
respectively.
$Lower$ $panel$: Wavelength-dependent planet-to-star flux ratios for HD 83443b,
HD 114762b, and $\epsilon$ Eri b.}
\label{fig_remainder}
\end{figure}

\newpage

\begin{figure}
\vspace*{6.0in}
\hbox to\hsize{\hfill\includegraphics{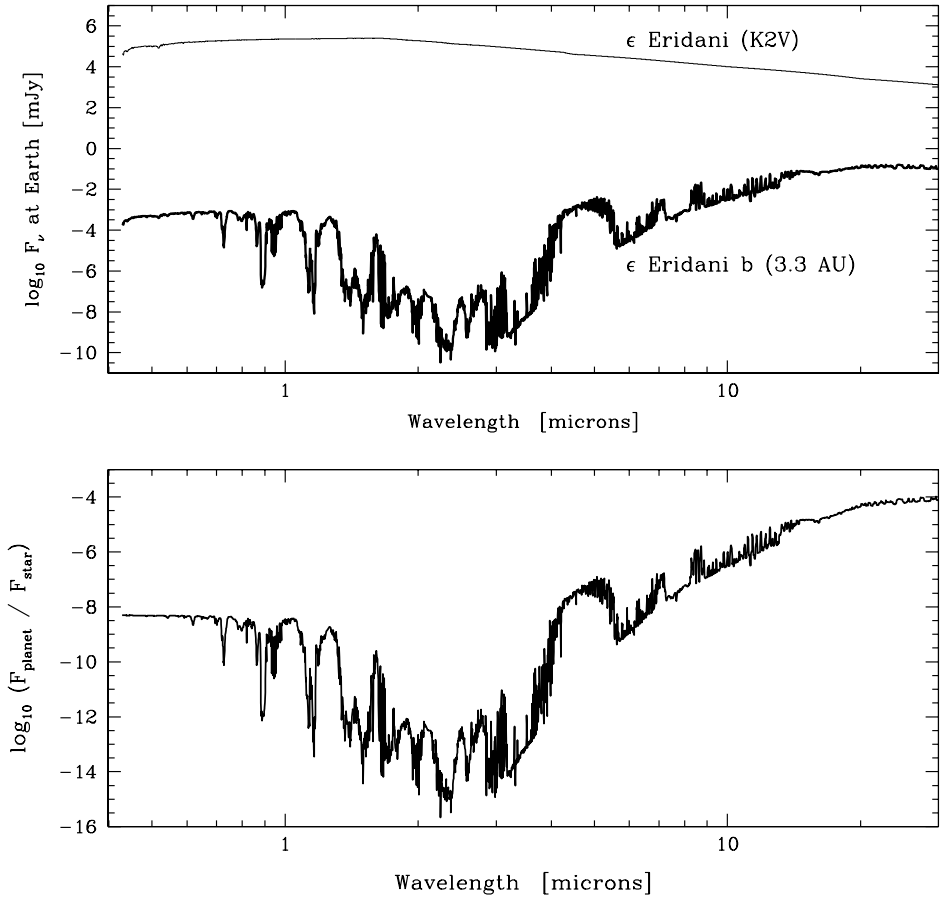}\kern+5.9in\hfill}
\caption{$Upper$ $panel$: Model spectrum of $\epsilon$ Eri b from 0.4 to 30 $\mu$m.  $\epsilon$
Eri b is a Class II EGP with a water cloud layer near a pressure of 1 bar.
For this fiducial model, a cloud particle size distribution peaked at 5 $\mu$m is used, and
10\% of the available H$_2$O is assumed to condense.
$Lower$ $panel$: Wavelength-dependent planet-to-star flux ratios for $\epsilon$
Eri b.}
\label{fig_spectrumEpsErib}
\end{figure}

\newpage

\begin{figure}
\vspace*{6.0in}
\hbox to\hsize{\hfill\includegraphics{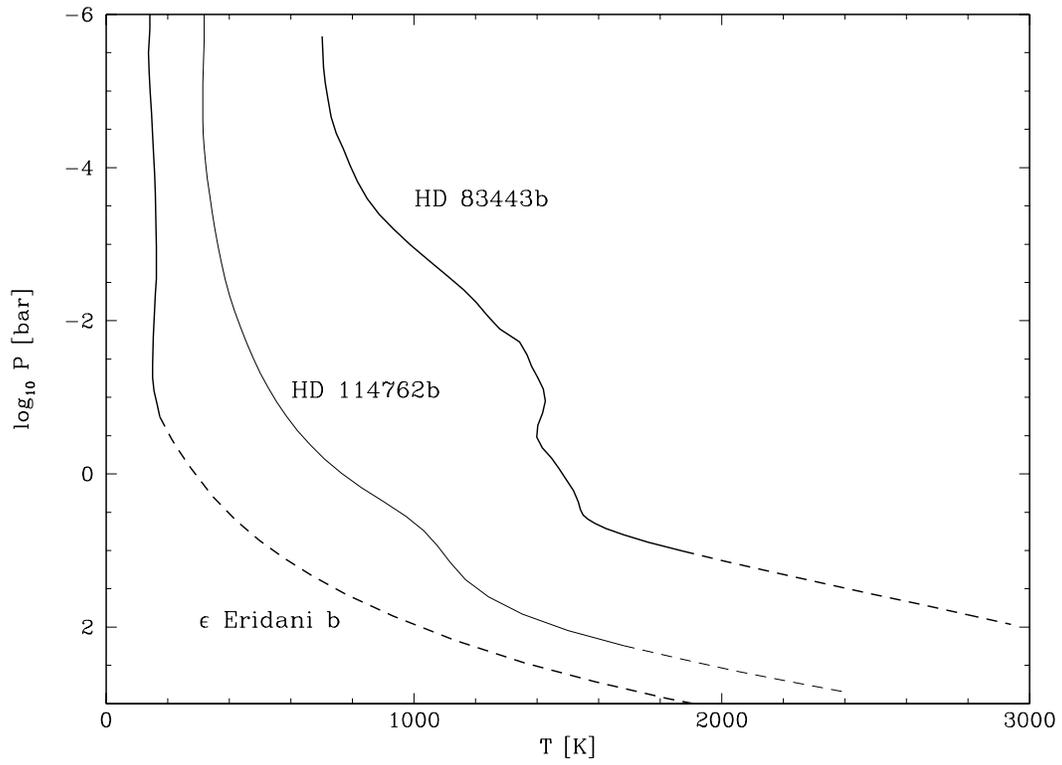}\kern+5.9in\hfill}
\caption{T-P profiles of HD 83443b, HD 114762b, and $\epsilon$ Eri b.  The dashed portions
of the profiles indicate convective regions.}
\label{fig_TPremainder}
\end{figure}